\documentclass[11pt,reqno]{amsart}
\usepackage{amsmath,amssymb,amsthm}
\usepackage[letterpaper,margin=1.2in]{geometry}
\usepackage[hidelinks]{hyperref}
\hypersetup{pdftitle={A proof of the Aaronson\textendash{}Ambainis conjecture},
  pdfauthor={David Turturean}}
\usepackage{etoolbox}
\usepackage[dvipsnames]{xcolor}
\newcommand{\modelname}{a non-public Anthropic model}

\makeatletter
\patchcmd{\@setauthors}{\endtrivlist}%
  {\par\vspace{3pt}{\normalfont Physical Superintelligence\par\texttt{david@psi.inc}\par}\endtrivlist}%
  {}{\PackageError{paper}{the author block of the title page could not be extended}{}}
\makeatother

\newtheorem{theorem}{Theorem}[section]
\newtheorem{lemma}[theorem]{Lemma}
\newtheorem{proposition}[theorem]{Proposition}
\newtheorem{corollary}[theorem]{Corollary}
\theoremstyle{definition}
\newtheorem{definition}[theorem]{Definition}
\newtheorem{example}[theorem]{Example}
\newtheorem*{definition*}{Definition}
\theoremstyle{remark}
\newtheorem{remark}[theorem]{Remark}
\numberwithin{equation}{section}

\newtheoremstyle{named}{9.5pt plus 3pt minus 2pt}{9.5pt plus 3pt minus 2pt}{\itshape}{}{\bfseries}{.}{.5em}{\thmnote{#3}}
\theoremstyle{named}
\newtheorem*{named}{}
\theoremstyle{plain}

\newcommand{\introhead}[1]{\par\medskip\noindent\textsc{#1.}\ }

\makeatletter
\appto\th@plain{\thm@preskip=9.5pt plus 3pt minus 2pt\relax\thm@postskip=\thm@preskip}
\appto\th@definition{\thm@preskip=9.5pt plus 3pt minus 2pt\relax\thm@postskip=\thm@preskip}
\appto\th@remark{\thm@preskip=9.5pt plus 3pt minus 2pt\relax\thm@postskip=\thm@preskip}
\renewcommand{\section}{\@startsection{section}{1}%
  \z@{1.3\linespacing\@plus.7\linespacing}{.65\linespacing}{\normalfont\scshape\centering}}
\renewcommand{\subsection}{\@startsection{subsection}{2}%
  \z@{1.0\linespacing\@plus.4\linespacing\@minus.15\linespacing}{-.5em}{\normalfont\bfseries}}
\makeatother

\newcommand{\R}{\mathbb{R}}
\newcommand{\Z}{\mathbb{Z}}
\newcommand{\E}{\mathbb{E}}
\newcommand{\Var}{\operatorname{Var}}
\newcommand{\Inf}{\operatorname{Inf}}
\newcommand{\I}{\operatorname{I}}
\newcommand{\Av}{\operatorname{Av}}
\newcommand{\ag}{\operatorname{ag}}
\newcommand{\Up}{\operatorname{Up}}
\newcommand{\cnt}{\operatorname{cnt}}
\newcommand{\sgn}{\operatorname{sgn}}
\newcommand{\cO}{\mathcal{O}}
\newcommand{\cN}{\mathcal{N}}
\newcommand{\one}{\mathbf{1}}
\newcommand{\eps}{\varepsilon}
\newcommand{\norm}[1]{\lVert #1\rVert_1}
\newcommand{\abs}[1]{\lvert #1\rvert}

\begin{document}

\title[A proof of the Aaronson--Ambainis conjecture]{A proof of the Aaronson--Ambainis conjecture}

\author{David Turturean}
\date{November 2026}


\begin{abstract}
Aaronson and Ambainis conjectured that every bounded polynomial of low degree on the discrete cube
has an influential variable. They showed that the conjecture implies that the acceptance probability
of every quantum algorithm making $T$ queries can be approximated on most inputs by a classical
algorithm making $\mathrm{poly}(T)$ queries. We prove their conjecture: every function
$f\colon\{-1,1\}^n\to[-1,1]$ of degree at most $d\ge1$ has a variable of influence at least
$c\,\Var(f)^{5/2}/d^{6}$, where $c>0$ is an absolute constant. In the original formulation, for
polynomials of degree at most $d$ with values in $[0,1]$ on $\{0,1\}^n$ and variance at least
$\eps$, this gives a variable of influence at least $(\eps/d)^{7}$. The proof is elementary, with
steps inspired by works of others. Thus, as Aaronson and Ambainis anticipated, superpolynomial quantum
speedups in the query model require structure, such as a promise on the inputs.
\end{abstract}

\maketitle

\clearpage
\vspace*{\fill}
\section*{AI disclosure}

The mathematical statements in this paper were produced by AI models.
David Turturean scoped the problem, developed a research harness internally at Physical
Superintelligence (PSI), and ran it on PSI's infrastructure. The run that produced this mathematical
approach and proof was done with instances of
\modelname, serving within the harness to search for existing ideas, to propose new ideas, to combine
and test these ideas, and to supervise the work and the communication between other instances.
Appendix~\ref{app:timeline} gives the timeline of the discovery.
The proof of Theorem~\ref{thm:main} and of Corollary~\ref{cor:AA01} has been formalized in the
Lean proof assistant and checked by its kernel; the formalization is in the repository
\url{https://github.com/physical-superintelligence/aaronson-ambainis}.

The author prompted an instance of \modelname{} to package the results into a first draft of this
paper, reviewed and edited the draft personally, and is responsible for its correctness.

\section*{Acknowledgements}

We thank Jim Halverson for discussions, auditing the proof, and comments on a draft, and Phiala Shanahan for comments on the
Lean formalization. We are especially grateful to 
Guy Blanc, Jordan Docter, Carmen Strassle and Li-Yang Tan for their comments and for writing a forthcoming
shorter exposition of the proof for experts, and to Scott Aaronson for referring us to them.

\vspace*{\fill}

\newpage

{\small\tableofcontents}

\linespread{1.08}\selectfont
\section{Introduction}\label{sec:intro}

Which problems can a quantum algorithm solve with exponentially fewer queries to its input than any
classical algorithm? For decision problems this is possible only if the input is subject to a
promise, since for total Boolean functions the quantum and the classical query complexity are
polynomially related \cite{BBCMW}. In Simon's problem \cite{Sim} and in Shor's period finding
\cite{Sho}, for instance, the input is promised to encode a periodic function, a structure that only a
minuscule fraction of all inputs have \cite{AA14}.

Aaronson and Ambainis made precise the idea that exponential speedups need such structure, through a
conjecture which they describe as folklore:

\begin{named}[Conjecture 1.5 {\normalfont(folklore) \cite{AA14}}]
Let $Q$ be a quantum algorithm that makes $T$ queries to a Boolean input $X=(x_1,\dots,x_N)$, and
let $\eps>0$. Then there exists a deterministic classical algorithm that makes
$\mathrm{poly}(T,1/\eps,1/\delta)$ queries to the $x_i$, and that approximates $Q$'s acceptance
probability to within an additive error $\eps$ on a $1-\delta$ fraction of inputs.
\end{named}
Aaronson
and Ambainis show that it is equivalent to the following statement \cite[Theorem~7.3]{AA14}: if a
problem has a fast quantum algorithm and its promise is satisfied by a constant fraction of all
inputs, then a fast classical algorithm solves it on most of these inputs.

Aaronson and Ambainis also reduced this conjecture to a statement about polynomials that does not
mention quantum algorithms: bounded polynomials of low degree have influential variables. This is the
Aaronson--Ambainis conjecture, which Section~\ref{subsec:conj} states in their words. The acceptance
probability of a quantum algorithm that makes $T$ queries is such a polynomial, of degree at most $2T$
in the bits of its input and with values in $[0,1]$ \cite{BBCMW}. If the conjecture holds, a classical
algorithm can simulate the quantum algorithm on most inputs by querying a variable of large influence,
fixing it, and repeating \cite[Theorem~3.3]{AA14}.

The conjecture goes back to a lecture of Aaronson in 2008 \cite{Aar08}. O'Donnell's book on the
analysis of Boolean functions states it in the notes to Chapter~8 and observes that ``if true, this
conjecture would have significant consequences for the limitations of efficient quantum
computation'' \cite[Section~8.7]{OD}. Before the present paper, the known bounds for general bounded
polynomials were exponential in the degree \cite{AA14,DFKO,DMP,OZ}, and the conjecture had been
proved only for special classes of polynomials; Section~\ref{subsec:earlier} describes this earlier
work. An elegant attempt at the conjecture was announced in \cite{KK19}, which unfortunately was later
withdrawn. We take a different approach.

In this paper we prove the conjecture of Aaronson and Ambainis: every function $f$ on $\{-1,1\}^{n}$
with values in $[-1,1]$ and of degree at most $d\ge1$ has a variable $j$ with
\[
   \Inf_j(f)\;\ge\;c\,\frac{\Var(f)^{5/2}}{d^{6}},
\]
where $\Inf_j(f)$ is the influence of the $j$th variable and $c>0$ is an absolute constant
(Theorem~\ref{thm:main}). In the normalisation of \cite{AA14}, a polynomial of degree at most $d$
with values in $[0,1]$ on $\{0,1\}^{n}$ and variance at least $\eps$ has a variable of influence at
least $(\eps/d)^{7}$ (Corollary~\ref{cor:AA01}). The proof is elementary, with steps inspired by work of Tal \cite{Tal}; of Gilmer, Saks and
Srinivasan \cite{GSS}; of Agarwal and Ben-David \cite{ABD}; of Reimer \cite{Rei}; and of Lovett and
Zhang \cite{LZ}.

Consequently the folklore conjecture holds: the acceptance probability of every quantum algorithm
that makes $T$ queries can be approximated to within $\eps$ on all but a $\delta$ fraction of the
inputs by a deterministic classical algorithm that makes at most $(4T)^{8}\eps^{-14}\delta^{-8}$
queries (Section~\ref{subsec:motivation}).
The other consequences that Aaronson and Ambainis derive
from their conjecture hold as well. For instance, separating $\mathsf{BQP}$ from $\mathsf{AvgP}$, the
class of problems solvable in polynomial time on most inputs, relative to a random oracle would
require a proof that $\mathsf{P}\neq\mathsf{P}^{\#\mathsf{P}}$ \cite[Theorem~3.5]{AA14}.

 We now state the conjecture and our results precisely;
Section~\ref{subsec:overview} outlines the proof.

\subsection{The conjecture}\label{subsec:conj}

For a real polynomial $p\colon\R^{N}\to\R$, Aaronson and Ambainis \cite{AA14} let
\[
   \Inf_i[p]\;:=\;\E_{X\in\{0,1\}^{N}}\bigl[(p(X)-p(X^{i}))^{2}\bigr]
\]
be the influence of the $i$th variable, where $X^{i}$ means $X$ with the $i$th bit flipped, and they state
the following conjecture \cite[Conjecture~1.7]{AA14}, which we quote.

\begin{quote}\sloppy
\textbf{Conjecture 1.7} (Bounded Polynomials Have Influential Variables)\textbf{.}
\emph{Let $p\colon\R^{N}\to\R$ be a polynomial of degree $d$. Suppose that $0\le p(X)\le 1$ for all
$X\in\{0,1\}^{N}$, and
\[
   \E_{X\in\{0,1\}^{N}}\bigl[(p(X)-\E[p])^{2}\bigr]\;\ge\;\eps .
\]
Then there exists an $i$ such that $\Inf_i[p]\ge(\eps/d)^{O(1)}$.}
\end{quote}

Corollary~\ref{cor:AA01} below proves it with the explicit exponent $7$ in place of $O(1)$.

\subsection{Results}\label{subsec:results}

It is convenient to work on the cube $Q_n=\{-1,1\}^{n}$ with the uniform probability measure; $\E$ and
$\Var$ denote expectation and variance with respect to it.

For $x\in Q_n$ and $j\in[n]=\{1,\dots,n\}$ let
$x^{j}$ be the point obtained from $x$ by changing the sign of the $j$th coordinate. For
$f\colon Q_n\to\R$ put
\[
   \Inf_j(f)\;:=\;\E_x\Bigl[\Bigl(\frac{f(x)-f(x^{j})}{2}\Bigr)^{2}\Bigr].
\]

Every function $f\colon Q_n\to\R$ can be written in exactly one way as
$f=\sum_{T\subseteq[n]}\hat f(T)\chi_T$ with real coefficients $\hat f(T)$, where
$\chi_T(x)=\prod_{i\in T}x_i$.
We say that $f$ has \emph{degree at most $d$} if $\hat f(T)=0$ whenever $\abs{T}>d$.

\begin{theorem}[Main theorem]\label{thm:main}
Let $n\ge 1$ and $d\ge 1$ be integers and let $f\colon\{-1,1\}^{n}\to[-1,1]$ be a function of degree at
most $d$. Then there is a coordinate $j\in[n]$ with
\[
   \Inf_j(f)\;\ge\;\frac{\sqrt{15}}{500\,\pi^{4}}\cdot\frac{\Var(f)^{5/2}}{d^{6}} .
\]
\end{theorem}

\introhead{Remarks on Theorem~\ref{thm:main}}If $\Var(f)=0$ there is nothing to prove, and a function of degree at most $0$ is constant; so the content of the
theorem concerns non-constant functions, and for these the assumption $d\ge 1$ is automatic.

The constant $\sqrt{15}/(500\pi^{4})$ is approximately $7.95\cdot10^{-5}$.

It can be multiplied
by $4$ if the bound of Agarwal and Ben-David \cite{ABD} for the fractional block sensitivity
(recalled in Remark~\ref{rem:ABD}) is used in place of
Theorem~\ref{thm:row} below, and by a further factor $2$ if an inequality of Filmus, Hatami,
Keller and Lifshitz \cite{FHKL} is used in place of Corollary~\ref{cor:sens-point} below; see
Remark~\ref{rem:constant}.

\introhead{The original normalisation}The next corollary is the statement in the normalisation of \cite{AA14}. We write $\Var[p]$ for the
variance of $p(X)$ when $X$ is uniform on $\{0,1\}^{n}$.

\begin{corollary}\label{cor:AA01}
Let $d\ge0$ be an integer, and let $p\colon\R^{n}\to\R$ be a polynomial of degree at most $d$ such
that $0\le p(X)\le 1$ for all $X\in\{0,1\}^{n}$ and $\Var[p]>0$. Then $n\ge 1$, $d\ge 1$, and there
exists an $i$ such that
\[
   \Inf_i[p]\;\ge\;\frac{8\sqrt{15}}{125\,\pi^{4}}\cdot\frac{\Var[p]^{5/2}}{d^{6}}
   \;\ge\;\Bigl(\frac{\Var[p]}{d}\Bigr)^{7}.
\]
In particular, Conjecture~1.7 of \cite{AA14} holds in the following form: if\/
$\Var[p]\ge\eps>0$, then there exists an $i$ such that
\[
   \Inf_i[p]\;\ge\;\frac{8\sqrt{15}}{125\,\pi^{4}}\cdot\frac{\eps^{5/2}}{d^{6}}
   \;\ge\;\Bigl(\frac{\eps}{d}\Bigr)^{7}.
\]
\end{corollary}

\introhead{Remarks on Corollary~\ref{cor:AA01}}The corollary follows from Theorem~\ref{thm:main} by the change of variables $X_i=(1-x_i)/2$ and a
numerical estimate; the proof is given at the end of Section~\ref{sec:avg}.

The exponent $7$ only
absorbs the constant: since $\Var[p]\le\frac14$, we have
$\Var[p]^{5/2}/d^{6}\ge128\,(\Var[p]/d)^{6}$, and
$128\cdot8\sqrt{15}/(125\pi^{4})>\frac14\ge\Var[p]/d$.

\introhead{Rows and mean absolute increments}The proof of Theorem~\ref{thm:main} uses two properties of the function $f$, which are expressed
with the following notions. For $S\subseteq[n]$ let $x^{S}$ be the point obtained from $x$ by
changing the signs of the coordinates in $S$.

\begin{definition*}[Rows]
A \emph{row} of a function $f\colon Q_n\to\R$ at a
point $x\in Q_n$ is a vector $a\in[0,\infty)^{n}$ such that
\[
   \abs{f(x)-f(x^{S})}\;\le\;a(S):=\sum_{i\in S}a_i\qquad\text{for every }S\subseteq[n].
\]
The number $\norm{a}=\sum_ia_i$ is the \emph{cost} of the row.
\end{definition*}

\begin{definition*}[Mean absolute increments]
The \emph{mean absolute increment}
of $f$ in the coordinate $j$ is
\[
   L_j(f)\;:=\;\E_x\abs{f(x)-f(x^{j})},
\]
and $L(f)=\sum_{j=1}^{n}L_j(f)$.
\end{definition*}

The number $\frac12L_j(f)$ is known as the $L_1$ influence of the
coordinate $j$ on $f$ \cite{BB,FHKL}. For a polynomial $p$ on $\{0,1\}^{N}$, Aaronson and Ambainis
write $\Inf^{1}_i[p]=\E\abs{p(X)-p(X^{i})}$ for the corresponding quantity
\cite[Section~1.3]{AA14}.

If a row $a^{x}$ is chosen at every point $x\in Q_n$, the vectors $a^{x}$ are the rows of a matrix
with non-negative entries, with one row for each point of $Q_n$ and one column for each coordinate;
hence the name, and the cost of a row is its row sum. Taking $S=\{j\}$ in the definition of a row
gives $a^{x}_j\ge\abs{f(x)-f(x^{j})}$, so the mean of the $j$th column is at least $L_j(f)$.

A function $f\colon Q_n\to[-1,1]$ of degree at most $d$ has a row of cost at most $\pi^{2}d^{2}$
at every point (Theorem~\ref{thm:row}), and it satisfies $L(f)\le4d^{2}$
(Corollary~\ref{cor:sens-point}). This row bound is due to Agarwal and Ben-David
\cite[Theorem~1.7]{ABD}, with the better constant $\pi^{2}/2$ in place of $\pi^{2}$
(see Remark~\ref{rem:ABD}). That $L(f)$ is
bounded by a polynomial in $d$ was first proved by Ba\v{c}kurs and Bavarian \cite{BB}, and Filmus,
Hatami, Keller and Lifshitz proved $L(f)\le2d^{2}$ \cite[Theorem~3.3]{FHKL}.

The next two theorems concern arbitrary real functions with a
row of bounded cost at every point; no bound for the values of $f$ is assumed in them.

\begin{theorem}[Mean absolute increments]\label{thm:incr-intro}
Let $n\ge0$, let $f\colon Q_n\to\R$ be a function with $\Var(f)>0$, and let $K$ be a real number
such that for every $x\in Q_n$ there is a row $a^{x}$ of $f$ at $x$ with $\norm{a^{x}}\le K$. Then
$n\ge1$ and $K>0$, and there is a coordinate $j\in[n]$ with
\[
   \E_x\abs{f(x)-f(x^{j})}\;\ge\;\frac{\Var(f)^{3/2}}{6\,K^{2}} .
\]
\end{theorem}

\begin{theorem}[Influences]\label{thm:infl-intro}
Let $n\ge0$, let $f\colon Q_n\to\R$ be a function with $\Var(f)>0$, and let $K$ be a real number
such that for every $x\in Q_n$ there is a row $a^{x}$ of $f$ at $x$ with $\norm{a^{x}}\le K$. Let
$L(f)=\sum_{k=1}^{n}L_k(f)$. Then $n\ge1$ and $0<L(f)\le K$, and there is a coordinate $j\in[n]$
with
\[
   \Inf_j(f)\;\ge\;\frac{\sqrt{15}}{125}\cdot\frac{\Var(f)^{5/2}}{L(f)\,K^{2}} .
\]
In particular, $\Inf_j(f)\ge(\sqrt{15}/125)\,\Var(f)^{5/2}/K^{3}$ for this coordinate $j$.
\end{theorem}

\introhead{Remarks on Theorems~\ref{thm:incr-intro} and~\ref{thm:infl-intro}}These are Theorems~\ref{thm:incr} and~\ref{thm:infl} below, where the inequalities are also stated
in a form without square roots.

Theorem~\ref{thm:main} follows from
Theorem~\ref{thm:infl-intro} with $K=\pi^{2}d^{2}$ (Theorem~\ref{thm:row}) and $L(f)\le4d^{2}$
(Corollary~\ref{cor:sens-point}).

\introhead{Boolean functions}Theorems~\ref{thm:incr-intro} and~\ref{thm:infl-intro} are derived from an inequality for Boolean functions with a row at every point,
the all-pairs inequality (Proposition~\ref{prop:pairs}), which is described in Section~\ref{subsec:overview}. For Boolean
functions themselves this inequality has a further consequence, Theorem~\ref{thm:two-intro} below.
For a Boolean function $g\colon Q_n\to\{0,1\}$ the difference $\abs{g(x)-g(x^{S})}$ is $1$ or $0$,
so a row of $g$ at $x$ is a vector $\lambda\in[0,\infty)^{n}$ with $\lambda(S)\ge1$ for every set $S$
that changes the value of $g$, that is, with $g(x^{S})\ne g(x)$.

Rows of Boolean functions are also called fractional certificates \cite{Tal,GSS,KT}: the rows with
entries in $\{0,1\}$ are exactly the
indicator vectors of the certificates of $g$ at $x$, that is, of the sets $C\subseteq[n]$ such that $g$
is constant on $\{y : y_i=x_i \text{ for all } i\in C\}$.

We write $\I_j(g)=\Pr_x[g(x)\ne g(x^{j})]$ for the influence of the
coordinate $j$ on the Boolean function $g$; thus $\I_j(g)=4\Inf_j(g)$.

Here and below,
$1_{\{g(x)=1\}}$ denotes the number which is $1$ if $g(x)=1$ and $0$ otherwise, and similarly for other
conditions in place of $g(x)=1$.

\begin{theorem}[Two-sided inequality]\label{thm:two-intro}
Let $g\colon Q_n\to\{0,1\}$ be non-constant, and for every $x\in Q_n$ let $\lambda^{x}$ be a row
of $g$ at $x$. Then
\[
   4\;\E_x\Bigl[\frac{1_{\{g(x)=1\}}}{\norm{\lambda^{x}}}\Bigr]\;
      \E_x\Bigl[\frac{1_{\{g(x)=0\}}}{\norm{\lambda^{x}}}\Bigr]\;\le\;\max_{j\in[n]}\I_j(g).
\]
\end{theorem}

\introhead{Remarks on Theorem~\ref{thm:two-intro}}This is Theorem~\ref{thm:two} below.

There is
equality if $g(x)=1$ exactly when $x_1=1$ and
$\lambda^{x}=(1,0,\dots,0)$ for all $x$ (Example~\ref{ex:dictator}).

We call the inequality two-sided because the sets $g^{-1}(1)$ and $g^{-1}(0)$ enter in the same
way.

For a non-constant Boolean function $g$ with a row of cost at most $K$ at every point it gives
a coordinate $j$ with $\Inf_j(g)\ge\Var(g)/K^{2}$ (Section~\ref{subsec:open}). If every point of
$g^{-1}(1)$ has a certificate of size at most $c_1$ and every point of $g^{-1}(0)$ one of size at
most $c_0$, the theorem, applied to the indicator vectors of these certificates, gives
$\max_j\I_j(g)\ge4\Pr[g=1]\Pr[g=0]/(c_0c_1)$; this special case also follows from the inequality of O'Donnell, Saks, Schramm and Servedio \cite{OSSS} for decision trees and the
classical bound $D(g)\le c_0c_1$ for the least depth $D(g)$ of a decision tree that computes $g$.
Section~\ref{subsec:cert} compares
Theorem~\ref{thm:two-intro} with the inequality of \cite{OSSS}.

\subsection{Motivation and consequences}\label{subsec:motivation}

The conjecture comes from quantum query complexity. By a lemma of Beals, Buhrman, Cleve, Mosca and
de~Wolf \cite{BBCMW} (stated as Lemma~3.2 in \cite{AA14}), the acceptance probability of a quantum
algorithm that makes $T$ queries to a Boolean input $X\in\{0,1\}^{N}$ is a real multilinear polynomial
in $X$ of degree at most $2T$, and it takes values in $[0,1]$.

Aaronson and Ambainis relate their
Conjecture~1.7 to the following conjecture, which they describe as folklore
\cite[Conjecture~1.5]{AA14}:

\begin{quote}
\emph{Let $Q$ be a quantum algorithm that makes $T$ queries to a Boolean input $X=(x_1,\dots,x_N)$, and
let $\eps>0$. Then there exists a deterministic classical algorithm that makes
$\mathrm{poly}(T,1/\eps,1/\delta)$ queries to the $x_i$, and that approximates $Q$'s acceptance
probability to within an additive error $\eps$ on a $1-\delta$ fraction of inputs.}
\end{quote}

Aaronson and Ambainis derive three consequences from Conjecture~1.7 \cite[Theorem~1.8]{AA14}. To
state them, let $D_{\eps}(f)$ be the minimum number of queries made by a deterministic
algorithm that evaluates a Boolean function $f\colon\{0,1\}^{N}\to\{0,1\}$ correctly on at least a
$1-\eps$ fraction of the inputs. Let $Q_{\eps}(f)$ be the corresponding minimum for quantum
algorithms, which evaluate $f$ with bounded error on such a fraction of the inputs, as in the proof of
\cite[Corollary~3.4]{AA14}.

Finally, let
$\mathsf{AvgP}$ be the class of languages for which there exists a polynomial-time
algorithm that solves a $1-o(1)$ fraction of the instances of each size
\cite[Section~1.1]{AA14}.

Theorem~1.8 of \cite{AA14} reads:

\begin{quote}
\emph{Assume Conjecture~1.7. Then}

\smallskip
\emph{\textup{(i)} Conjecture~1.5 holds.}

\smallskip
\emph{\textup{(ii)} $D_{\eps+\delta}(f)\le(Q_{\eps}(f)/\delta)^{O(1)}$ for all Boolean functions
$f\colon\{0,1\}^{N}\to\{0,1\}$ and all $\eps,\delta>0$.}

\smallskip
\emph{\textup{(iii)} If\/ $\mathsf{P}=\mathsf{P}^{\#\mathsf{P}}$, then
$\mathsf{BQP}^{A}\subset\mathsf{AvgP}^{A}$ with probability $1$ for a random oracle $A$.}
\end{quote}

\noindent These three implications are proved in \cite{AA14} as Theorem~3.3, Corollary~3.4 and
Theorem~3.5.

Because we prove Conjecture~1.7 (Corollary~\ref{cor:AA01}), the three statements (i), (ii) and
(iii) hold.

For (i), the proof of \cite[Theorem~3.3]{AA14} with the exponent $7$ gives a deterministic classical
algorithm that makes at most $(4T)^{8}\eps^{-14}\delta^{-8}$ queries. Indeed, the acceptance
probability has degree at most $2T$, so its total influence is at most $2T$ \cite[Lemma~3.1]{AA14};
as long as the variance of the polynomial obtained by fixing the bits queried so far exceeds
$\eps^{2}\delta/2$, Corollary~\ref{cor:AA01} gives a variable of influence greater than
$(\eps^{2}\delta/(4T))^{7}$, which the algorithm queries next; and by Markov's inequality the
algorithm stops within $4T/(\delta(\eps^{2}\delta/(4T))^{7})=(4T)^{8}\eps^{-14}\delta^{-8}$ queries,
except with probability at most $\delta/2$.

As is noted in \cite[Section~1.1]{AA14}, statement (ii) is a quantum counterpart of a theorem
of Smyth, which bounds the approximate decision-tree complexity of a Boolean function by a quantity
that is quadratic in its approximate certificate complexity \cite{Smy02,Smy}.

Girish, Mittal, Nehoran and Raz \cite{GMNR} replace the uniform distribution on the inputs by other
distributions: for a fixed $p\in(0,1)$ and every $n\ge1$ they let $\mu_p$ be the product distribution
on $\{0,1\}^{n}$ under which the coordinates are independent and each is equal to $1$ with
probability $p$, and $\nu_p$ the uniform distribution on the strings in $\{0,1\}^{n}$ of Hamming
weight $\lfloor pn\rfloor$. Let $\mathcal{D}$ stand for $\mu_p$ or for $\nu_p$.

In our notation, the statement $\mathrm{AA}(\mathcal{D})$ of \cite[Definition~4.1]{GMNR} reads as
follows.

\begin{quote}
There are
constants $c,C>0$ and a function $S\colon(0,\frac14]\times\{1,2,\dots\}\to[0,1]$ with
$S(v,d)\ge c\,(v/d)^{C}$ such that for all $n,d\ge 1$ and all $v\in(0,\frac14]$, every
$f\colon\{0,1\}^{n}\to[0,1]$ that is given by a multilinear polynomial of degree at most $d$ and whose
variance with respect to $\mathcal{D}$ is at least $v$ has a coordinate $i$ with
\[
   \E_{X\sim\mathcal{D}}\bigl[(f(X)-f(X^{i}))^{2}\bigr]\;\ge\;S(v,d).
\]
\end{quote}

It is proved in
\cite[Theorem~4.3]{GMNR} that the statements $\mathrm{AA}(\mu_p)$ and $\mathrm{AA}(\nu_q)$ are
equivalent for all $p,q\in(0,1)$.

It is proved in \cite[Theorem~4.14]{GMNR} that each of them
implies the
following, again in our notation, for every $p\in(0,1)$, for $\mathcal{D}=\mu_p$ as well as for
$\mathcal{D}=\nu_p$.

\begin{quote}
There is a
polynomial $P$ such that for all $n,T\ge1$, all $\eps\in(0,1)$ and every quantum algorithm that makes
at most $T$ queries to an input $X\in\{0,1\}^{n}$, with acceptance probability $Q(X)$, there is a
deterministic classical algorithm that makes at most $P(T,1/\eps)$ queries and whose output
$\widetilde{Q}(X)\in[0,1]$ satisfies
\[
   \E_{X\sim\mathcal{D}}\abs{\widetilde{Q}(X)-Q(X)}\;\le\;\eps .
\]
\end{quote}

The statement
$\mathrm{AA}(\mu_{1/2})$ concerns the uniform distribution, and it holds by
Corollary~\ref{cor:AA01}, with $S(v,d)=(v/d)^{7}$.

Hence, by \cite[Theorem~4.3]{GMNR}, the
statements $\mathrm{AA}(\mu_p)$ and $\mathrm{AA}(\nu_p)$ hold for every $p\in(0,1)$, and so does the
conclusion of \cite[Theorem~4.14]{GMNR}.

In summary, by Corollary~\ref{cor:AA01} the statements (i)--(iii) above hold, and for every
$p\in(0,1)$ so do $\mathrm{AA}(\mu_p)$, $\mathrm{AA}(\nu_p)$ and the simulation of
\cite[Theorem~4.14]{GMNR}. Moreover, by \cite[Claim~A.1]{BDST}, Corollary~\ref{cor:AA01} implies the conjecture
of Blanc, Docter, Strassle and Tan on query weights \cite[Conjecture~1]{BDST}; so their
Theorem~11 holds without that assumption: if $\mathsf{PromiseBPP}=\mathsf{PromiseBQP}$, then
$\mathsf{PromiseBQP}^{A}\subseteq\mathsf{PromiseHeurBPP}^{A}$ with probability $1$ for a random
oracle $A$ \cite[Theorem~11]{BDST}.

\subsection{Earlier work}\label{subsec:earlier}

We recall what was known, in the notation of the papers cited.

\medskip
\emph{Boolean functions.} O'Donnell, Saks, Schramm and Servedio \cite{OSSS} proved, in the formulation
of \cite[Theorem~1.9]{AA14}, that a Boolean function $f\colon\{0,1\}^{N}\to\{0,1\}$ with
$\Pr[{f=1}]\allowbreak\Pr[{f=0}]\ge\eps$ has a variable with $\Inf_i[f]\ge 4\eps/D(f)$, where $D(f)$ is the
decision-tree complexity of $f$.

By a theorem of Nisan and Szegedy \cite{NS} the decision-tree complexity
of a Boolean function is bounded by a polynomial in its degree, so the Aaronson--Ambainis conjecture was known for
polynomials that take only the values $0$ and $1$ on the cube; see the introduction of \cite{LZ}.

As
is pointed out in \cite{AA14}, this does not settle the case of bounded polynomials, because the
acceptance probability of a quantum algorithm need not approximate a total Boolean function.

\medskip
\emph{Exponential dependence on the degree.} As observed in \cite{AA14}, a theorem of Dinur, Friedgut,
Kindler and O'Donnell \cite{DFKO} implies the conjecture with the weaker bound
$\Inf_i[p]\ge\eps^{3}/2^{O(d)}$ in place of $(\eps/d)^{O(1)}$.

Defant, Masty{\l}o and P\'erez
\cite{DMP} derive the bound
\[
   \max_{j}\Inf_j(f)\;\ge\;\frac{\Var(f)^{2}}{d^{4}e^{4d}}
\]
for functions $f\colon\{-1,1\}^{n}\to[-1,1]$ of degree $d$; see also \cite{OZ}.

\medskip
\emph{Special classes.} Montanaro \cite{Mon} proved the Aaronson--Ambainis conjecture for multilinear forms all of whose
coefficients have the same absolute value; here a multilinear form is a polynomial in $k$ blocks of
variables in which every monomial is a product of $k$ variables, one from each block (see also
\cite[Section~1.3]{AA14}).

Ivanisvili \cite{Iva} proved it for polynomials whose Fourier coefficients
have the same absolute value on each level, as an application of inequalities of
Bohnenblust--Hille type on the cube whose constants grow polynomially in the degree; see also
\cite{STV}, and \cite{DMP} for the relation between such inequalities and the conjecture.

Bansal, Sinha and de~Wolf \cite{BSW} proved that every completely bounded block-multilinear form
of degree $d$ with constant variance has a variable with influence at least $1/\mathrm{poly}(d)$.
Escudero Guti\'errez \cite{EG} gave another proof of this result and proved the conjecture for
homogeneous Fourier completely bounded polynomials.

O'Donnell and Zhao \cite{OZ} showed that the conjecture holds if and only if it
holds for one-block decoupled functions.

Bhattacharya \cite[Theorem~6.4]{Bha} showed that the conjecture is true for
a non-negligible fraction of the random restrictions of a given polynomial, assuming that its variance
is not too low.

\medskip
\emph{Restricted quantum algorithms.} Two recent papers prove statements of the kind of
Conjecture~1.5 of \cite{AA14}, which is quoted above, for quantum algorithms with limited
adaptivity, without the use of Conjecture~1.7.

Blanc, Docter, Strassle and Tan \cite{BDST} prove
that for every quantum algorithm that makes $t$ queries in $d$ rounds, the queries of each round
being made in parallel, and for all $\eps,\delta>0$ there is a classical algorithm that makes
$2^{O(d^{2})}\cdot(t\log(1/\delta)/\eps)^{O(d)}$ queries and approximates the acceptance
probability of the quantum algorithm to within an additive error $\eps$ on a $1-\delta$ fraction of
the inputs \cite[Theorem~1]{BDST}. For a bounded number $d$ of rounds this number of queries is
polynomial in $t$, $1/\eps$ and $\log(1/\delta)$, whereas the number obtained from
Corollary~\ref{cor:AA01} through \cite[Theorem~3.3]{AA14} is polynomial in $t$, $1/\eps$ and
$1/\delta$, for every quantum algorithm.

Liu and Mutreja \cite{LM} prove Conjecture~1.5 for three
classes of quantum algorithms (see \cite[Theorem~1.2]{LM}, which is stated there informally):
algorithms that make a single layer of parallel quantum queries; algorithms in which such a layer is
preceded by a bounded, slowly growing number of adaptive quantum queries or by polynomially many
adaptive classical queries; and algorithms with a constant number of adaptive layers of parallel
queries.

\medskip
\emph{Fractional certificates.} Fractional certificates of Boolean functions were introduced independently
by Tal \cite{Tal} and by Gilmer, Saks and Srinivasan \cite{GSS}; see the introduction
of \cite{KT}. Up to a constant factor, the least cost of a fractional certificate at a point is the
randomized certificate complexity which Aaronson had introduced earlier \cite{Aar}; see
\cite[Appendix~A]{GSS}.

For a function $f\colon\{0,1\}^{n}\to[0,1]$, a point $x$ and
$\eps>0$, Lovett and Zhang \cite{LZ} define the $\eps$-fractional certificate complexity of $f$ at
$x$ as the least number $k$ for which there is a probability distribution $\pi$ on $[n]$ such that
\[
   \pi(B)\;\ge\;\frac1k\qquad\text{for every set $B$ of coordinates with }
   \abs{f(x\oplus B)-f(x)}\ge\eps,
\]
where
$x\oplus B$ is $x$ with the bits in $B$ flipped.

They prove that it is at most
$\mathrm{poly}(d,1/\eps,\log n)$ at every point if $f$ is computed by a polynomial of degree $d$
\cite[Theorem~2]{LZ}.

The fractional certificates of the present paper are the rows of Definition~\ref{def:row}, which
is recalled in Section~\ref{subsec:results}. A row $a$ of cost $K>0$ at $x$ gives, for every
$\eps>0$, an $\eps$-fractional certificate in the sense of Lovett and Zhang, namely $\pi=a/K$ with
$k=K/\eps$; the relation of rows to the other notions is described after Definition~\ref{def:row}.

For rows a bound in terms of the
degree alone is known: Agarwal and Ben-David \cite[Theorem~3.3]{ABD} proved that the fractional block
sensitivity (recalled in Remark~\ref{rem:ABD}) of every $f\colon\{-1,1\}^{n}\to[-1,1]$ is at most
$\frac{\pi^{2}}{4}\deg(f)^{2}$,
adapting a method of Anshu, Ben-David and Kundu \cite{ABK} which expands on earlier work of Tal
\cite{Tal}.

For Boolean functions the bound $\deg(f)^{2}$ is due to Kulkarni and Tal \cite{KT}.

By
linear programming duality the bound of Agarwal and Ben-David gives rows of cost at most
$\frac{\pi^{2}}{2}\deg(f)^{2}$ at every point (Remark~\ref{rem:ABD}). The statement about rows is \cite[Theorem~1.7]{ABD}, where it is
formulated for functions with values in $[0,1]$, with the constant $\frac{\pi^{2}}{4}$.

Section~\ref{sec:row} contains a complete proof of the bound with a
constant that is larger by a factor $2$, and discusses the relation with \cite{ABD}.

\medskip
\emph{From fractional certificates to the Aaronson--Ambainis conjecture.} Lovett and Zhang also show
that the Aaronson--Ambainis conjecture would follow from their bound together with a conjectured extension of Talagrand's concentration
inequality to the $L_\infty$ norm \cite[Section~5]{LZ}.

Their conjectured inequality \cite[Conjecture~37]{LZ} concerns two subsets
$X$ and $Y$ of the cube neither of which has an influential variable, and it asserts that some point of
$X$ is close in the $L_\infty$ norm to the convex hull of $Y$.

Their derivation applies it to a sublevel
set and a superlevel set of $f$; it uses that a fractional certificate at a point bounds from below the
$L_\infty$ distance from this point to the convex hull of the set of points at which the value of $f$
differs from its value there by at least $\eps$ \cite[Lemma~28]{LZ}.

Conjecture~37 of Lovett and Zhang is not true as stated: for the tribes function $g$ of width $w$ with
$2^{w}$ tribes, every point of $X=g^{-1}(1)$ has $L_\infty$ distance $1/w$ from the convex hull of
$Y=g^{-1}(0)$, while all the influences of $X$ and of $Y$ in the sense of \cite[Definition~36]{LZ}
are at most $1/(2^{w}-1)$ (Section~\ref{subsec:cert}). The proof of Theorem~\ref{thm:main} follows
the plan of Lovett and Zhang in passing to the level sets of $f$, but it replaces their
Conjecture~37 by the all-pairs inequality for Boolean functions (Proposition~\ref{prop:pairs}), in which
the rows at the points of value $1$ and at the points of value $0$ enter in the same way; this
inequality is applied at every level and integrated over the levels, so that no single level is
chosen (Section~\ref{subsec:lz}).

\subsection{Overview of the proof}\label{subsec:overview}

In this overview, $g\colon Q_n\to\{0,1\}$ is a Boolean function, $\cO=g^{-1}(1)$, $\cN=g^{-1}(0)$, and
\[
   b_j(g)\;:=\;\#\{x\in\cO:\ x^{j}\in\cN\}\;=\;2^{n-1}\,\I_j(g)\qquad(j\in[n]).
\]
The antipode of $x$ is $\bar x=-x$, and $Z(g)=\{u\in\cO:\bar u\in\cN\}$.

The proof of
Theorem~\ref{thm:main} is a chain of six links. The first three concern Boolean functions, the
fourth concerns bounded functions of low degree, and the last two concern real functions with a
row of bounded cost at every point.

\medskip\noindent\emph{Link 1: a bijection theorem and the determinant inequality
\textup{(Section~\ref{sec:det})}.}
A \emph{$1$-wing} at $x\in\cO$ is a set $A\subseteq[n]$ such that $x^{S}\in\cO$ for every
$S\subseteq A$.

Theorem~\ref{thm:bij} states that for every choice of a $1$-wing $A_x$ at each
$x\in\cO$ there is a bijection $\tau\colon\cO\to\cO$ such that $\tau(x)$ differs from $x$ in every
coordinate of $A_x$, and such that for every $j$ the number of $x\in\cO$ with $\tau(x)_j=x_j$ is at
most $b_j(g)$.

Without the last condition, the existence of such a bijection is equivalent to the
Butterfly Lemma of Reimer \cite{Rei}; see Remark~\ref{rem:butterfly}.

The proof studies the
determinant of the matrix with entries
$\prod_{i:\,x_i=z_i}s_i$ $(x,z\in\cO)$, a polynomial in indeterminates $s_1,\dots,s_n$: we show that
it is not zero, that every monomial of lowest total degree occurring in it has degree at most $b_j(g)$
in $s_j$, and that at least one such monomial comes from a permutation of $\cO$ compatible with the
wings.

If positive lengths $\ell_1,\dots,\ell_n$ are assigned to the coordinates, the bijection yields
the \emph{determinant inequality} (Corollary~\ref{cor:det})
\[
   \sum_{u\in Z(g)}D_g(u,A_u;\ell)\;\le\;\sum_{j=1}^{n}\ell_j\,b_j(g),
\]
where $D_g(u,A;\ell)$ is the least total length of a set $G$ of coordinates, disjoint from $A$, such
that $\bar u^{\,G}\in\cO$.

\medskip\noindent\emph{Link 2: the antipodal inequality \textup{(Section~\ref{sec:anti})}.}
Let a row $\lambda^{x}$ of $g$ be given at every point $x$, and let $w\in(0,\infty)^{n}$.
Theorem~\ref{thm:anti} states that
\[
   \sum_{u\in Z(g)}\Bigl(\sum_{j=1}^{n}w_j\lambda^{u}_j\lambda^{\bar u}_j\Bigr)^{-1}
   \;\le\;\sum_{j=1}^{n}\frac{b_j(g)}{w_j}.
\]

It follows from the determinant inequality with $\ell_j=1/w_j$: the two rows at $u$ and at $\bar u$
single out a $1$-wing $A$ at $u$ for which $D_g(u,A;\ell)$ is at least the reciprocal on the left.

\medskip\noindent\emph{Link 3: from antipodal pairs to all pairs
\textup{(Section~\ref{sec:two})}.}
Every pair $(x,y)\in\cO\times\cN$ is a pair of antipodal points of exactly one subcube of $Q_n$.
Applying the antipodal inequality to the restrictions of $g$ to all subcubes and adding up gives the
\emph{all-pairs inequality} (Proposition~\ref{prop:pairs})
\[
   \sum_{x\in\cO}\sum_{y\in\cN}\Bigl(\sum_{j:\,x_j\ne y_j}w_j\lambda^{x}_j\lambda^{y}_j\Bigr)^{-1}
   \;\le\;2^{n-1}\sum_{j=1}^{n}\frac{b_j(g)}{w_j}.
\]

Two applications of the Cauchy--Schwarz inequality and an explicit choice of the weights $w$
remove the weights from an inequality of this kind (Lemma~\ref{lem:weights}).

For the all-pairs
inequality itself the result is Theorem~\ref{thm:two-intro}. That theorem is not used in the
proof of Theorem~\ref{thm:main}; the lemma is used again in Link~5.

\medskip\noindent\emph{Link 4: the row bound \textup{(Section~\ref{sec:row})}.}
If $f\colon Q_n\to[-1,1]$ has degree at most $d$, then for every $x$ there is a vector
$a\in[0,\infty)^{n}$ with $\norm{a}\le\pi^{2}d^{2}$ and
\[
   \abs{f(x)-f(y)}\;\le\;\sum_{i:\,x_i\ne y_i}a_i\qquad\text{for all }y
\]
(Theorem~\ref{thm:row}).

The proof combines
Markov's inequality for the derivative of a polynomial, a composition with Chebyshev polynomials, and
linear programming duality.

The bound is known with a better constant, and the proof follows the
method of Agarwal and Ben-David \cite{ABD}; see Remark~\ref{rem:ABD}.

An intermediate step of this proof, the block inequality (Lemma~\ref{lem:block}), also shows that
$\sum_j\abs{f(x)-f(x^{j})}\le4d^{2}$ at every point $x$ (Corollary~\ref{cor:sens-point}).

\medskip\noindent\emph{Link 5: integration over the level \textup{(Section~\ref{sec:level})}.}
Let $f\colon Q_n\to\R$ be a function with a row $a^{x}$ of cost at most $K$ at every point $x$.
For $w\in(0,\infty)^{n}$ put
\[
   s_{xy}\;=\;\sum_{j:\,x_j\ne y_j}w_j\,a^{x}_j\,a^{y}_j .
\]

For a level $t$ let $g_t=1_{\{f>t\}}$. If
$f(x)\ne t$, the vector $a^{x}$, divided by $\abs{f(x)-t}$, is a row of $g_t$ at $x$. With these
rows, and with the row $(1,\dots,1)$ at the points $x$ with $f(x)=t$, the all-pairs inequality for
$g_t$ gives
\[
   \sum_{x,y:\ f(y)<t<f(x)}\frac{(f(x)-t)\,(t-f(y))}{s_{xy}}
   \;\le\;2^{n-1}\sum_{j=1}^{n}\frac{b_j(g_t)}{w_j}.
\]

We integrate this inequality over $t$. On the left, the integral of $(f(x)-t)(t-f(y))$ over the
levels $t$ between $f(y)$ and $f(x)$ is $(f(x)-f(y))^{3}/6$. On the right, the integral of
$b_j(g_t)$ is $2^{n-1}L_j(f)$.

The result is the \emph{integrated all-pairs inequality}
(Proposition~\ref{prop:pairs-real})
\[
   \sum_{x,y:\ f(y)<f(x)}\frac{(f(x)-f(y))^{3}}{6\,s_{xy}}
   \;\le\;4^{\,n-1}\sum_{j=1}^{n}\frac{L_j(f)}{w_j}.
\]

We keep the pairs with $f(y)\le r<f(x)$, for a real number $r$, and remove the weights with the
lemma of Link~3. This gives the \emph{third-moment inequality} (Theorem~\ref{thm:third})
\[
   \E\bigl[(\max(f-r,0))^{3}\bigr]\cdot\Pr[f\le r]\;\le\;\frac32\,K^{2}\,\max_{j\in[n]}L_j(f).
\]

Applied at a median $m$ of $f$, and also to $-f$, it shows that
\[
   \E\abs{f-m}^{3}\;\le\;6K^{2}\max_{j\in[n]}L_j(f).
\]
Since $\Var(f)^{3/2}\le\E\abs{f-m}^{3}$, this is
Theorem~\ref{thm:incr-intro}.

\medskip\noindent\emph{Link 6: averaging \textup{(Section~\ref{sec:avg})}.}
To pass from the numbers $L_j(f)$ to the influences, consider the coordinates $k$ with
\[
   5\,L(f)\,\Inf_k(f)\;<\;2\,\Var(f)\,L_k(f).
\]
The sum of their influences is at most $\frac25\Var(f)$. Let $h$
be the function obtained from $f$ by averaging over these coordinates.

Then the function $h$ has a row of cost at most $K$ at every point, and
\begin{gather*}
   \Var(h)\ge\tfrac35\Var(f),\qquad L_j(h)\le L_j(f)\ \text{ for every }j,\\
   L_j(h)=0\ \text{ for the coordinates $j$ over which we have averaged.}
\end{gather*}

By Theorem~\ref{thm:incr-intro}, applied to $h$, there is a coordinate $j$ with
$L_j(h)\ge\Var(h)^{3/2}/(6K^{2})$. Since $L_j(h)>0$, we have not averaged over $j$, and so
\[
   5\,L(f)\,\Inf_j(f)\;\ge\;2\,\Var(f)\,L_j(f)\;\ge\;2\,\Var(f)\,L_j(h).
\]
This gives Theorem~\ref{thm:infl-intro}.

With the two results of Link~4, Theorem~\ref{thm:infl-intro} gives Theorem~\ref{thm:main}.

\smallskip
In summary,
\begin{align*}
&\text{bijection theorem (Theorem~\ref{thm:bij})}\;\Longrightarrow\;
\text{determinant inequality (Corollary~\ref{cor:det})}\\
&\qquad\Longrightarrow\;\text{antipodal inequality (Theorem~\ref{thm:anti})}\;\Longrightarrow\;
\text{all-pairs inequality (Proposition~\ref{prop:pairs})}\\
&\qquad\Longrightarrow\;\text{integrated all-pairs inequality (Proposition~\ref{prop:pairs-real})}\\
&\qquad\Longrightarrow\;\text{third-moment inequality (Theorem~\ref{thm:third})}\;\Longrightarrow\;
\text{Theorem~\ref{thm:incr-intro} (${}=\,$Theorem~\ref{thm:incr})}\\
&\qquad\Longrightarrow\;\text{Theorem~\ref{thm:infl-intro} (${}=\,$Theorem~\ref{thm:infl})},\\
&\text{Theorem~\ref{thm:infl-intro}}\;+\;\text{row bound (Theorem~\ref{thm:row})}\;+\;
\text{Corollary~\ref{cor:sens-point}}\;\Longrightarrow\;\text{Theorem~\ref{thm:main}}.
\end{align*}

The arguments are elementary: besides linear algebra over a polynomial ring and the Cauchy--Schwarz
inequality, we use Markov's inequality for the derivative of a polynomial at an endpoint of an
interval, the separation of a compact convex set from the non-negative orthant, and integrals of
polynomials over intervals.

\section{Notation and definitions}\label{sec:notation}

\subsection{The cube}\label{subsec:cube}

Let $n\ge 0$ be an integer, $[n]=\{1,\dots,n\}$, and $Q_n=\{-1,1\}^{n}$; the cube $Q_0$ consists of
one point.

For $x\in Q_n$ and $S\subseteq[n]$ the point $x^{S}\in Q_n$ is defined by
\[
   (x^{S})_i=-x_i\quad(i\in S),\qquad (x^{S})_i=x_i\quad(i\notin S).
\]
We write $x^{j}=x^{\{j\}}$ and $\bar x=x^{[n]}=-x$, the \emph{antipode} of $x$. Clearly
$(x^{S})^{T}=x^{S\triangle T}$, where $\triangle$ is the symmetric difference.

For $x,z\in Q_n$ put
\[
   \Delta(x,z)=\{i\in[n]: x_i\ne z_i\},\quad \ag(x,z)=\{i\in[n]:x_i=z_i\}=[n]\setminus\Delta(x,z),
\]
the sets of coordinates in which $x$ and $z$ differ and agree. Thus
\begin{equation}\label{eq:delta-ag}
   z=x^{\Delta(x,z)}=\bar x^{\,\ag(x,z)},\qquad \Delta(x,x^{S})=S,\qquad \ag(\bar x,x^{S})=S .
\end{equation}

For a vector $a\in\R^{n}$ and $S\subseteq[n]$ we write $a(S)=\sum_{i\in S}a_i$, and
$\norm{a}=\sum_{i=1}^{n}\abs{a_i}$; for $a\in[0,\infty)^{n}$ this is $a([n])$.

Probabilities, expectations and variances ($\Pr$, $\E$, $\Var$) are taken over a uniformly random
point of the cube, each of the $2^{n}$ points having probability $2^{-n}$; thus
$\E f=2^{-n}\sum_{x\in Q_n}f(x)$ and $\Var(f)=\E[(f-\E f)^{2}]$ for $f\colon Q_n\to\R$. For every
$S\subseteq[n]$ the map $x\mapsto x^{S}$ is a bijection of $Q_n$ which preserves the measure.

\medskip
\emph{Conventions.} An empty sum is $0$ and an empty product is $1$; the determinant of the matrix
with no rows and no columns is $1$. We take reciprocals of positive real numbers only; whenever a
reciprocal is written, the positivity of the number is proved first. The cardinality of a finite set
$X$ is written $\abs{X}$ or $\#X$.

\subsection{Functions and degree}

For $T\subseteq[n]$ let $\chi_T(x)=\prod_{i\in T}x_i$. Since $x_i^{2}=1$ on $Q_n$, we have
$\chi_T\chi_U=\chi_{T\triangle U}$, and $\E\chi_V=0$ for $V\ne\emptyset$ (the map $x\mapsto x^{i}$
with $i\in V$ changes the sign of $\chi_V$), while $\chi_\emptyset=1$. Hence
\[
   \E[\chi_T\chi_U]\;=\;\begin{cases}1,&\text{if }T=U,\\ 0,&\text{otherwise.}\end{cases}
\]

The $2^{n}$ functions $\chi_T$ are
therefore linearly independent in the space of all functions $Q_n\to\R$, which has dimension $2^{n}$,
and so they form a basis: every $f\colon Q_n\to\R$ has a unique expansion
\begin{equation}\label{eq:expansion}
   f=\sum_{T\subseteq[n]}\hat f(T)\,\chi_T,\qquad \hat f(T)=\E[f\chi_T]\in\R .
\end{equation}

Let $d\ge0$ be an integer. The function $f$ has \emph{degree at most $d$} if $\hat f(T)=0$ for all $T$ with
$\abs{T}>d$.

The functions of degree at most $d$ form a linear space. If $f$ has degree at most $d$ and $h$ has degree at most
$e$, then $fh$ has degree at most $d+e$, because $\abs{T\triangle U}\le\abs{T}+\abs{U}$. A function of
degree at most $0$ is constant.

The \emph{influence} of the coordinate $j$ on $f\colon Q_n\to\R$ is
\[
   \Inf_j(f)=\E_x\Bigl[\Bigl(\frac{f(x)-f(x^{j})}{2}\Bigr)^{2}\Bigr].
\]

The \emph{mean absolute increment} of $f$ in the coordinate $j$ is
\[
   L_j(f)=\E_x\abs{f(x)-f(x^{j})},
\]
and we write $L(f)=\sum_{j=1}^{n}L_j(f)$ for the sum of these numbers.

We shall use the following identity several times. For every function $h\colon Q_n\to\R$ and
every real number $c$,
\begin{equation}\label{eq:var-shift}
   \E\bigl[(h-c)^{2}\bigr]\;=\;\Var(h)+(\E h-c)^{2}\;\ge\;\Var(h),
\end{equation}
because $h-c=(h-\E h)+(\E h-c)$ and $\E[h-\E h]=0$.

The influences and the mean absolute increments are related as follows. For
$h(x)=\abs{f(x)-f(x^{j})}$ and $c=0$ the
equality in \eqref{eq:var-shift} and $\Var(h)\ge0$ give $(\E h)^{2}\le\E[h^{2}]$, that is,
\begin{equation}\label{eq:L-Inf}
   L_j(f)^{2}\;\le\;4\,\Inf_j(f)\qquad\text{for every }j\in[n].
\end{equation}

\subsection{Boolean functions}

A \emph{Boolean function} is a function $g\colon Q_n\to\{0,1\}$. For such a function we put
\[
   \cO=\cO(g)=\{x\in Q_n:g(x)=1\},\qquad \cN=\cN(g)=\{x\in Q_n:g(x)=0\},
\]
and, for $j\in[n]$,
\[
   \I_j(g)=\Pr_x\bigl[g(x)\ne g(x^{j})\bigr],\qquad b_j(g)=\#\{x\in\cO:\ x^{j}\in\cN\}.
\]
The \emph{zone} of $g$ is the set $Z(g)=\{u\in\cO:\bar u\in\cN\}$ of points of $\cO$ whose antipode
lies in $\cN$.

\begin{lemma}\label{lem:bj}
For every Boolean function $g$ on $Q_n$ and every $j\in[n]$ we have $b_j(g)=2^{n-1}\,\I_j(g)$.
\end{lemma}

\begin{proof}
Let $E_j=\{x\in Q_n:g(x)\ne g(x^{j})\}$, so that $\I_j(g)=2^{-n}\abs{E_j}$. The involution
$x\mapsto x^{j}$ maps $E_j$ onto itself and exchanges $E_j\cap\cO$ and $E_j\cap\cN$. Hence
$\abs{E_j}=2\abs{E_j\cap\cO}=2\,b_j(g)$.
\end{proof}

\begin{definition}[Wings]\label{def:wing}
Let $g$ be a Boolean function on $Q_n$ and $x\in\cO$. A set $A\subseteq[n]$ is a \emph{$1$-wing of
$g$ at $x$} if $x^{S}\in\cO$ for every $S\subseteq A$.
\end{definition}

The empty set is a $1$-wing at every $x\in\cO$, and a subset of a $1$-wing at $x$ is a $1$-wing at
$x$.

\begin{definition}[Rows]\label{def:row}
Let $f\colon Q_n\to\R$ and $x\in Q_n$. A \emph{row of $f$ at $x$} is a vector $a\in[0,\infty)^{n}$
such that
\[
   \abs{f(x)-f(x^{S})}\;\le\;a(S)\qquad\text{for every }S\subseteq[n].
\]
The number $\norm{a}$ is the \emph{cost} of the row.
\end{definition}

By \eqref{eq:delta-ag}, the condition says that $\abs{f(x)-f(y)}\le a(\Delta(x,y))$ for every
$y\in Q_n$.

If $g$ is a Boolean function and $x\in Q_n$, a set $S\subseteq[n]$ with $g(x^{S})\ne g(x)$ is called
a \emph{sensitive block} of $g$ at $x$. The difference $\abs{g(x)-g(x^{S})}$ is $1$ if $S$ is a
sensitive block at $x$ and $0$ otherwise; so a vector $\lambda\in[0,\infty)^{n}$ is a row of $g$ at
$x$ if and only if
\begin{equation}\label{eq:row-bool}
   \lambda(S)\ge 1\qquad\text{for every sensitive block $S$ of $g$ at $x$,}
\end{equation}
in agreement with the definition given in the introduction.

For Boolean functions, rows are the fractional certificates introduced by Tal \cite{Tal} and by
Gilmer, Saks and Srinivasan \cite{GSS}. For a function $f$ with values in $[-1,1]$, a vector $a$ is a
row of $f$ at $x$ exactly when $a/2$ is a fractional certificate of $f$ at $x$ in the sense of
Agarwal and Ben-David \cite[Section~3.1]{ABD}. A row bounds every change $\abs{f(x)-f(x^{S})}$ at
once, whereas the $\eps$-fractional certificates of Lovett and Zhang \cite{LZ} constrain only the
sets $S$ with $\abs{f(x)-f(x^{S})}\ge\eps$.

We record three simple facts.

\begin{lemma}\label{lem:rows-basic}
Let $g$ be a Boolean function on $Q_n$.
\begin{itemize}
\item[(i)] The vector $(1,\dots,1)\in\R^{n}$ is a row of $g$ at every point.
\item[(ii)] If $g$ is not constant, then $n\ge 1$ and every row $\lambda$ of $g$ at every point
satisfies $\norm{\lambda}\ge 1$.
\item[(iii)] If $f\colon Q_n\to\R$, $t\in\R$, $g=1_{\{f>t\}}$, $x\in Q_n$ with $f(x)\ne t$, and $a$ is
a row of $f$ at $x$, then $a/\abs{f(x)-t}$ is a row of $g$ at $x$.
\end{itemize}
\end{lemma}

\begin{proof}
(i) A sensitive block is not empty, so it contains some $i$, and the sum in \eqref{eq:row-bool} is at
least $1$.

(ii) The cube $Q_0$ has one point, so $n\ge1$. Let $x\in Q_n$. Since $g$ is not constant there is
$y$ with $g(y)\ne g(x)$, and $S=\Delta(x,y)$ is a sensitive block at $x$. Hence
$\norm{\lambda}\ge\lambda(S)\ge 1$.

(iii) Let $S$ be a sensitive block of $g$ at $x$. Then exactly one of the numbers $f(x)$, $f(x^{S})$
is greater than $t$, and $f(x)\ne t$; so $t$ lies between them and
$\abs{f(x)-f(x^{S})}\ge\abs{f(x)-t}>0$. Therefore
$a(S)\ge\abs{f(x)-f(x^{S})}\ge\abs{f(x)-t}$, which is \eqref{eq:row-bool} for the vector
$a/\abs{f(x)-t}$; its entries are non-negative.
\end{proof}

\begin{example}\label{ex:and}
Let $n=2$ and let $g$ be the function with $\cO=\{(1,1)\}$. At $x=(1,1)$ every non-empty set is a
sensitive block, so the rows at $x$ are the vectors $\lambda$ with $\lambda_1\ge1$ and $\lambda_2\ge1$;
the least cost is $2$, and the only $1$-wing at $x$ is the empty set.

At $(-1,1)$ the only sensitive
block is $\{1\}$, and $(1,0)$ is a row of cost $1$; similarly $(0,1)$ is a row at $(1,-1)$. At
$(-1,-1)$ the only sensitive block is $\{1,2\}$, and $(\frac12,\frac12)$ is a row of cost $1$.

Further
$Z(g)=\{(1,1)\}$, $b_1(g)=b_2(g)=1$ and $\I_1(g)=\I_2(g)=\frac12$. With these rows the left-hand side
of the inequality of Theorem~\ref{thm:two-intro} is
\[
   4\cdot\Bigl(\frac14\cdot\frac12\Bigr)\cdot\Bigl(\frac34\cdot 1\Bigr)\;=\;\frac38,
\]
and the right-hand
side is $\frac12$.
\end{example}

\section{A bijection theorem and the determinant inequality}\label{sec:det}

This section contains the combinatorial core of the paper: a theorem about permutations of the set
$\cO=g^{-1}(1)$ of a Boolean function $g$ (Theorem~\ref{thm:bij}), and its consequence, the
determinant inequality (Corollary~\ref{cor:det}), which is the first link of the proof chain.

At every point $x$ of $\cO$ we fix a $1$-wing $A_x$, a set of coordinates that can be changed at $x$
in any combination without leaving $\cO$. The bijection theorem permutes the points of $\cO$ so that
every point changes all the coordinates of its wing (property (a)), while for every coordinate $j$ at
most $b_j(g)$ points keep their $j$th coordinate (property (b)); here $b_j(g)$ is the number of edges
in the direction $j$ along which $g$ changes. Summing over the points gives the determinant
inequality: from the antipode of a point $u$ whose antipode lies outside $\cO$, the coordinates that
$u$ keeps lead back into $\cO$ without using the wing at $u$, and property (b) bounds how often each
coordinate is used. Section~\ref{sec:anti} turns this into the antipodal inequality.

The permutation of the bijection theorem is read off from a term of lowest degree of a determinant.
The matrix $M$ has rows
and columns indexed by $\cO$, and its entry at $(x,z)$ is the product of indeterminates $s_i$ over
the coordinates $i$ in which $x$ and $z$ agree; a permutation of $\cO$ contributes to $\det M$ a
monomial in which the exponent of $s_j$ is the number of points that keep their $j$th coordinate.
Combining the rows of pairs of points that differ in one coordinate shows that $\det M$ has many
factors $1-s_j^{2}$, so that in its terms of lowest degree the exponent of $s_j$ is at most $b_j(g)$
(Section~\ref{subsec:agreement}). Combining the rows of the points of each wing gives a matrix with
zeros at the entries that property (a) forbids, without changing the terms of lowest degree; a
permutation that produces such a term in the new determinant has both properties
(Section~\ref{subsec:wings}). Section~\ref{subsec:butterfly} contains examples and the relation to
Reimer's Butterfly Lemma.

\subsection{Statements}

\begin{theorem}[Bijection theorem]\label{thm:bij}
Let $n\ge 0$, let $g$ be a Boolean function on $Q_n$, and for every $x\in\cO=g^{-1}(1)$ let $A_x$ be
a $1$-wing of $g$ at $x$. Then there is a bijection $\tau\colon\cO\to\cO$ such that
\begin{itemize}
\item[\textup{(a)}] $\tau(x)_i\ne x_i$ for every $x\in\cO$ and every $i\in A_x$, and
\item[\textup{(b)}] $\#\{x\in\cO:\ \tau(x)_j=x_j\}\le b_j(g)$ for every $j\in[n]$.
\end{itemize}
\end{theorem}

Let $\ell\in(0,\infty)^{n}$; we think of $\ell_j$ as the length of the coordinate $j$. For
$u\in Z(g)$ and a $1$-wing $A$ of $g$ at $u$ put
\[
   D_g(u,A;\ell)\;:=\;\min\bigl\{\ell(G):\ G\subseteq[n]\setminus A,\ \bar u^{\,G}\in\cO\bigr\}.
\]
The minimum is taken over a non-empty finite family of sets, because
$\bar u^{\,[n]\setminus A}=u^{A}\in\cO$.

In words, $D_g(u,A;\ell)$ is the least total length of a
set of coordinates outside the wing whose change brings the antipode of $u$ back into $\cO$.

\begin{corollary}[Determinant inequality]\label{cor:det}
Let $g$ be a Boolean function on $Q_n$, let $\ell\in(0,\infty)^{n}$, and for every $u\in Z(g)$ let
$A_u$ be a $1$-wing of $g$ at $u$. Then
\[
   \sum_{u\in Z(g)}D_g(u,A_u;\ell)\;\le\;\sum_{j=1}^{n}\ell_j\,b_j(g).
\]
\end{corollary}

\begin{proof}[Proof of Corollary~\ref{cor:det} from Theorem~\ref{thm:bij}]
Put $A_x=\emptyset$ for $x\in\cO\setminus Z(g)$; the empty set is a $1$-wing. Let $\tau$ be the
bijection of Theorem~\ref{thm:bij} for the wings $A_x$ $(x\in\cO)$, and for $y\in\cO$ put
$G_y=\ag(y,\tau(y))$.

By (a), $G_y\cap A_y=\emptyset$, and by \eqref{eq:delta-ag},
$\bar y^{\,G_y}=\tau(y)\in\cO$. Hence for $y\in Z(g)$ the set $G_y$ belongs to the family over which
$D_g(y,A_y;\ell)$ is the minimum, so that $D_g(y,A_y;\ell)\le\ell(G_y)$. For $y\in\cO\setminus Z(g)$
we use only $\ell(G_y)\ge0$. Therefore
\begin{align*}
   \sum_{u\in Z(g)}D_g(u,A_u;\ell)&\;\le\;\sum_{y\in\cO}\ell\bigl(\ag(y,\tau(y))\bigr)\\
   &\;=\;\sum_{j=1}^{n}\ell_j\cdot\#\{y\in\cO:\ \tau(y)_j=y_j\}\;\le\;\sum_{j=1}^{n}\ell_j\,b_j(g),
\end{align*}
where we exchanged the order of summation and then used (b) and $\ell_j\ge0$.
\end{proof}

The rest of this section is the proof of Theorem~\ref{thm:bij}.

\subsection{Lowest monomials}

Let $s_1,\dots,s_n$ be indeterminates and let $\Z[s]$ be the ring of polynomials in them with
integer coefficients. For a vector $m$ of non-negative integers and for a set $S\subseteq[n]$ we
write
\[
   s^{m}=\prod_{i=1}^{n}s_i^{m_i},\qquad \abs{m}=\sum_{i=1}^{n}m_i,\qquad s^{S}=\prod_{i\in S}s_i .
\]

A monomial $s^{m}$ \emph{occurs} in $q\in\Z[s]$ if its coefficient in $q$ is not zero. A
\emph{lowest monomial} of $q$ is a monomial $s^{m}$ which occurs in $q$ and satisfies
$\abs{m}\le\abs{m'}$ for every monomial $s^{m'}$ occurring in $q$. A non-zero polynomial has at
least one lowest monomial, and the zero polynomial has none.

We write $q(0)$ for the constant
coefficient of $q$, and $\deg_jq$ for the degree of a non-zero polynomial $q$ in the variable $s_j$.
The ring $\Z[s]$ has no zero divisors, and for non-zero $p,q\in\Z[s]$ we have
\[
   \deg_j(pq)=\deg_jp+\deg_jq .
\]

The matrices of this section have their rows and their columns indexed by the same finite set $X$,
and entries in a commutative ring, which is $\Z[s]$ or $\R$. We use the Leibniz formula
\begin{equation}\label{eq:leibniz}
   \det B\;=\;\sum_{\tau}\sgn(\tau)\prod_{x\in X}B[x,\tau(x)],
\end{equation}
in which the sum is over all permutations $\tau$ of $X$ and $\sgn(\tau)\in\{1,-1\}$ is the sign of
$\tau$, and the rule $\det(BC)=\det B\cdot\det C$; see \cite[Chapter~XIII, \S4]{Lang}.

By
\eqref{eq:leibniz}, the determinant of a diagonal matrix is the product of its diagonal entries, and
a ring homomorphism, applied to all entries of a matrix, maps the determinant of the matrix to the
determinant of the new matrix.

The first lemma says that a factor with constant coefficient $1$ does not change the lowest
monomials. It will be applied to powers of $1-s_j^{2}$ (Lemma~\ref{lem:lowest}) and to the
determinants of the matrices $P$ and $U$ of Lemma~\ref{lem:PMUQ}.

\begin{lemma}\label{lem:unit}
Let $u,q\in\Z[s]$ with $u(0)=1$. Then $uq$ and $q$ have the same lowest monomials.
\end{lemma}

\begin{proof}
Write $u_a$ and $q_b$ for the coefficients of $s^{a}$ in $u$ and of $s^{b}$ in $q$. We first show:
\begin{equation}\label{eq:unit-comp}
   \parbox{0.82\textwidth}{\emph{if $\abs{m}\le\abs{m'}$ for every monomial $s^{m'}$ occurring in
   $q$, then $s^{m}$ has the same coefficient in $uq$ as in $q$.}}
\end{equation}
Indeed, the coefficient of $s^{m}$ in $uq$ is $\sum_{a+b=m}u_a\,q_b$. If $a\ne0$ then
$\abs{b}<\abs{m}$, so $s^{b}$ does not occur in $q$ and $q_b=0$. Only the term with $a=0$ and $b=m$
remains, and it equals $u_0\,q_m=q_m$.

Let $s^{m}$ be a lowest monomial of $q$. By \eqref{eq:unit-comp} its coefficient in $uq$ is
$q_m\ne0$. Suppose that a monomial $s^{m'}$ with $\abs{m'}<\abs{m}$ occurs in $uq$. Every monomial
occurring in $q$ has total degree at least $\abs{m}>\abs{m'}$, so \eqref{eq:unit-comp} applies to
$m'$: the coefficient of $s^{m'}$ in $uq$ is $q_{m'}$, which is $0$ because $\abs{m'}<\abs{m}$. This
is a contradiction. So $s^{m}$ is a lowest monomial of $uq$.

Conversely, let $s^{m}$ be a lowest monomial of $uq$. Then $uq\ne0$, so $q\ne0$, and $q$ has a lowest
monomial $s^{m_0}$. By the previous paragraph $s^{m_0}$ occurs in $uq$; hence
$\abs{m}\le\abs{m_0}$, and $\abs{m_0}\le\abs{m'}$ for every $s^{m'}$ occurring in $q$. By
\eqref{eq:unit-comp} the coefficient of $s^{m}$ in $q$ equals its coefficient in $uq$, which is not
zero. So $s^{m}$ occurs in $q$ and has the least total degree there.
\end{proof}

\subsection{The agreement matrix}\label{subsec:agreement}

Fix a Boolean function $g$ on $Q_n$. The \emph{agreement matrix} of $g$ is the matrix $M$ with rows
and columns indexed by $\cO$ and entries
\[
   M[x,z]\;=\;s^{\ag(x,z)}\;=\;\prod_{i:\,x_i=z_i}s_i\qquad(x,z\in\cO).
\]

If $\cO=\emptyset$, all matrices in this section are empty and have determinant $1$; nothing below
requires $\cO\ne\emptyset$.

Since $M[x,z]=\prod_{i}s_i\cdot\prod_{i:\,x_i\ne z_i}s_i^{-1}$, the matrix $M$ is, up to the factor
$\prod_is_i$ and the substitution of $1/s_i$ for the weights, a principal submatrix of the Varchenko
matrix of the arrangement of the $n$ coordinate hyperplanes in $\R^{n}$ \cite{Var}: the regions of
this arrangement correspond to the points of $Q_n$, and the entry of the Varchenko matrix at two
regions is the product of the weights of the hyperplanes that separate them.

Varchenko proved that the determinant of his matrix is a product of factors of the form
$\bigl(1-\prod_{i\in I}w_i^{2}\bigr)^{b}$, where $w_i$ is the weight of the $i$th hyperplane, $I$
runs over certain sets of hyperplanes, and the exponents $b$ are non-negative integers; see
\cite{HKY}, where the determinants of such submatrices are studied for the vertex sets of partial
cubes.

The divisibility of $\det M$ in
Lemma~\ref{lem:factor} below is of this kind. We use only the properties of $M$ that are proved
below.

We first check that $\det M$ is not the zero polynomial, so that it has lowest monomials.

\begin{lemma}\label{lem:nonzero}
Let $g$ be a Boolean function on $Q_n$ and $M$ its agreement matrix. Then \mbox{$\det M\ne0$}.
\end{lemma}

\begin{proof}
Let $c=2^{n}$, and substitute $c$ for every $s_i$. The substitution is a ring homomorphism
$\Z[s]\to\R$, so it maps $\det M$ to the determinant of the real matrix $B$ with entries
$B[x,z]=c^{\abs{\ag(x,z)}}$.

The diagonal entries of $B$ are $c^{n}$, and for $z\ne x$ we have
$\abs{\ag(x,z)}\le n-1$. A row has $\abs{\cO}-1<2^{n}=c$ off-diagonal entries. So for every $x\in\cO$
\[
   \sum_{z\in\cO,\ z\ne x}B[x,z]\;\le\;(\abs{\cO}-1)\,c^{\,n-1}\;<\;c^{n}\;=\;B[x,x].
\]

A real square matrix $B$ with non-negative entries in which every diagonal entry exceeds the sum
of the other entries of its row has $\det B\ne0$: if $Bv=0$ with $v\ne0$, choose $x$ with $\abs{v_x}$
largest; then
\[
   B[x,x]\abs{v_x}\;=\;\Bigl\lvert\sum_{z\ne x}B[x,z]v_z\Bigr\rvert\;\le\;\sum_{z\ne x}B[x,z]\abs{v_x}
   \;<\;B[x,x]\abs{v_x},
\]
which is impossible.

Hence the image of $\det M$ is not zero, and so $\det M\ne0$.
\end{proof}

The next two lemmas bound the exponent of $s_j$ in a lowest monomial of $\det M$. For $j\in[n]$ let
\[
   \Up_j=\{x\in\cO:\ x_j=1,\ x^{j}\in\cO\},\qquad e_j=\abs{\Up_j}.
\]

\begin{lemma}\label{lem:factor}
For every $j\in[n]$ there is $R_j\in\Z[s]$ with $\det M=(1-s_j^{2})^{e_j}R_j$.
\end{lemma}

\begin{proof}
\emph{Step 1: an identity for the entries.} For all $x,z\in Q_n$,
\begin{equation}\label{eq:factor}
   s^{\ag(x,z)}=s_j\,s^{\ag(x^{j},z)}\ \ \text{if }x_j=z_j,\qquad\quad
   s^{\ag(x^{j},z)}=s_j\,s^{\ag(x,z)}\ \ \text{if }x_j\ne z_j ,
\end{equation}
because $x$ and $x^{j}$ agree with $z$ in the same coordinates $i\ne j$, and exactly one of the two
agrees with $z$ in the coordinate $j$.

\smallskip
\emph{Step 2: the matrix $E$.} Let $E$ be the matrix indexed by $\cO$ with $E[x,y]=1$ if
$x\in\Up_j$ and $y=x^{j}$, and $E[x,y]=0$ otherwise. Then $E^{2}=0$: a non-zero product
$E[x,y]E[y,z]$ would need $y=x^{j}$ with $x_j=1$, so $y_j=-1$, and also $y\in\Up_j$, so $y_j=1$.
Hence, with $I$ the identity matrix,
\begin{equation}\label{eq:unipotent}
   (I+s_jE)(I-s_jE)=I .
\end{equation}

\smallskip
\emph{Step 3: the row operation.} Next, $(EM)[x,z]=M[x^{j},z]$ if $x\in\Up_j$, and $(EM)[x,z]=0$
otherwise. So the row of a point
$x\notin\Up_j$ in $(I-s_jE)M$ is its row in $M$, and for $x\in\Up_j$, by \eqref{eq:factor},
\[
   \bigl((I-s_jE)M\bigr)[x,z]=M[x,z]-s_jM[x^{j},z]=
   \begin{cases}0,& \text{if }z_j=x_j,\\[2pt] (1-s_j^{2})\,M[x,z],& \text{if }z_j\ne x_j.\end{cases}
\]

\smallskip
\emph{Step 4: determinants.} In matrix form, $(I-s_jE)M=\Lambda\widetilde M$, where $\Lambda$ is
the diagonal matrix with entry
$1-s_j^{2}$ at the points of $\Up_j$ and $1$ at the other points, and where $\widetilde M[x,z]=0$ if
$x\in\Up_j$ and $z_j=x_j$, and $\widetilde M[x,z]=M[x,z]$ otherwise.

Taking determinants,
\[
   \det(I-s_jE)\cdot\det M\;=\;(1-s_j^{2})^{e_j}\det\widetilde M .
\]
Multiplying by $\det(I+s_jE)$ and using
\eqref{eq:unipotent} we get $\det M=(1-s_j^{2})^{e_j}R_j$ with
\[
   R_j\;=\;\det(I+s_jE)\cdot\det\widetilde M . \qedhere
\]
\end{proof}

\begin{lemma}\label{lem:lowest}
Every lowest monomial $s^{m}$ of $\det M$ satisfies $m_j\le b_j(g)$ for every $j\in[n]$.
\end{lemma}

\begin{proof}
Fix $j$ and let $R_j$ be as in Lemma~\ref{lem:factor}.

\smallskip
\emph{Step 1: $\deg_j\det M\le\abs{\cO}$.} Every entry of $M$ has degree at most $1$ in
$s_j$, and by the Leibniz formula $\det M$ is a signed sum of products of $\abs{\cO}$ entries; so
$\deg_j\det M\le\abs{\cO}$.

\smallskip
\emph{Step 2: $\abs{\cO}=2e_j+b_j(g)$.} The set $\cO$ is the disjoint union of $\Up_j$, of
$\{x\in\cO:x_j=-1,\ x^{j}\in\cO\}$, which is the image of $\Up_j$ under the injective map
$x\mapsto x^{j}$, and of $\{x\in\cO:x^{j}\in\cN\}$, which has $b_j(g)$ elements; so
$\abs{\cO}=2e_j+b_j(g)$.

\smallskip
\emph{Step 3: $\deg_jR_j\le b_j(g)$.} By Lemma~\ref{lem:nonzero}, $\det M\ne0$ and hence
$R_j\ne0$. Since
$\deg_j(1-s_j^{2})^{e_j}=2e_j$, we get
\[
   2e_j+\deg_jR_j\;=\;\deg_j\det M\;\le\;\abs{\cO}\;=\;2e_j+b_j(g),
\]
that is, $\deg_jR_j\le b_j(g)$.

\smallskip
\emph{Step 4: lowest monomials.} The polynomial $(1-s_j^{2})^{e_j}$ has constant coefficient $1$. By
Lemma~\ref{lem:unit}, $s^{m}$ is a lowest monomial of $R_j$; in particular it occurs in $R_j$, and
therefore $m_j\le\deg_jR_j\le b_j(g)$.
\end{proof}

\subsection{Row operations along the wings}\label{subsec:wings}

The following identity describes the effect, on one entry of $M$, of combining the rows of the
points $x^{S}$, $S\subseteq A$.

\begin{lemma}[Wing identity]\label{lem:wing}
For all $x,z\in Q_n$ and every set $A\subseteq[n]$,
\[
  \sum_{S\subseteq A}\ \prod_{i\in S}(-s_i)\cdot s^{\ag(x^{S},z)}\;=\;
  \begin{cases}
     \prod_{i\in A}(1-s_i^{2})\cdot s^{\ag(x,z)}, & \text{if }\ag(x,z)\cap A=\emptyset,\\[3pt]
     0, & \text{if }\ag(x,z)\cap A\ne\emptyset.
  \end{cases}
\]
\end{lemma}

\begin{proof}
Put $\varphi_i=s_i$ and $\gamma_i=1$ if $x_i=z_i$, and put $\varphi_i=1$ and $\gamma_i=s_i$ if
$x_i\ne z_i$. Then $s^{\ag(x,z)}=\prod_{i\in[n]}\varphi_i$. Changing the sign of the $i$th coordinate
of $x$ exchanges agreement and disagreement with $z$ in that coordinate, so
\[
   s^{\ag(x^{S},z)}\;=\;\prod_{i\in S}\gamma_i\cdot\prod_{i\notin S}\varphi_i .
\]

For $S\subseteq A$ the term
of $S$ on the left-hand side is therefore
\[
   \prod_{i\in S}(-s_i\gamma_i)\cdot\prod_{i\in A\setminus S}\varphi_i\cdot\prod_{i\notin A}\varphi_i ,
\]
and the sum over all $S\subseteq A$ is
\[
   \prod_{i\in A}\bigl(\varphi_i-s_i\gamma_i\bigr)\cdot\prod_{i\notin A}\varphi_i .
\]

For $i\in A$ the factor $\varphi_i-s_i\gamma_i$ equals $s_i-s_i=0$ if $x_i=z_i$, and $1-s_i^{2}$ if
$x_i\ne z_i$. If some $i\in A$ has $x_i=z_i$, the product is $0$. Otherwise every factor with
$i\in A$ is $1-s_i^{2}$, and $\varphi_i=1$ for all $i\in A$, so that
$\prod_{i\notin A}\varphi_i=s^{\ag(x,z)}$.
\end{proof}

From now on a $1$-wing $A_x$ of $g$ at $x$ is fixed for every $x\in\cO$, as in
Theorem~\ref{thm:bij}.

Define matrices $P$, $U$, $Q$ indexed by $\cO$, with entries in $\Z[s]$, by
\begin{align*}
  P[x,w]&=\begin{cases}\prod_{i\in\Delta(x,w)}(-s_i),&\text{if }\Delta(x,w)\subseteq A_x,\\
                       0,&\text{otherwise,}\end{cases}\\[2pt]
  U[x,x]&=\prod_{i\in A_x}(1-s_i^{2}),\qquad U[x,w]=0\ \text{ for }w\ne x,\\[2pt]
  Q[x,z]&=\begin{cases}s^{\ag(x,z)},&\text{if }\ag(x,z)\cap A_x=\emptyset,\\
                        0,&\text{otherwise.}\end{cases}
\end{align*}
In matrix form the wing identity reads $PM=UQ$.

\begin{lemma}\label{lem:PMUQ}
Let $g$, the wings $A_x$ and the matrices $P$, $U$, $Q$ be as above.
\begin{itemize}
\item[\textup{(i)}] $PM=UQ$.
\item[\textup{(ii)}] $\det P$ and $\det U$ have constant coefficient $1$.
\item[\textup{(iii)}] $\det Q\ne0$, and every lowest monomial $s^{m}$ of $\det Q$ satisfies
$m_j\le b_j(g)$ for every $j\in[n]$.
\end{itemize}
\end{lemma}

\begin{proof}
(i) Fix $x,z\in\cO$. By definition,
\[
   (PM)[x,z]\;=\;\sum_{w}\ \prod_{i\in\Delta(x,w)}(-s_i)\cdot M[w,z],
\]
where the sum is over the points
$w\in\cO$ with $\Delta(x,w)\subseteq A_x$.

The map $w\mapsto\Delta(x,w)$ is a bijection from this set
of points onto the set of all subsets of $A_x$: it is injective because $w=x^{\Delta(x,w)}$, and it
is onto because for $S\subseteq A_x$ the point $x^{S}$ lies in $\cO$ and $\Delta(x,x^{S})=S$. That
$x^{S}\in\cO$ is the wing property of $A_x$; this is the only place in the proof of
Theorem~\ref{thm:bij} where it is used. Hence
\[
   (PM)[x,z]\;=\;\sum_{S\subseteq A_x}\ \prod_{i\in S}(-s_i)\cdot s^{\ag(x^{S},z)},
\]
and by Lemma~\ref{lem:wing} with $A=A_x$ this equals $U[x,x]\cdot Q[x,z]=(UQ)[x,z]$.

(ii) The map $q\mapsto q(0)$ is a ring homomorphism $\Z[s]\to\Z$, so $(\det P)(0)$ is the
determinant of the integer matrix with entries $P[x,w](0)$. Now $P[x,x]=1$, the empty product (note
that $\Delta(x,x)=\emptyset\subseteq A_x$), and for $w\ne x$ the entry $P[x,w]$ is $0$ or the product
of the $-s_i$ over the non-empty set $\Delta(x,w)$, so that $P[x,w](0)=0$. Thus
$(\det P)(0)=\det I=1$.

Also
\[
   \det U\;=\;\prod_{x\in\cO}\ \prod_{i\in A_x}(1-s_i^{2})
\]
has constant
coefficient $1$.

(iii) By (i),
\[
   \det P\cdot\det M\;=\;\det U\cdot\det Q .
\]
Here $\det P\ne0$ by (ii) and $\det M\ne0$ by Lemma~\ref{lem:nonzero}, so the left-hand side is not
zero, and therefore $\det Q\ne0$.

Let $s^{m}$ be a lowest monomial of $\det Q$. By
Lemma~\ref{lem:unit} with $u=\det U$ it is a lowest monomial of $\det U\cdot\det Q=\det P\cdot\det M$,
and by Lemma~\ref{lem:unit} with $u=\det P$ it is a lowest monomial of $\det M$. Now
Lemma~\ref{lem:lowest} gives $m_j\le b_j(g)$.
\end{proof}

\begin{proof}[Proof of Theorem~\ref{thm:bij}]
By the Leibniz formula \eqref{eq:leibniz},
\[
   \det Q\;=\;\sum_{\tau}\sgn(\tau)\prod_{x\in\cO}Q[x,\tau(x)],
\]
where the sum is over all permutations $\tau$ of $\cO$.

Every entry of $Q$ is $0$ or a monomial with
coefficient $1$. So the product belonging to $\tau$ is $0$ unless
\begin{equation}\label{eq:roword}
   \ag(x,\tau(x))\cap A_x=\emptyset\qquad\text{for every }x\in\cO,
\end{equation}
and if \eqref{eq:roword} holds, it is the single monomial $\prod_{x\in\cO}s^{\ag(x,\tau(x))}$, in
which $s_j$ has the exponent $\#\{x\in\cO:\tau(x)_j=x_j\}$.

By Lemma~\ref{lem:PMUQ}(iii), $\det Q$
has a lowest monomial $s^{m}$, and $m_j\le b_j(g)$ for all $j$. The coefficient of $s^{m}$ in
$\det Q$ is not zero, so for at least one permutation $\tau$ the product belonging to $\tau$ is not
zero and is equal to $s^{m}$.

This $\tau$ satisfies \eqref{eq:roword}, which is (a), and
\[
   \#\{x\in\cO:\ \tau(x)_j=x_j\}\;=\;m_j\;\le\;b_j(g)\qquad\text{for every }j\in[n],
\]
which is (b).
\end{proof}

\subsection{Examples and the relation to Reimer's Butterfly Lemma}\label{subsec:butterfly}

\begin{example}\label{ex:det-small}
(1) Let $g$ be the constant function $1$ on $Q_n$. Then $b_j(g)=0$ for all $j$, and (b) says that
$\tau(x)_j\ne x_j$ for all $x$ and $j$; so $\tau$ is the antipodal map $x\mapsto\bar x$, whatever the
wings are.

For $n=1$ the agreement matrix has diagonal entries $s_1$ and off-diagonal entries $1$;
its determinant is $s_1^{2}-1$, and the only lowest monomial is $1$, the monomial of the antipodal
map.

(2) Let $n=2$ and $g(x)=1$ if and only if $x_1=1$. Then $\cO=\{p,q\}$ with $p=(1,1)$ and
$q=(1,-1)$, the set $\{2\}$ is a $1$-wing at $p$ and at $q$, and $b_1(g)=2$, $b_2(g)=0$.

The
agreement matrix has diagonal entries $s_1s_2$ and off-diagonal entries $s_1$, so that
$\det M=s_1^{2}(s_2^{2}-1)$. The only lowest monomial is $s_1^{2}$; it is the monomial of the
transposition $\tau$ of $p$ and $q$. This $\tau$ satisfies (a) for the wings $A_p=A_q=\{2\}$, and in
(b) there is equality for both coordinates.

Moreover $Z(g)=\cO$, and
$D_g(p,\{2\};\ell)=D_g(q,\{2\};\ell)=\ell_1$, since the only admissible set is $G=\{1\}$. So the
determinant inequality reads $2\ell_1\le 2\ell_1$.

(3) For the function $g$ of Example~\ref{ex:and} the matrix $M$ has the single entry $s_1s_2$, the
bijection is the identity, and in (b) there is equality: $1\le b_j(g)=1$ for $j=1,2$.
\end{example}

\begin{remark}[Relation to Reimer's Butterfly Lemma]\label{rem:butterfly}
Part (a) of Theorem~\ref{thm:bij}, without part (b), is equivalent to the Butterfly Lemma, which
Reimer used to prove the van den Berg--Kesten conjecture \cite{Rei}.

We recall the lemma in the
formulation of Smyth \cite[Lemma~2.4]{Smy}, transcribed to
$Q_n$ and specialised to the case in which the ambient subcube is the whole cube.

A \emph{butterfly}
is a pair $B=(a,b)$ of points of $Q_n$ with \emph{body} $a$ and \emph{tip} $b$; its \emph{red} and
\emph{yellow} sets are
\begin{align*}
   \mathrm{Red}(B)&=\{b^{S}:\ S\subseteq\Delta(a,b)\},\\
   \mathrm{Yellow}(B)&=\{y\in Q_n:\ y_i\ne b_i\ \text{for all }i\in\Delta(a,b)\},
\end{align*}
the subcubes spanned by $a$ and $b$, and by $a$ and $\bar b$.

The lemma states that if
$\mathcal B$ is a family of butterflies with distinct tips, and $\mathrm{Red}(\mathcal B)$ and
$\mathrm{Yellow}(\mathcal B)$ are the unions of the red and of the yellow sets of its members, then
$\abs{\mathcal B}\le\abs{\mathrm{Red}(\mathcal B)\cap\mathrm{Yellow}(\mathcal B)}$.

\smallskip
\emph{The lemma follows from \textup{(a)}.} Let $\cO=\mathrm{Red}(\mathcal B)$ and $g=1_{\cO}$. If
$b$ is the tip of $B=(a,b)\in\mathcal B$, then $A_b=\Delta(a,b)$ is a $1$-wing of $g$ at $b$, because
$b^{S}\in\mathrm{Red}(B)\subseteq\cO$ for $S\subseteq A_b$; this is well defined since the tips are
distinct. Put $A_x=\emptyset$ for the other points $x$ of $\cO$.

A bijection $\tau$ with (a) maps
the tip $b$ of $B$ to a point of $\cO$ which differs from $b$ in every coordinate of $\Delta(a,b)$,
that is, to a point of $\mathrm{Red}(\mathcal B)\cap\mathrm{Yellow}(B)$. Since $\tau$ is injective,
$\abs{\mathcal B}\le\abs{\mathrm{Red}(\mathcal B)\cap\mathrm{Yellow}(\mathcal B)}$.

\smallskip
\emph{\textup{(a)} follows from the lemma.} Let $g$ and the wings $A_x$ be as in
Theorem~\ref{thm:bij}, and for $x\in\cO$ let $B_x$ be the butterfly with tip $x$ and body $x^{A_x}$.
Then $\mathrm{Red}(B_x)=\{x^{S}:S\subseteq A_x\}\subseteq\cO$, and
$Y_x:=\cO\cap\mathrm{Yellow}(B_x)$ is the set of points of $\cO$ which differ from $x$ in every
coordinate of $A_x$.

For every $T\subseteq\cO$ the lemma, applied to the family $\{B_x:x\in T\}$,
gives $\abs{T}\le\bigl\lvert\bigcup_{x\in T}Y_x\bigr\rvert$. By Hall's theorem on distinct
representatives \cite{Hall} there is an injective map $\tau\colon\cO\to\cO$ with $\tau(x)\in Y_x$
for all $x$. It is a bijection because $\cO$ is finite, and it satisfies (a).

Reimer's proof of the lemma \cite[Section~2]{Rei} is also by linear algebra. In the presentation of Borgs, Chayes and
Randall \cite[Section~5]{BCR}, where the lemma is stated in terms of cylinders
\cite[Lemma~4.1]{BCR}, vectors which are tensor products over the coordinates are attached
to the points, and the lemma is deduced from the linear independence of these vectors; the main
step is carried out over the field with two elements.

The proof of Theorem~\ref{thm:bij} given
above works instead with a matrix over a polynomial ring, and it obtains the bijection from a term
of lowest degree of the determinant.

Part (b), the control of the number of points which keep their $j$th coordinate, is what
Theorem~\ref{thm:bij} adds to the Butterfly Lemma. The determinant inequality (Corollary~\ref{cor:det}) uses both parts.
\end{remark}

\section{The antipodal inequality}\label{sec:anti}

The determinant inequality (Corollary~\ref{cor:det}) bounds a sum, over the points $u$ of the zone, of the distances
$D_g(u,A_u;\ell)$, for an arbitrary choice of wings. In this section the wings are chosen with the
help of rows of $g$ in the sense of Definition~\ref{def:row}.

A row at $u$ shows that a set of
coordinates of small total weight is a $1$-wing at
$u$, and a row at the antipode $\bar u$ shows that it is expensive to return from $\bar u$ to $\cO$
without using the coordinates of this wing. The result is a lower bound for $D_g(u,A_u;\ell)$ by
the reciprocal of a bilinear expression in the two rows (Lemma~\ref{lem:point}). Summing it over $u$
and applying the determinant inequality gives the antipodal inequality (Theorem~\ref{thm:anti}).

This use of the rows at a point and at its antipode has a precedent in the proof of the dual
inequality by Kahn, Saks and Smyth \cite[Section~2]{KSS}, where a certificate at a point and a
certificate at its antipode in a subcube give the two wings of a butterfly.

The antipodal inequality (Theorem~\ref{thm:anti}) is used in Section~\ref{sec:two}: applied to the
restrictions of $g$ to all subcubes, it gives the all-pairs inequality (Proposition~\ref{prop:pairs}).

We first check that the expressions whose reciprocals are taken are positive. The lemma which
shows this is due to Kulkarni and Tal \cite[Lemma~6.2]{KT}; we include its short proof.

\begin{lemma}[Overlap]\label{lem:overlap}
Let $g$ be a Boolean function on $Q_n$, let $p,q\in Q_n$ with $g(p)\ne g(q)$, let $\lambda$ be a row
of $g$ at $p$ and $\mu$ a row of $g$ at $q$. Then
\[
   \sum_{j\in\Delta(p,q)}\min(\lambda_j,\mu_j)\;\ge\;1 .
\]
In particular, $\lambda_j\mu_j>0$ for some $j\in\Delta(p,q)$, and
$\sum_{j\in\Delta(p,q)}w_j\lambda_j\mu_j>0$ for every $w\in(0,\infty)^{n}$.
\end{lemma}

\begin{proof}
Let $\Delta=\Delta(p,q)$ and $S=\{j\in\Delta:\lambda_j\le\mu_j\}$. Then $\min(\lambda_j,\mu_j)$ is
$\lambda_j$ for $j\in S$ and $\mu_j$ for $j\in\Delta\setminus S$, so
\[
   \sum_{j\in\Delta}\min(\lambda_j,\mu_j)\;=\;\lambda(S)+\mu(\Delta\setminus S).
\]
Since $q=p^{\Delta}$, we have $q^{\Delta\setminus S}=p^{S}$.

If $g(p^{S})\ne g(p)$,
then $S$ is a sensitive block at $p$ and $\lambda(S)\ge1$. Otherwise
$g(q^{\Delta\setminus S})=g(p^{S})=g(p)\ne g(q)$, so $\Delta\setminus S$ is a sensitive block at $q$
and $\mu(\Delta\setminus S)\ge1$. In both cases the sum is at least $1$.

In particular
$\min(\lambda_j,\mu_j)>0$ for some $j\in\Delta$, and the remaining statements follow because all
$\lambda_i\mu_i$ are non-negative.
\end{proof}

\begin{lemma}[Pointwise step]\label{lem:point}
Let $g$ be a Boolean function on $Q_n$, $u\in Z(g)$, $\lambda$ a row of $g$ at $u$, $\mu$ a row of
$g$ at $\bar u$, and $\ell\in(0,\infty)^{n}$. Then $\sum_{j}\lambda_j\mu_j/\ell_j>0$, and with
\[
   \omega:=\Bigl(\sum_{j=1}^{n}\frac{\lambda_j\mu_j}{\ell_j}\Bigr)^{-1},\qquad
   A:=\{j\in[n]:\ \ell_j<\omega\,\mu_j\},
\]
the set $A$ is a $1$-wing of $g$ at $u$ and $D_g(u,A;\ell)\ge\omega$.
\end{lemma}

\begin{proof}
We have $g(u)=1\ne0=g(\bar u)$ and $\Delta(u,\bar u)=[n]$, so the sum is positive by
Lemma~\ref{lem:overlap} with $w_j=1/\ell_j$. By the definition of $\omega$,
\begin{equation}\label{eq:point-one}
   \sum_{j=1}^{n}\lambda_j\cdot\frac{\omega\mu_j}{\ell_j}=1,\qquad\text{and}\qquad
   \frac{\omega\mu_j}{\ell_j}>1\ \text{ for }j\in A .
\end{equation}

\smallskip
\emph{Step 1: $\lambda(A)<1$.} If $\lambda_j=0$ for all $j\in A$ this is clear. Otherwise, by
\eqref{eq:point-one},
\[
   \lambda(A)\;\overset{\text{Sec.~\ref{subsec:cube}}}{=}\;\sum_{j\in A}\lambda_j
   \;\overset{\eqref{eq:point-one}}{<}\;\sum_{j\in A}\lambda_j\,\frac{\omega\mu_j}{\ell_j}
   \;\overset{\text{Def.~\ref{def:row}}}{\le}\;\sum_{j\in[n]}\lambda_j\,\frac{\omega\mu_j}{\ell_j}
   \;\overset{\eqref{eq:point-one}}{=}\;1 .
\]

\smallskip
\emph{Step 2: $A$ is a $1$-wing at $u$.} For $S\subseteq A$ we have $\lambda(S)\le\lambda(A)<1$.
Since $\lambda$ is a row at $u$, the set $S$ is not a sensitive block at $u$; that is,
$g(u^{S})=g(u)=1$.

\smallskip
\emph{Step 3: $D_g(u,A;\ell)\ge\omega$.} Let $G\subseteq[n]\setminus A$ with $\bar u^{\,G}\in\cO$. Since
$g(\bar u)=0$, the set $G$ is a sensitive block at $\bar u$, so $\mu(G)\ge1$. Since $G$ is disjoint
from $A$, we have $\ell_j\ge\omega\mu_j$ for all $j\in G$. Hence $\ell(G)\ge\omega\,\mu(G)\ge\omega$.
\end{proof}

\begin{theorem}[Antipodal inequality]\label{thm:anti}
Let $g$ be a Boolean function on $Q_n$, let $w\in(0,\infty)^{n}$, and for every $x\in Q_n$ let
$\lambda^{x}$ be a row of $g$ at $x$. Then $\sum_{j}w_j\lambda^{u}_j\lambda^{\bar u}_j>0$ for every
$u\in Z(g)$, and
\[
   \sum_{u\in Z(g)}\Bigl(\sum_{j=1}^{n}w_j\,\lambda^{u}_j\,\lambda^{\bar u}_j\Bigr)^{-1}
   \;\le\;\sum_{j=1}^{n}\frac{b_j(g)}{w_j}.
\]
\end{theorem}

\begin{proof}
Put $\ell_j=1/w_j$. For each $u\in Z(g)$, Lemma~\ref{lem:point} with $\lambda=\lambda^{u}$ and
$\mu=\lambda^{\bar u}$ gives the positivity and a $1$-wing $A_u$ of $g$ at $u$ with
\[
   D_g(u,A_u;\ell)\;\ge\;\Bigl(\sum_{j=1}^{n}\frac{\lambda^{u}_j\,\lambda^{\bar u}_j}{\ell_j}\Bigr)^{-1}
   \;=\;\Bigl(\sum_{j=1}^{n}w_j\,\lambda^{u}_j\,\lambda^{\bar u}_j\Bigr)^{-1}.
\]

We sum over $u\in Z(g)$ and apply Corollary~\ref{cor:det} to the wings $A_u$:
\[
   \sum_{u\in Z(g)}\Bigl(\sum_{j=1}^{n}w_j\,\lambda^{u}_j\,\lambda^{\bar u}_j\Bigr)^{-1}
   \;\overset{\text{Lem.~\ref{lem:point}}}{\le}\;\sum_{u\in Z(g)}D_g(u,A_u;\ell)
   \;\overset{\text{Cor.~\ref{cor:det}}}{\le}\;\sum_{j=1}^{n}\ell_j\,b_j(g)
   \;=\;\sum_{j=1}^{n}\frac{b_j(g)}{w_j}.\qedhere
\]
\end{proof}

\begin{example}\label{ex:anti-dictator}
Let $n\ge1$ and $g(x)=1$ if and only if $x_1=1$. A set $S$ is a sensitive block, at any point, if
and only if $1\in S$; so the vector $(1,0,\dots,0)$ is a row of $g$ at every point. Take
$\lambda^{x}=(1,0,\dots,0)$ for all $x$. Then $Z(g)=\cO$ has $2^{n-1}$ points,
$\sum_jw_j\lambda^{u}_j\lambda^{\bar u}_j=w_1$, $b_1(g)=2^{n-1}$ and $b_j(g)=0$ for $j\ge2$. Both
sides of the antipodal inequality are equal to $2^{n-1}/w_1$.
\end{example}

\section{The all-pairs inequality and the two-sided inequality}\label{sec:two}

In the first part of this section, we apply the antipodal inequality (Theorem~\ref{thm:anti}) to the restrictions of $g$ to all
subcubes; the sum of the resulting inequalities is an inequality for all pairs
$(x,y)\in\cO\times\cN$. This reduction of a statement about all pairs of points to statements about
the antipodal pairs of the subcubes is the one by which the van den Berg--Kesten--Reimer inequality
is derived from the Butterfly Lemma \cite[Section~3]{Rei} (see also \cite[Section~4]{BCR}), and by
which Kahn, Saks and Smyth derive the dual inequality \cite[Section~2]{KSS}.

In the second part the weights $w$ are removed: the Cauchy--Schwarz
inequality, applied to a sum over the pairs, leads to an expression in $w$ which can be minimised
explicitly.

The all-pairs inequality (Proposition~\ref{prop:pairs}) and the removal of the weights
(Lemma~\ref{lem:weights}) are used in Section~\ref{sec:level}.

\subsection{From antipodal pairs to all pairs}

For $F\subseteq[n]$ and $z,u\in Q_n$ let $\gamma_F(z,u)\in Q_n$ be the point with
\[
   \gamma_F(z,u)_i=u_i\quad(i\in F),\qquad \gamma_F(z,u)_i=z_i\quad(i\notin F).
\]
For fixed $z$ the points $\gamma_F(z,u)$, $u\in Q_n$, form the subcube through $z$ in which the
coordinates in $F$ are free.

\begin{lemma}[Gluing]\label{lem:glue}
Let $F\subseteq[n]$.
\begin{itemize}
\item[\textup{(i)}] $\gamma_F(z,u^{S})=\gamma_F(z,u)^{S\cap F}$ for all $z,u\in Q_n$ and
$S\subseteq[n]$. In particular $\gamma_F(z,\bar u)=\gamma_F(z,u)^{F}$, and
$\gamma_F(z,u^{j})$ is equal to $\gamma_F(z,u)^{j}$ if $j\in F$ and to $\gamma_F(z,u)$ if
$j\notin F$.
\item[\textup{(ii)}] For every function $\varphi\colon Q_n\to\R$,
\[
   \sum_{z\in Q_n}\ \sum_{u\in Q_n}\varphi\bigl(\gamma_F(z,u)\bigr)\;=\;2^{n}\sum_{x\in Q_n}\varphi(x).
\]
\end{itemize}
\end{lemma}

\begin{proof}
(i) For $i\notin F$ both points have the $i$th coordinate $z_i$. For $i\in F$ both have the $i$th
coordinate $-u_i$ if $i\in S$, and $u_i$ if $i\notin S$. The special cases are $S=[n]$ and
$S=\{j\}$.

(ii) Consider the map $\Phi(z,u)=\bigl(\gamma_F(z,u),\gamma_F(u,z)\bigr)$ from $Q_n\times Q_n$ to
itself. Let $x=\gamma_F(z,u)$ and $x'=\gamma_F(u,z)$. The point $\gamma_F(x,x')$ agrees with $x'$,
hence with $z$, in the coordinates of $F$, and with $x$, hence with $z$, in the other coordinates;
so $\gamma_F(x,x')=z$. In the same way $\gamma_F(x',x)=u$.

Thus $\Phi(\Phi(z,u))=(z,u)$, and $\Phi$
is a bijection. Therefore
\[
   \sum_{(z,u)\in Q_n\times Q_n}\varphi\bigl(\gamma_F(z,u)\bigr)
   \;=\;\sum_{(x,x')\in Q_n\times Q_n}\varphi(x)\;=\;2^{n}\sum_{x\in Q_n}\varphi(x). \qedhere
\]
\end{proof}

\begin{proposition}[All-pairs inequality]\label{prop:pairs}
Let $g$ be a Boolean function on $Q_n$, let $w\in(0,\infty)^{n}$, and for every $x\in Q_n$ let
$\lambda^{x}$ be a row of $g$ at $x$. Then
$\sum_{j\in\Delta(x,y)}w_j\lambda^{x}_j\lambda^{y}_j>0$ for all $x\in\cO$ and $y\in\cN$, and
\begin{equation}\label{eq:pairs}
   \sum_{x\in\cO}\ \sum_{y\in\cN}\Bigl(\sum_{j\in\Delta(x,y)}w_j\,\lambda^{x}_j\,\lambda^{y}_j\Bigr)^{-1}
   \;\le\;2^{n-1}\sum_{j=1}^{n}\frac{b_j(g)}{w_j}.
\end{equation}
\end{proposition}

\begin{proof}
The positivity is contained in Lemma~\ref{lem:overlap}.

\smallskip
\emph{Step 1: one set of free coordinates.} Fix $F\subseteq[n]$. For $x\in Q_n$ put
\[
   T_F(x)=\Bigl(\sum_{j\in F}w_j\,\lambda^{x}_j\,\lambda^{x^{F}}_j\Bigr)^{-1}
   \ \ \text{if $g(x)=1$ and $g(x^{F})=0$,}\qquad T_F(x)=0\ \ \text{otherwise.}
\]
In the first case the sum is positive by Lemma~\ref{lem:overlap}, because $\Delta(x,x^{F})=F$.

We
show that
\begin{equation}\label{eq:pairsF}
   \sum_{x\in Q_n}T_F(x)\;\le\;\sum_{j\in F}\frac{b_j(g)}{w_j}.
\end{equation}

Fix $z\in Q_n$. Let $g_{z}$ be the Boolean function $g_{z}(u)=g(\gamma_F(z,u))$ on $Q_n$ (this is
the restriction of $g$ to the subcube through $z$ with free coordinates $F$, written as a function
on the whole cube which does not depend on the coordinates outside $F$; so the antipodal
inequality applies without renumbering the coordinates), and
for $u\in Q_n$ let $\mu^{u}\in[0,\infty)^{n}$ be the vector with
\[
   \mu^{u}_j=\lambda^{\gamma_F(z,u)}_j\quad(j\in F),\qquad \mu^{u}_j=0\quad(j\notin F).
\]

Then $\mu^{u}$ is a row of $g_{z}$ at $u$. Indeed, let $S$ be a sensitive block of $g_{z}$ at $u$
and $x=\gamma_F(z,u)$. By Lemma~\ref{lem:glue}(i),
$g(x^{S\cap F})=g_{z}(u^{S})\ne g_{z}(u)=g(x)$; so $S\cap F$ is a sensitive block of $g$ at $x$, and
$\mu^{u}(S)=\lambda^{x}(S\cap F)\ge1$.

We apply Theorem~\ref{thm:anti} to $g_{z}$ with the weights
$w$ and the rows $\mu^{u}$:
\begin{equation}\label{eq:anti-z}
   \sum_{u\in Z(g_{z})}\Bigl(\sum_{j=1}^{n}w_j\,\mu^{u}_j\,\mu^{\bar u}_j\Bigr)^{-1}
   \;\le\;\sum_{j=1}^{n}\frac{b_j(g_{z})}{w_j}.
\end{equation}

\smallskip
\emph{The left-hand side of \eqref{eq:anti-z}.} Let $u\in Q_n$ and $x=\gamma_F(z,u)$. By
Lemma~\ref{lem:glue}(i), $\gamma_F(z,\bar u)=x^{F}$. Hence $u\in Z(g_{z})$ if and only if $g(x)=1$
and $g(x^{F})=0$, and
\[
   \sum_{j}w_j\,\mu^{u}_j\,\mu^{\bar u}_j\;=\;\sum_{j\in F}w_j\,\lambda^{x}_j\,\lambda^{x^{F}}_j .
\]
So the left-hand
side is $\sum_{u\in Q_n}T_F(\gamma_F(z,u))$.

\smallskip
\emph{The right-hand side of \eqref{eq:anti-z}.} For $x\in Q_n$ let $\beta_j(x)=1$ if $g(x)=1$ and
$g(x^{j})=0$, and $\beta_j(x)=0$ otherwise; thus $b_j(g)=\sum_{x\in Q_n}\beta_j(x)$.

If $j\notin F$,
then $g_{z}(u^{j})=g_{z}(u)$ for all $u$ by Lemma~\ref{lem:glue}(i), and so $b_j(g_{z})=0$. If
$j\in F$, then $g_{z}(u^{j})=g(\gamma_F(z,u)^{j})$, and so
\[
   b_j(g_{z})\;=\;\sum_{u\in Q_n}\beta_j\bigl(\gamma_F(z,u)\bigr).
\]

Therefore \eqref{eq:anti-z} reads
\[
   \sum_{u\in Q_n}T_F\bigl(\gamma_F(z,u)\bigr)\;\le\;
   \sum_{j\in F}\frac1{w_j}\sum_{u\in Q_n}\beta_j\bigl(\gamma_F(z,u)\bigr).
\]

This holds for every $z\in Q_n$. Adding over all $z$, we get
\begin{align*}
   2^{n}\sum_{x\in Q_n}T_F(x)
   &\;\overset{\text{Lem.~\ref{lem:glue}(ii)}}{=}\;\sum_{z\in Q_n}\sum_{u\in Q_n}T_F\bigl(\gamma_F(z,u)\bigr)
   \;\le\;\sum_{j\in F}\frac1{w_j}\sum_{z\in Q_n}\sum_{u\in Q_n}\beta_j\bigl(\gamma_F(z,u)\bigr)\\
   &\;\overset{\text{Lem.~\ref{lem:glue}(ii)}}{=}\;2^{n}\sum_{j\in F}\frac{b_j(g)}{w_j},
\end{align*}
which is \eqref{eq:pairsF} multiplied by $2^{n}$.

\smallskip
\emph{Step 2: all sets of free coordinates.} We add \eqref{eq:pairsF} over all $F\subseteq[n]$.
On the left, fix $x\in Q_n$. The map $F\mapsto x^{F}$ is a bijection from the set of all subsets of
$[n]$ onto $Q_n$, with inverse $y\mapsto\Delta(x,y)$. So $\sum_{F}T_F(x)=0$ if $g(x)=0$, and if
$g(x)=1$, then
\[
   \sum_{F\subseteq[n]}T_F(x)\;=\;\sum_{y\in\cN}
   \Bigl(\sum_{j\in\Delta(x,y)}w_j\,\lambda^{x}_j\,\lambda^{y}_j\Bigr)^{-1}.
\]
Summing over $x$ gives the left-hand side of \eqref{eq:pairs}.

On the right, every $j\in[n]$ belongs
to exactly $2^{n-1}$ of the subsets of $[n]$, because $F\mapsto F\triangle\{j\}$ exchanges the
subsets which contain $j$ and those which do not. Hence
\[
   \sum_{F\subseteq[n]}\ \sum_{j\in F}\frac{b_j(g)}{w_j}\;=\;2^{n-1}\sum_{j}\frac{b_j(g)}{w_j}. \qedhere
\]
\end{proof}

\subsection{Removing the weights}

We use the Cauchy--Schwarz inequality for finite families of real numbers,
\[
   \Bigl(\sum_{i}a_ib_i\Bigr)^{2}\;\le\;\Bigl(\sum_ia_i^{2}\Bigr)\Bigl(\sum_ib_i^{2}\Bigr).
\]

The following lemma contains the step in which the weights are removed. It is formulated for two
finite families of non-negative vectors, because it is applied twice: to the rows of a Boolean
function in the proof of Theorem~\ref{thm:two}, and to multiples of the rows of a real function in
the proof of Theorem~\ref{thm:third}.

\begin{lemma}[Removing the weights]\label{lem:weights}
Let $A$ and $B$ be non-empty finite sets. For every $x\in A$ let $\lambda^{x}\in[0,\infty)^{n}$,
for every $y\in B$ let $\mu^{y}\in[0,\infty)^{n}$, and for every $(x,y)\in A\times B$ let $J(x,y)$
be a subset of $[n]$. Let $C\ge0$ be a real number. Suppose that for every $w\in(0,\infty)^{n}$ the
following two conditions are satisfied:
\begin{itemize}
\item[\textup{(i)}] $\sum_{j\in J(x,y)}w_j\lambda^{x}_j\mu^{y}_j>0$ for all $(x,y)\in A\times B$;
\item[\textup{(ii)}] $\displaystyle\sum_{x\in A}\ \sum_{y\in B}
\Bigl(\sum_{j\in J(x,y)}w_j\,\lambda^{x}_j\,\mu^{y}_j\Bigr)^{-1}\;\le\;C\sum_{j=1}^{n}\frac1{w_j}$.
\end{itemize}
Then $\norm{\lambda^{x}}>0$ for every $x\in A$, $\norm{\mu^{y}}>0$ for every $y\in B$, and
\[
   \Bigl(\sum_{x\in A}\frac1{\norm{\lambda^{x}}}\Bigr)
   \Bigl(\sum_{y\in B}\frac1{\norm{\mu^{y}}}\Bigr)\;\le\;C .
\]
\end{lemma}

\begin{proof}
Let $x\in A$. Since $B$ is not empty, there is a $y\in B$, and by (i) with $w=(1,\dots,1)$ there is
a $j$ with $\lambda^{x}_j\mu^{y}_j>0$; so $\norm{\lambda^{x}}\ge\lambda^{x}_j>0$. In the same way
$\norm{\mu^{y}}>0$ for every $y\in B$. Put
\[
   v(x)=\frac1{\norm{\lambda^{x}}}\quad(x\in A),\qquad u(y)=\frac1{\norm{\mu^{y}}}\quad(y\in B).
\]

Further let
\[
   V_{A}=\sum_{x\in A}v(x),\quad V_{B}=\sum_{y\in B}u(y),\quad
   X_j=\sum_{x\in A}v(x)^{2}\lambda^{x}_j,\quad Y_j=\sum_{y\in B}u(y)^{2}\mu^{y}_j
   \quad(j\in[n]).
\]
The numbers $X_j$ and $Y_j$ are non-negative, and $V_{A}>0$ and $V_{B}>0$, because $A$ and $B$ are
not empty.

Since $v(x)^{2}\norm{\lambda^{x}}=v(x)$ for every $x\in A$, and
$u(y)^{2}\norm{\mu^{y}}=u(y)$ for every $y\in B$,
\begin{equation}\label{eq:XY}
   \sum_{j=1}^{n}X_j=\sum_{x\in A}v(x)^{2}\norm{\lambda^{x}}=\sum_{x\in A}v(x)=V_{A},\qquad
   \sum_{j=1}^{n}Y_j=V_{B}.
\end{equation}

\smallskip
\emph{Step 1: the Cauchy--Schwarz inequality over the pairs.} Fix $w\in(0,\infty)^{n}$, and for
$(x,y)\in A\times B$ let $s_{xy}=\sum_{j\in J(x,y)}w_j\lambda^{x}_j\mu^{y}_j$, which is
positive by (i). Writing
$v(x)u(y)=v(x)u(y)\sqrt{s_{xy}}\cdot(1/\sqrt{s_{xy}})$, we get from the Cauchy--Schwarz inequality
\[
   (V_{A}V_{B})^{2}=\Bigl(\sum_{x\in A}\sum_{y\in B}v(x)\,u(y)\Bigr)^{2}
   \;\le\;\Bigl(\sum_{x\in A}\sum_{y\in B}v(x)^{2}u(y)^{2}\,s_{xy}\Bigr)
   \Bigl(\sum_{x\in A}\sum_{y\in B}\frac1{s_{xy}}\Bigr).
\]

By (ii), the second factor is at most $C\sum_j1/w_j$.

In the first factor we have
$s_{xy}\le\sum_{j=1}^{n}w_j\lambda^{x}_j\mu^{y}_j$, because the terms with $j\notin J(x,y)$
are non-negative; hence
\begin{align*}
   \sum_{x\in A}\sum_{y\in B}v(x)^{2}u(y)^{2}\,s_{xy}
   &\;\le\;\sum_{x\in A}\sum_{y\in B}v(x)^{2}u(y)^{2}\sum_{j=1}^{n}w_j\lambda^{x}_j\mu^{y}_j\\
   &\;=\;\sum_{j=1}^{n}w_j\Bigl(\sum_{x\in A}v(x)^{2}\lambda^{x}_j\Bigr)
                 \Bigl(\sum_{y\in B}u(y)^{2}\mu^{y}_j\Bigr)\;=\;\sum_{j=1}^{n}w_jX_jY_j .
\end{align*}

All the factors are non-negative, so that
\begin{equation}\label{eq:allw}
   (V_{A}V_{B})^{2}\;\le\;C\,\Bigl(\sum_{j=1}^{n}w_jX_jY_j\Bigr)\Bigl(\sum_{j=1}^{n}\frac1{w_j}\Bigr)
   \qquad\text{for every }w\in(0,\infty)^{n}.
\end{equation}

\smallskip
\emph{Step 2: the choice of the weights.} Let $r_j=\sqrt{X_jY_j}$ and $R=\sum_{j}r_j$. For $\eps>0$
the weights $w_j=1/(r_j+\eps)$ are positive, and
$w_jX_jY_j=r_j^{2}/(r_j+\eps)\le r_j$ and $1/w_j=r_j+\eps$. So \eqref{eq:allw} gives
\[
   (V_{A}V_{B})^{2}\;\le\;C\,R\,(R+n\eps)\qquad\text{for every }\eps>0,
\]
and letting $\eps\to0$ we obtain
\[
   (V_{A}V_{B})^{2}\;\le\;C\,R^{2}.
\]

(The weights $1/r_j$, which minimise the right-hand side of \eqref{eq:allw} when all $r_j$ are
positive, are not available if some $r_j$ is $0$.)

\smallskip
\emph{Step 3: the Cauchy--Schwarz inequality over the coordinates.} By the Cauchy--Schwarz
inequality and \eqref{eq:XY},
\[
   R^{2}=\Bigl(\sum_{j=1}^{n}\sqrt{X_j}\,\sqrt{Y_j}\Bigr)^{2}\;\le\;
   \Bigl(\sum_{j=1}^{n}X_j\Bigr)\Bigl(\sum_{j=1}^{n}Y_j\Bigr)=V_{A}V_{B}.
\]

\smallskip
\emph{Conclusion.} Steps 2 and 3 give $(V_{A}V_{B})^{2}\le C\,V_{A}V_{B}$. Since
$V_{A}V_{B}>0$ we may divide, and get $V_{A}V_{B}\le C$, which is the assertion.
\end{proof}

For the rows of a Boolean function, Lemma~\ref{lem:weights} and the all-pairs inequality (Proposition~\ref{prop:pairs}) give the two-sided
inequality.

\begin{theorem}[Two-sided inequality]\label{thm:two}
Let $g\colon Q_n\to\{0,1\}$ be non-constant, and for every $x\in Q_n$ let $\lambda^{x}$ be a row of
$g$ at $x$. Then $\norm{\lambda^{x}}\ge1$ for every $x$, and
\[
   4\;\E_x\Bigl[\frac{1_{\{g(x)=1\}}}{\norm{\lambda^{x}}}\Bigr]\;
      \E_x\Bigl[\frac{1_{\{g(x)=0\}}}{\norm{\lambda^{x}}}\Bigr]\;\le\;\max_{j\in[n]}\I_j(g).
\]
\end{theorem}

\begin{proof}
By Lemma~\ref{lem:rows-basic}(ii), $n\ge1$ and $\norm{\lambda^{x}}\ge1$ for every $x$. Put
\[
   \I_{\max}=\max_{j\in[n]}\I_j(g),\qquad C=4^{\,n-1}\I_{\max} .
\]
By Lemma~\ref{lem:bj}, $b_j(g)=2^{n-1}\I_j(g)\le2^{n-1}\I_{\max}$ for every $j$.

We apply Lemma~\ref{lem:weights} to the sets $A=\cO$ and $B=\cN$, which are not empty because $g$
is not constant, to the vectors $\lambda^{x}$ $(x\in\cO)$ and $\mu^{y}=\lambda^{y}$ $(y\in\cN)$, to
the sets $J(x,y)=\Delta(x,y)$, and to the constant $C\ge0$.

Let $w\in(0,\infty)^{n}$. Condition (i) of
the lemma holds by Proposition~\ref{prop:pairs}. By \eqref{eq:pairs} and
$b_j(g)\le2^{n-1}\I_{\max}$,
\[
   \sum_{x\in\cO}\ \sum_{y\in\cN}\Bigl(\sum_{j\in\Delta(x,y)}w_j\,\lambda^{x}_j\,\lambda^{y}_j\Bigr)^{-1}
   \;\le\;2^{n-1}\sum_{j=1}^{n}\frac{2^{n-1}\,\I_{\max}}{w_j}\;=\;C\sum_{j=1}^{n}\frac1{w_j},
\]
which is condition (ii). The lemma therefore gives
\[
   \Bigl(\sum_{x\in\cO}\frac1{\norm{\lambda^{x}}}\Bigr)
   \Bigl(\sum_{y\in\cN}\frac1{\norm{\lambda^{y}}}\Bigr)\;\le\;C\;=\;4^{\,n-1}\,\I_{\max}.
\]

Finally,
\begin{align*}
   4\;\E_x\Bigl[\frac{1_{\{g(x)=1\}}}{\norm{\lambda^{x}}}\Bigr]\;
      \E_x\Bigl[\frac{1_{\{g(x)=0\}}}{\norm{\lambda^{x}}}\Bigr]
   &\;=\;4\cdot4^{-n}\Bigl(\sum_{x\in\cO}\frac1{\norm{\lambda^{x}}}\Bigr)
     \Bigl(\sum_{y\in\cN}\frac1{\norm{\lambda^{y}}}\Bigr)\\
   &\;\le\;4\cdot4^{-n}\cdot4^{\,n-1}\,\I_{\max}\;=\;\I_{\max}. \qedhere
\end{align*}
\end{proof}

\begin{example}[Equality]\label{ex:dictator}
Let $n\ge1$ and $g(x)=1$ if and only if $x_1=1$, with the rows $\lambda^{x}=(1,0,\dots,0)$
of Example~\ref{ex:anti-dictator}. Then $\norm{\lambda^{x}}=1$ for all $x$, the two expectations in
Theorem~\ref{thm:two} are equal to $\frac12$, and $\I_1(g)=1$; so both sides of the two-sided
inequality are equal to $1$. In \eqref{eq:pairs} there is equality as well: every pair
$(x,y)\in\cO\times\cN$ has $1\in\Delta(x,y)$ and
$\sum_{j\in\Delta(x,y)}w_j\lambda^{x}_j\lambda^{y}_j=w_1$, so the left-hand side is
$2^{n-1}\cdot2^{n-1}/w_1$, and the right-hand side is $2^{n-1}\,b_1(g)/w_1$ with $b_1(g)=2^{n-1}$.
\end{example}

\section{The row bound}\label{sec:row}

In this section we prove that a function $f\colon Q_n\to[-1,1]$ of degree at most $d$ has, at every
point $x$, a row of cost at most $\pi^{2}d^{2}$ (Theorem~\ref{thm:row}). The idea, which is that of the
proof of Agarwal and Ben-David \cite{ABD} (see Remark~\ref{rem:ABD}), is as follows.

By linear programming
duality, it suffices to bound the total sensitivity $\sum_S\kappa_S\abs{f(x){-}f(x^{S})}$ of every
\emph{fractional packing} $\kappa$ of blocks, in which the blocks through any one coordinate have total
weight at most $1$. A fractional packing is approximated by a list of blocks in which every coordinate
occurs at most $N$ times.

For such a list we construct a bounded function $G$ of low degree in new
variables, one for each block of the list, such that, at the point where all new variables are equal
to $1$, changing the sign of one new variable has the same effect as flipping the corresponding block
in $f$.

The construction composes $f$ with a Chebyshev
polynomial, which takes the value $1$ at a coordinate through which no block is flipped and the value
$-1$ at a coordinate through which exactly one block is flipped, and stays in $[-1,1]$ otherwise.
Markov's inequality then bounds the sum of the changes of $G$.

The inequality for lists of blocks (Lemma~\ref{lem:block}) has a second consequence, which is
recorded at the end of the section: at every point $x$, the sum of the numbers
$\abs{f(x)-f(x^{j})}$ is at most $4d^{2}$ (Corollary~\ref{cor:sens-point}). It is used in
Section~\ref{sec:avg}.

Both results are known, with better constants (Remark~\ref{rem:ABD}); they are proved here to make
the paper self-contained. A reader who accepts Theorem~\ref{thm:row} and
Corollary~\ref{cor:sens-point} may skip this section: nothing else from it is used in the proof of
Theorem~\ref{thm:main}.

\subsection{Chebyshev polynomials and Markov's inequality at an endpoint}

The \emph{Chebyshev polynomials} $T_0,T_1,T_2,\dots$ are defined by
\[
   T_0(t)=1,\qquad T_1(t)=t,\qquad T_{k+1}(t)=2t\,T_k(t)-T_{k-1}(t)\quad(k\ge1).
\]
We use the following standard facts (see \cite{BE}), valid for every integer $k\ge0$.
\begin{itemize}
\item[(C1)] $T_k$ is a polynomial with real coefficients of degree at most $k$.
\item[(C2)] $T_k(\cos\theta)=\cos(k\theta)$ for every real $\theta$.
\item[(C3)] $\abs{T_k(t)}\le1$ for $t\in[-1,1]$; $T_k(1)=1$; and $T_k(\cos(\pi/k))=-1$ if $k\ge1$.
\item[(C4)] $T_k'(1)=k^{2}$.
\end{itemize}
Indeed, (C1) is clear from the recurrence.

(C2) holds for $k=0$ and $k=1$, and the step from $k-1$ and
$k$ to $k+1$ is the identity
\[
   \cos((k+1)\theta)+\cos((k-1)\theta)\;=\;2\cos\theta\cos(k\theta).
\]

(C3)
follows from (C2), because every $t\in[-1,1]$ is of the form $\cos\theta$, $1=\cos0$, and
$\cos(k\cdot\pi/k)=-1$.

For (C4), differentiate the recurrence and put $t=1$: by (C3),
\[
   T_{k+1}'(1)\;=\;2+2\,T_k'(1)-T_{k-1}'(1)\qquad\text{for }k\ge1.
\]
The numbers $k^{2}$ satisfy the same recurrence,
\[
   (k+1)^{2}\;=\;2+2k^{2}-(k-1)^{2},
\]
and the initial values agree: $T_0'(1)=0$ and $T_1'(1)=1$.

Markov's inequality states that a real polynomial $p$ of degree at most $D$ with $\abs{p}\le1$ on
$[-1,1]$ satisfies $\abs{p'}\le D^{2}$ on $[-1,1]$; see \cite{BE}. We need it only at an endpoint,
where it has the following short proof.

\begin{lemma}[Markov's inequality at an endpoint]\label{lem:markov}
Let $D\ge0$ be an integer and let $p$ be a polynomial with real coefficients of degree at most $D$ such
that $\abs{p(t)}\le1$ for all $t\in[-1,1]$. Then $\abs{p'(1)}\le D^{2}$. Consequently, if $q$ is a
polynomial with real coefficients of degree at most $D$ such that $\abs{q(s)}\le1$ for all
$s\in[0,2]$, then $\abs{q'(0)}\le D^{2}$.
\end{lemma}

\begin{proof}
\emph{Step 1: the nodes.} If $D=0$ then $p$ is constant and $p'(1)=0$. Let $D\ge1$, and put
$\eta_k=\cos(k\pi/D)$ for
$k=0,1,\dots,D$. Since the cosine is strictly decreasing on $[0,\pi]$,
\[
   1=\eta_0>\eta_1>\dots>\eta_D=-1 .
\]

\smallskip
\emph{Step 2: the interpolation formula.} For $k=0,\dots,D$ let
\[
   \psi_k(t)=\prod_{i\ne k}\frac{t-\eta_i}{\eta_k-\eta_i},
\]
where $i$ runs over $\{0,\dots,D\}\setminus\{k\}$. These are polynomials of degree $D$ with
$\psi_k(\eta_k)=1$ and $\psi_k(\eta_i)=0$ for $i\ne k$.

If $r$ is a polynomial of degree at most $D$, then
$r-\sum_kr(\eta_k)\psi_k$ has degree at most $D$ and vanishes at the $D+1$ points $\eta_0,\dots,\eta_D$,
so it is the zero polynomial. Hence
\begin{equation}\label{eq:lagrange}
   r'(1)=\sum_{k=0}^{D}r(\eta_k)\,\psi_k'(1)\qquad\text{for every polynomial $r$ of degree at most $D$.}
\end{equation}

\smallskip
\emph{Step 3: the signs.} Next we show that
\begin{equation}\label{eq:Lsign}
   (-1)^{k}\psi_k'(1)>0\qquad\text{for every }k\in\{0,\dots,D\}.
\end{equation}

For $k=0$, all factors of $\psi_0$ are equal to $1$ at
$t=1$, so
\[
   \psi_0'(1)\;=\;\sum_{i=1}^{D}\frac1{1-\eta_i}\;>\;0 .
\]

For $k\ge1$ the polynomial $\psi_k$ has the factor
$t-\eta_0=t-1$, and therefore
\[
   \psi_k'(1)=\frac{\prod_{i\ne0,k}(1-\eta_i)}{\prod_{i\ne k}(\eta_k-\eta_i)} .
\]
The numerator is positive. In the denominator the factor $\eta_k-\eta_i$ is negative for the $k$
indices $i<k$ and positive for the indices $i>k$, so the denominator has the sign $(-1)^{k}$. This
proves \eqref{eq:Lsign}.

\smallskip
\emph{Step 4: comparison with $T_D$.} By (C1) and (C2), the polynomial $T_D$ has degree at most
$D$ and
$T_D(\eta_k)=\cos(k\pi)=(-1)^{k}$. Applying \eqref{eq:lagrange} to $p$ and to $T_D$, and using
$\abs{p(\eta_k)}\le1$, \eqref{eq:Lsign} and (C4), we get
\[
   \abs{p'(1)}\;\le\;\sum_{k=0}^{D}\abs{\psi_k'(1)}\;=\;\sum_{k=0}^{D}(-1)^{k}\psi_k'(1)\;=\;T_D'(1)\;=\;D^{2}.
\]

\smallskip
\emph{Step 5: the interval $[0,2]$.} For the last statement apply the first one to $p(t)=q(1-t)$:
its degree is at most $D$, the number
$1-t$ runs over $[0,2]$ when $t$ runs over $[-1,1]$, and $p'(1)=-q'(0)$.
\end{proof}

\subsection{Bounded functions and their multilinear extensions}

Let $h\colon Q_n\to\R$ have the expansion $h=\sum_{V\subseteq[n]}\hat h(V)\chi_V$ of
\eqref{eq:expansion}. Its \emph{multilinear extension} is the function
\[
   \tilde h(u)=\sum_{V\subseteq[n]}\hat h(V)\prod_{i\in V}u_i\qquad(u\in\R^{n}),
\]
which agrees with $h$ on $Q_n$.

The extension is bounded on the solid cube, because its values there are averages of values of
$h$.

\begin{lemma}[Extension]\label{lem:extension}
If\/ $\abs{h(z)}\le1$ for all $z\in Q_n$, then $\abs{\tilde h(u)}\le1$ for all $u\in[-1,1]^{n}$.
\end{lemma}

\begin{proof}
Fix $u\in[-1,1]^{n}$ and put $W_u(z)=\prod_{i=1}^{n}\frac{1+z_iu_i}{2}$ for $z\in Q_n$; these numbers
are non-negative.

Let $V\subseteq[n]$. The sum over $z\in Q_n$ of the product $W_u(z)\chi_V(z)$ is the
product of the sums over the single coordinates $z_i\in\{-1,1\}$:
\[
   \sum_{z\in Q_n}W_u(z)\chi_V(z)
   =\prod_{i\in V}\Bigl(\frac{1+u_i}{2}-\frac{1-u_i}{2}\Bigr)\,
    \prod_{i\notin V}\Bigl(\frac{1+u_i}{2}+\frac{1-u_i}{2}\Bigr)
   =\prod_{i\in V}u_i .
\]

For $V=\emptyset$ this gives $\sum_zW_u(z)=1$. Multiplying by $\hat h(V)$ and summing over $V$ we get
\[
   \tilde h(u)\;=\;\sum_{z}W_u(z)\,h(z).
\]
Hence
\[
   \abs{\tilde h(u)}\;\le\;\sum_zW_u(z)\abs{h(z)}\;\le\;\sum_zW_u(z)\;=\;1. \qedhere
\]
\end{proof}

\subsection{The sensitivity lemma}

In the next two lemmas new variables are indexed by $[k]$, and $\one=(1,\dots,1)\in Q_k$. Thus
$\one^{b}$ is the point of $Q_k$ whose $b$th coordinate is $-1$ and whose other coordinates are $1$.

The next lemma is the form in which Markov's inequality is used.

\begin{lemma}[Sensitivity]\label{lem:sens}
Let $k\ge0$ and $D\ge0$ be integers and let $G\colon Q_k\to[-1,1]$ be a function of degree at most
$D$. Then
\[
   \sum_{b=1}^{k}\abs{G(\one)-G(\one^{b})}\;\le\;4D^{2}.
\]
\end{lemma}

\begin{proof}
Let $G=\sum_{U\subseteq[k]}\widehat G(U)\chi_U$, with $\widehat G(U)=0$ for $\abs{U}>D$. Since $\chi_U(\one)=1$,
while $\chi_U(\one^{b})$ equals $-1$ if $b\in U$ and $1$ if $b\notin U$, we have
$G(\one)-G(\one^{b})=2\sum_{U\ni b}\widehat G(U)$. Hence, for every $R\subseteq[k]$,
\begin{equation}\label{eq:sens-R}
   \sum_{b\in R}\bigl(G(\one)-G(\one^{b})\bigr)=2\sum_{U\subseteq[k]}\widehat G(U)\,\abs{U\cap R}.
\end{equation}

Fix $R$, and for real $s$ let $u(s)\in\R^{k}$ be the point with $u(s)_b=1-s$ for $b\in R$ and
$u(s)_b=1$ for $b\notin R$. Put
\[
   \varphi_R(s)=\tilde G(u(s))=\sum_{U\subseteq[k]}\widehat G(U)\,(1-s)^{\abs{U\cap R}} .
\]

This is a polynomial in $s$ of degree at most $D$, because $\abs{U\cap R}\le\abs{U}\le D$ whenever
$\widehat G(U)\ne0$. For $s\in[0,2]$ the point $u(s)$ lies in $[-1,1]^{k}$, so $\abs{\varphi_R(s)}\le1$ by
Lemma~\ref{lem:extension}.

Lemma~\ref{lem:markov} gives $\abs{\varphi_R'(0)}\le D^{2}$. Since
$\varphi_R'(0)=-\sum_U\widehat G(U)\abs{U\cap R}$, we conclude from \eqref{eq:sens-R} that
\[
   \Bigl|\sum_{b\in R}\bigl(G(\one)-G(\one^{b})\bigr)\Bigr|\le2D^{2}
   \qquad\text{for every }R\subseteq[k].
\]

Applying this to the set $R_+$ of all $b$ with $G(\one)-G(\one^{b})\ge0$ and to its complement $R_-$,
we obtain
\[
   \sum_{b=1}^{k}\abs{G(\one)-G(\one^{b})}
   =\sum_{b\in R_+}\bigl(G(\one)-G(\one^{b})\bigr)-\sum_{b\in R_-}\bigl(G(\one)-G(\one^{b})\bigr)
   \le4D^{2}. \qedhere
\]
\end{proof}

\subsection{The block inequality}

We now carry out the construction described at the beginning of the section.

\begin{lemma}[Block inequality]\label{lem:block}
Let $d\ge0$ be an integer, let $f\colon Q_n\to[-1,1]$ be a function of degree at most $d$, and let
$x\in Q_n$. Let
$S_1,\dots,S_k$ be subsets of $[n]$, not necessarily distinct, and let $N\ge0$ and $m\ge1$ be integers
such that every $i\in[n]$ lies in at most $N$ of the sets $S_1,\dots,S_k$ and
\[
   N\bigl(1-\cos(\pi/m)\bigr)\le2 .
\]
Then
\[
   \sum_{b=1}^{k}\abs{f(x)-f(x^{S_b})}\;\le\;4\,d^{2}m^{2}.
\]
\end{lemma}

\begin{proof}
Put $c=1-\cos(\pi/m)$, so that $c\ge0$ and $Nc\le2$. For $w\in Q_k$ and $i\in[n]$ let
\[
   \cnt_i(w)=\#\{b\in[k]:\ i\in S_b,\ w_b=-1\}=\sum_{b:\,i\in S_b}\frac{1-w_b}{2},
   \qquad \theta_i(w)=T_m\bigl(1-c\,\cnt_i(w)\bigr).
\]

Let $f=\sum_{V\subseteq[n]}\hat f(V)\chi_V$ with $\hat f(V)=0$ for $\abs{V}>d$, and define
$G\colon Q_k\to\R$ by
\[
   G(w)=\tilde f\bigl(x_1\theta_1(w),\dots,x_n\theta_n(w)\bigr)
       =\sum_{V\subseteq[n]}\hat f(V)\prod_{i\in V}x_i\,\theta_i(w).
\]

The function $G$ has the following four properties.

\smallskip
\emph{\textup{(a)} $\abs{G(w)}\le1$ for all $w\in Q_k$.} Since $0\le\cnt_i(w)\le N$, the number
$1-c\,\cnt_i(w)$ lies in $[1-Nc,1]\subseteq[-1,1]$, so $\abs{\theta_i(w)}\le1$ by (C3). Thus the
point $(x_i\theta_i(w))_{i\in[n]}$ lies in $[-1,1]^{n}$, and (a) follows from
Lemma~\ref{lem:extension} applied to $f$.

\smallskip
\emph{\textup{(b)} $G$ has degree at most $dm$.} As a function of $w\in Q_k$, $1-c\,\cnt_i$ has
degree at most $1$, so its $j$th power has degree at most $j$, and by (C1) the function $\theta_i$ has
degree at most $m$. For $\abs{V}\le d$ the product $\prod_{i\in V}x_i\theta_i$ is a constant multiple
of a product of at most $d$ functions of degree at most $m$, so it has degree at most $dm$.

\smallskip
\emph{\textup{(c)} $G(\one)=f(x)$.} Indeed $\cnt_i(\one)=0$ and $\theta_i(\one)=T_m(1)=1$ for every
$i$, so $G(\one)=\tilde f(x)=f(x)$.

\smallskip
\emph{\textup{(d)} $G(\one^{b})=f(x^{S_b})$ for every $b\in[k]$.} The only coordinate of $\one^{b}$
equal to $-1$ is the $b$th one, so $\cnt_i(\one^{b})$ is $1$ if $i\in S_b$ and $0$ otherwise. By
(C3), $\theta_i(\one^{b})=T_m(\cos(\pi/m))=-1$ if $i\in S_b$, and $\theta_i(\one^{b})=T_m(1)=1$
otherwise. Hence $x_i\theta_i(\one^{b})=(x^{S_b})_i$ for all $i$, and
$G(\one^{b})=\tilde f(x^{S_b})=f(x^{S_b})$.

By (a) and (b), Lemma~\ref{lem:sens} applies to $G$ with $D=dm$; by (c) and (d) its conclusion is the
assertion.
\end{proof}

\subsection{Fractional packings}

A \emph{fractional packing} on $[n]$ is a family $\kappa=(\kappa_S)_{S\subseteq[n]}$ of real numbers
such that
\[
   \kappa_S\ge0\quad\text{for all }S\subseteq[n],\qquad
   \sum_{S\ni i}\kappa_S\le1\quad\text{for all }i\in[n].
\]

A fractional packing is approximated by lists of blocks; in the limit the block inequality gives
the following bound.

\begin{lemma}[Packing bound]\label{lem:packing}
Let $d\ge0$ be an integer, let $f\colon Q_n\to[-1,1]$ be a function of degree at most $d$, let
$x\in Q_n$, and let $\kappa$ be a fractional packing on $[n]$. Then
\[
   \sum_{S\subseteq[n]}\kappa_S\,\abs{f(x)-f(x^{S})}\;\le\;\pi^{2}d^{2}.
\]
\end{lemma}

\begin{proof}
Write $t_S=\abs{f(x)-f(x^{S})}$, $\Lambda=\sum_S\kappa_St_S$ and $C_0=\sum_St_S$. Let $M\ge1$ be an
integer, and put
\[
   N=M^{2},\qquad k_S=\lfloor M^{2}\kappa_S\rfloor\quad(S\subseteq[n]),\qquad m=\lceil\pi M/2\rceil .
\]

\smallskip
\emph{Step 1: a list of blocks.} Consider the list of blocks in which every set $S\subseteq[n]$
occurs exactly $k_S$ times. A
coordinate $i$ lies in
\[
   \sum_{S\ni i}k_S\;\le\;M^{2}\sum_{S\ni i}\kappa_S\;\le\;N
\]
blocks of the list.

Further $m\ge1$ and
\[
   \pi M/2\;\le\;m\;<\;\pi M/2+1 .
\]
Since
$1-\cos\vartheta=2\sin^{2}(\vartheta/2)\le\vartheta^{2}/2$ for real $\vartheta$,
\[
   N\bigl(1-\cos(\pi/m)\bigr)\le\frac{M^{2}\pi^{2}}{2m^{2}}\le2 .
\]

\smallskip
\emph{Step 2: the block inequality.} Lemma~\ref{lem:block} therefore gives
\[
   \sum_{S\subseteq[n]}k_S\,t_S\;\le\;4d^{2}m^{2}\;\le\;d^{2}(\pi M+2)^{2}.
\]

As $M^{2}\kappa_S<k_S+1$ and $t_S\ge0$, it follows that
\[
   M^{2}\Lambda\le\sum_S(k_S+1)\,t_S\le d^{2}(\pi M+2)^{2}+C_0,
   \qquad\text{that is,}\qquad
   \Lambda\le d^{2}\Bigl(\pi+\frac2M\Bigr)^{2}+\frac{C_0}{M^{2}} .
\]

\smallskip
\emph{Step 3: the limit.} This holds for every integer $M\ge1$, and $C_0$ does not depend on $M$.
Letting $M\to\infty$ we get
$\Lambda\le\pi^{2}d^{2}$.
\end{proof}

\subsection{Duality}

The passage from packings to rows is an instance of linear programming duality. We derive it from the
following separation lemma, whose short proof we include.

For a finite set $\mathcal S$ we write
$\R^{\mathcal S}$ for the space of real vectors $v=(v_s)_{s\in\mathcal S}$, with the inner product
$\langle v,v'\rangle=\sum_sv_sv'_s$ and the Euclidean norm $\abs{v}=\langle v,v\rangle^{1/2}$.

\begin{lemma}[Separation from the orthant]\label{lem:separation}
Let $\mathcal S$ be a finite set and let $C\subseteq\R^{\mathcal S}$ be a non-empty compact convex set
which contains no vector with all coordinates non-negative. Then there is a vector
$\kappa\in[0,\infty)^{\mathcal S}$ such that
\[
   \sum_{s\in\mathcal S}\kappa_s\,v_s<0\qquad\text{for every }v\in C .
\]
\end{lemma}

\begin{proof}
\emph{Step 1: the distance to the orthant.} For a real number $\xi$ let $\xi^{-}=\max(-\xi,0)$, and
for $v\in\R^{\mathcal S}$ let
$\rho(v)=\sum_s(v_s^{-})^{2}$.

If $\xi$ is real and $\zeta\ge0$, then $(\xi-\zeta)^{2}\ge(\xi^{-})^{2}$:
this is clear for $\xi\ge0$, and for $\xi<0$ we have $\zeta-\xi\ge-\xi>0$. Hence
\begin{equation}\label{eq:rho}
   \rho(v)\le\abs{v-p}^{2}\qquad\text{for all }v\in\R^{\mathcal S}\text{ and }p\in[0,\infty)^{\mathcal S}.
\end{equation}

\smallskip
\emph{Step 2: a nearest point.} The function $\rho$ is continuous, and $\rho(v)=0$ if and only if
all coordinates of $v$ are
non-negative. Since $C$ is compact and not empty, $\rho$ attains its minimum on $C$ at some point
$y\in C$, and $\rho(y)>0$ by the hypothesis on $C$.

Let $\kappa$ be the vector with
$\kappa_s=y_s^{-}\ge0$. Then $y+\kappa$ has the non-negative coordinates $\max(y_s,0)$, and
\[
   \abs{\kappa}^{2}=\rho(y),\qquad \langle\kappa,y\rangle=-\rho(y),
\]
the latter because $y_s^{-}\,y_s=-(y_s^{-})^{2}$ for every $s$.

\smallskip
\emph{Step 3: the inequality.} Let $v\in C$ and $0<\eps\le1$. The
point $y+\eps(v-y)$ belongs to $C$, because $C$ is convex. By the minimality of $\rho(y)$ and by
\eqref{eq:rho} with $p=y+\kappa$,
\[
   \rho(y)\;\le\;\rho\bigl(y+\eps(v-y)\bigr)\;\le\;\abs{\eps(v-y)-\kappa}^{2}
   =\eps^{2}\abs{v-y}^{2}-2\eps\langle\kappa,v-y\rangle+\rho(y).
\]

Hence $\langle\kappa,v-y\rangle\le\frac{\eps}{2}\abs{v-y}^{2}$ for all $\eps\in(0,1]$, so that
$\langle\kappa,v-y\rangle\le0$, and
\[
   \langle\kappa,v\rangle\;\le\;\langle\kappa,y\rangle=-\rho(y)<0 . \qedhere
\]
\end{proof}

\begin{proposition}[Duality]\label{prop:duality}
Let $n\ge0$, let $(t_S)_{S\subseteq[n]}$ be a family of real numbers and let $K$ be a real number.
Suppose that
\[
   \sum_{S\subseteq[n]}\kappa_S\,t_S\le K\qquad\text{for every fractional packing $\kappa$ on $[n]$.}
\]
Then there is a vector $a\in[0,\infty)^{n}$ with $\norm{a}\le K$ and $t_S\le a(S)$ for every
$S\subseteq[n]$.
\end{proposition}

\begin{proof}
The family $\kappa=0$ is a fractional packing, so $K\ge0$. Let
\[
   P=\{a\in[0,\infty)^{n}:\ \norm{a}\le K\}.
\]
This set contains $0$, it is convex, and it is compact, being closed and contained in $[0,K]^{n}$.

Let $\mathcal S$ be the set of all subsets of $[n]$, and let $C\subseteq\R^{\mathcal S}$ be the image
of $P$ under the map $a\mapsto(a(S)-t_S)_{S\in\mathcal S}$. This map is the sum of a linear map and a
constant, so $C$ is non-empty, compact and convex.

If some vector of $C$ has all coordinates
non-negative, then the corresponding $a\in P$ has the required properties. Suppose that this is not
the case. By Lemma~\ref{lem:separation} there is a family $\kappa=(\kappa_S)_{S\subseteq[n]}$ of
non-negative numbers with
\begin{equation}\label{eq:sep}
   \sum_{S\subseteq[n]}\kappa_S\bigl(a(S)-t_S\bigr)<0\qquad\text{for every }a\in P .
\end{equation}

Put $\vartheta=\sum_S\kappa_St_S$ and $\sigma_i=\sum_{S\ni i}\kappa_S\ge0$ for $i\in[n]$. Since
\[
   \sum_S\kappa_S\,a(S)\;=\;\sum_{i=1}^{n}a_i\sigma_i ,
\]
the inequality \eqref{eq:sep} says that
\begin{equation}\label{eq:sep2}
   \sum_{i=1}^{n}a_i\sigma_i<\vartheta\qquad\text{for every }a\in P .
\end{equation}

For $a=0$ this gives $\vartheta>0$. We distinguish two cases.

\smallskip
\emph{Case A: $\sigma_i=0$ for all $i\in[n]$.} Then every non-negative multiple of $\kappa$ is a
fractional packing, in particular $\kappa'=\frac{K+1}{\vartheta}\,\kappa$. But
$\sum_S\kappa'_St_S=K+1>K$, contrary to the hypothesis.

\smallskip
\emph{Case B: $\sigma_i>0$ for some $i\in[n]$.} Choose $j\in[n]$ with $\sigma_j=\max_i\sigma_i>0$.
Then $\kappa'=\kappa/\sigma_j$ is a fractional packing, because
$\sum_{S\ni i}\kappa'_S=\sigma_i/\sigma_j\le1$ for all $i$. The hypothesis gives
$\vartheta/\sigma_j\le K$, that is, $\vartheta\le K\sigma_j$.

On the other hand, the vector $a$ with
$a_j=K$ and $a_i=0$ for $i\ne j$ belongs to $P$, and \eqref{eq:sep2} gives $K\sigma_j<\vartheta$, a
contradiction.

Both cases are impossible, and the proposition is proved.
\end{proof}

\subsection{The row bound}

\begin{theorem}[Row bound]\label{thm:row}
Let $n\ge0$ and $d\ge0$ be integers and let $f\colon Q_n\to[-1,1]$ be a function of degree at most
$d$. Then for every $x\in Q_n$ there is a row $a$ of $f$ at $x$ with cost $\norm{a}\le\pi^{2}d^{2}$.
\end{theorem}

\begin{proof}
Fix $x$ and apply Proposition~\ref{prop:duality} to the numbers $t_S=\abs{f(x)-f(x^{S})}$ and to
$K=\pi^{2}d^{2}$. Its hypothesis is Lemma~\ref{lem:packing}. The resulting vector
$a\in[0,\infty)^{n}$ satisfies $\norm{a}\le\pi^{2}d^{2}$ and $\abs{f(x)-f(x^{S})}\le a(S)$ for all
$S\subseteq[n]$, so it is a row of $f$ at $x$ (Definition~\ref{def:row}).
\end{proof}

\begin{remark}[The bound of Agarwal and Ben-David]\label{rem:ABD}
Theorem~\ref{thm:row} is known, with a better constant.

For $f\colon\{-1,1\}^{n}\to[-1,1]$, Agarwal
and Ben-David \cite{ABD} define the sensitivity of a block $B$ at $x$ as $\abs{f(x)-f(x^{B})}/2$, and
the fractional block sensitivity $\mathrm{fbs}(f)$ as the largest value of
$\sum_Bw_B\abs{f(x)-f(x^{B})}/2$ over all points $x$ and all fractional packings $w$.

They prove
\cite[Theorem~3.3]{ABD} that
\[
   \mathrm{fbs}(f)\;\le\;\frac{\pi^{2}}{4}\deg(f)^{2}.
\]

In the notation of Lemma~\ref{lem:packing} this is the bound $\frac{\pi^{2}}{2}d^{2}$ in place of
$\pi^{2}d^{2}$, and by Proposition~\ref{prop:duality} it gives rows of cost at most
$\frac{\pi^{2}}{2}d^{2}$.

The statement about rows is \cite[Theorem~1.7]{ABD}, where it is formulated
for functions $f\colon\{0,1\}^{n}\to[0,1]$ of degree at most $d$, with the bound
$\frac{\pi^{2}}{4}d^{2}$.

Their proof follows the strategy of Anshu, Ben-David and Kundu \cite{ABK}
for Boolean functions.

It composes $f$ with Chebyshev polynomials of affine functions of new
variables, $N$ of them for each coordinate, so that the blocks become disjoint, and then applies,
through block sensitivity, the inequality $\sum_i\abs{f(x)-f(x^{i})}/2\le\deg(f)^{2}$ for bounded
functions, which is due to Filmus, Hatami, Keller and Lifshitz \cite[Theorem~3.3]{FHKL}.

The
proof given above uses the same composition with Chebyshev polynomials, with one new variable for each
block, and it uses Lemma~\ref{lem:sens} in place of the inequality of \cite{FHKL}.

Lemma~\ref{lem:sens} is weaker than that inequality by a factor $2$, which accounts for the factor
$2$ between the two bounds. That the classical inequality of Markov, which is what
Lemma~\ref{lem:sens} uses, gives the inequality of \cite{FHKL} with this weaker constant is remarked
in \cite[Section~3]{FHKL}.
\end{remark}

\subsection{The increments at a point}\label{subsec:sens-point}

Lemma~\ref{lem:sens} bounds the sum of the numbers $\abs{G(\one){-}G(\one^{b})}$, the increments
of $G$ at the point $\one$. The block inequality (Lemma~\ref{lem:block}), applied to blocks with one element, gives the same bound at every
point.

This is the second property of bounded functions of low degree, besides the row bound (Theorem~\ref{thm:row}), that
enters the proof of Theorem~\ref{thm:main}.

\begin{corollary}[Sensitivity at every point]\label{cor:sens-point}
Let $n\ge0$ and $d\ge0$ be integers, and let $f\colon Q_n\to[-1,1]$ be a function of degree at most
$d$. Then
\[
   \sum_{j=1}^{n}\abs{f(x)-f(x^{j})}\;\le\;4\,d^{2}\qquad\text{for every }x\in Q_n .
\]
Consequently $L(f)=\sum_{j=1}^{n}L_j(f)\le4\,d^{2}$.
\end{corollary}

\begin{proof}
Fix $x\in Q_n$. We apply Lemma~\ref{lem:block} with $k=n$, with the sets $S_b=\{b\}$ for
$b\in[n]$, and with $N=1$ and $m=1$. Every $i\in[n]$ lies in exactly one of these sets, and
$N\bigl(1-\cos(\pi/m)\bigr)=1-\cos\pi=2$. Since $x^{S_b}=x^{b}$, the conclusion of the lemma is the
first assertion.

The second assertion follows by taking the expectation over $x$.
\end{proof}

Corollary~\ref{cor:sens-point} is the inequality of Filmus, Hatami, Keller and Lifshitz
\cite[Theorem~3.3]{FHKL}, which is recalled in Remark~\ref{rem:ABD}, with the constant $4$ in place
of $2$; if their inequality is used, the bound $4d^{2}$ can be replaced by $2d^{2}$ (see
Remark~\ref{rem:constant}).

That the sum of the $L_1$ influences of a bounded function is bounded by a polynomial in its degree
was asked by Aaronson and Ambainis \cite[Section~1.3]{AA14} and was first proved by Ba\v{c}kurs and
Bavarian \cite{BB}.

\section{Real functions: integration over the level}\label{sec:level}

The all-pairs inequality of Section~\ref{sec:two} concerns Boolean functions. In this section it is
transferred to real functions through their level functions, following the plan of Lovett and Zhang
\cite[Section~5]{LZ}, which is recalled in Section~\ref{subsec:earlier} and compared step by step
with the present argument in Section~\ref{subsec:lz}.

Let $f\colon Q_n\to\R$ be a function with a row at every point. For a real number $t$, the
\emph{level function} $g_t=1_{\{f>t\}}$ is Boolean, and at every point $x$ with $f(x)\ne t$ the row of
$f$ at $x$, divided by $\abs{f(x)-t}$, is a row of $g_t$. We apply the all-pairs inequality to every
$g_t$ and integrate over $t$. The result is the integrated all-pairs inequality
(Proposition~\ref{prop:pairs-real}): it bounds a sum over the pairs of points $x,y$ with
$f(y)<f(x)$, in which each pair contributes a multiple of $(f(x)-f(y))^{3}$, by a sum over the
coordinates, in which the coordinate $j$ contributes a multiple of the mean absolute increment
$L_j(f)$.

Removing the weights with Lemma~\ref{lem:weights}, we obtain an inequality between a third moment of
$f$ and the largest of the numbers $L_j(f)$ (Theorem~\ref{thm:third}), and from it a lower bound for
the largest $L_j(f)$ in terms of the variance of $f$ (Theorem~\ref{thm:incr}). The arguments use only
the existence of rows of bounded cost: neither the degree of $f$ nor a bound for its values appears
before Corollary~\ref{cor:incr-degree}.

\subsection{Level functions and their rows}\label{subsec:level-rows}

Throughout this section, $n\ge0$ is an integer, $f\colon Q_n\to\R$ is a function, and for every
$x\in Q_n$ a row $a^{x}$ of $f$ at $x$ is given (Definition~\ref{def:row}).

For
$w\in(0,\infty)^{n}$ and $x,y\in Q_n$ we put
\begin{equation}\label{eq:sxy}
   s_{xy}(w)\;:=\;\sum_{j\in\Delta(x,y)}w_j\,a^{x}_j\,a^{y}_j .
\end{equation}

For a real number $\xi$ we write $\xi^{+}=\max(\xi,0)$. One of the numbers $\xi^{+}$ and
$(-\xi)^{+}$ is equal to $\abs{\xi}$, and the other is $0$; hence
\begin{equation}\label{eq:pospart}
   \abs{\xi}=\xi^{+}+(-\xi)^{+}\qquad\text{and}\qquad
   \abs{\xi}^{3}=(\xi^{+})^{3}+\bigl((-\xi)^{+}\bigr)^{3}\qquad\text{for every real }\xi .
\end{equation}

For a real number $t$ let $g_t=1_{\{f>t\}}$, that is,
\[
   g_t(x)=1\quad\text{if }f(x)>t,\qquad g_t(x)=0\quad\text{if }f(x)\le t .
\]

For $t\in\R$ and $x\in Q_n$ let
\[
   \lambda^{t,x}=\frac{a^{x}}{\abs{f(x)-t}}\quad\text{if }f(x)\ne t,\qquad
   \lambda^{t,x}=(1,\dots,1)\quad\text{if }f(x)=t .
\]
By Lemma~\ref{lem:rows-basic}(iii) and (i), the vector $\lambda^{t,x}$ is a row of $g_t$ at $x$.

At
a point $x$ with $f(x)=t$ any other row of $g_t$ would serve as well: the terms in which such a
point occurs are discarded in the proof of Lemma~\ref{lem:level-pairs}.

The first lemma evaluates the bilinear expression of the all-pairs inequality (Proposition~\ref{prop:pairs}) for these rows, and
it shows that the denominators which occur in this section are positive.

\begin{lemma}\label{lem:level-rows}
Let $w\in(0,\infty)^{n}$.
\begin{itemize}
\item[\textup{(i)}] If $t\in\R$ and $x,y\in Q_n$ satisfy $f(y)<t<f(x)$, then
\[
   \sum_{j\in\Delta(x,y)}w_j\,\lambda^{t,x}_j\,\lambda^{t,y}_j
   \;=\;\frac{s_{xy}(w)}{(f(x)-t)\,(t-f(y))} .
\]
\item[\textup{(ii)}] $s_{xy}(w)>0$ for all $x,y\in Q_n$ with $f(y)<f(x)$.
\end{itemize}
\end{lemma}

\begin{proof}
(i) Here $f(x)\ne t$ and $f(y)\ne t$, and $\abs{f(x)-t}=f(x)-t$ and $\abs{f(y)-t}=t-f(y)$. So
\[
   \lambda^{t,x}_j\,\lambda^{t,y}_j\;=\;\frac{a^{x}_j\,a^{y}_j}{(f(x)-t)\,(t-f(y))}
   \qquad\text{for every }j\in[n],
\]
and the identity follows from \eqref{eq:sxy}.

(ii) Let $f(y)<f(x)$, and put $t=(f(x)+f(y))/2$, so that $f(y)<t<f(x)$. Then
$g_t(x)=1\ne0=g_t(y)$, and $\lambda^{t,x}$ and $\lambda^{t,y}$ are rows of $g_t$ at $x$ and at $y$.

By Lemma~\ref{lem:overlap}, the left-hand side of the identity in (i) is positive. The denominator
on its right-hand side is positive as well, and therefore $s_{xy}(w)>0$.
\end{proof}

The next lemma is the all-pairs inequality for the level function $g_t$, written in terms of $f$.

\begin{lemma}[The all-pairs inequality at a level]\label{lem:level-pairs}
Let $w\in(0,\infty)^{n}$. Then for every real number $t$
\begin{equation}\label{eq:level-pairs}
   \sum_{\substack{x,y\in Q_n\\ f(y)<f(x)}}\frac{(f(x)-t)^{+}\,(t-f(y))^{+}}{s_{xy}(w)}
   \;\le\;2^{n-1}\sum_{j=1}^{n}\frac{b_j(g_t)}{w_j}.
\end{equation}
\end{lemma}

\begin{proof}
The denominators are positive by Lemma~\ref{lem:level-rows}(ii).

Fix $t$. For the function $g_t$
we have $\cO=\{x:f(x)>t\}$ and $\cN=\{y:f(y)\le t\}$. Proposition~\ref{prop:pairs}, applied to
$g_t$ with the weights $w$ and the rows $\lambda^{t,x}$, states that
\[
   \sum_{x:\,f(x)>t}\ \ \sum_{y:\,f(y)\le t}
   \Bigl(\sum_{j\in\Delta(x,y)}w_j\,\lambda^{t,x}_j\,\lambda^{t,y}_j\Bigr)^{-1}
   \;\le\;2^{n-1}\sum_{j=1}^{n}\frac{b_j(g_t)}{w_j},
\]
and that all the terms on the left-hand side are positive.

We discard the terms with $f(y)=t$. For the remaining terms we have $f(y)<t<f(x)$, and by
Lemma~\ref{lem:level-rows}(i) the term of the pair $(x,y)$ is equal to
$(f(x)-t)(t-f(y))/s_{xy}(w)$. Hence
\begin{equation}\label{eq:level-pairs-2}
   \sum_{\substack{x,y\in Q_n\\ f(y)<t<f(x)}}\frac{(f(x)-t)\,(t-f(y))}{s_{xy}(w)}
   \;\le\;2^{n-1}\sum_{j=1}^{n}\frac{b_j(g_t)}{w_j}.
\end{equation}

The left-hand sides of \eqref{eq:level-pairs} and \eqref{eq:level-pairs-2} are equal. Indeed, let
$f(y)<f(x)$. If $f(y)<t<f(x)$, then $(f(x)-t)^{+}=f(x)-t$ and $(t-f(y))^{+}=t-f(y)$. Otherwise
$t\le f(y)$ or $t\ge f(x)$; then $(t-f(y))^{+}=0$ or $(f(x)-t)^{+}=0$, and the pair $(x,y)$
contributes $0$ to the sum in \eqref{eq:level-pairs}.
\end{proof}

\subsection{Integration over the level}\label{subsec:integration}

We now integrate \eqref{eq:level-pairs} over $t$. By definition, $b_j(g_t)$ is the number of points
$x$ with $g_t(x)=1$ and $g_t(x^{j})=0$, that is,
\begin{equation}\label{eq:bj-level}
   b_j(g_t)\;=\;\sum_{x\in Q_n}1_{\{f(x^{j})\le t<f(x)\}}\qquad(j\in[n],\ t\in\R).
\end{equation}

So, as functions of $t$, the left-hand side of \eqref{eq:level-pairs} is a finite linear combination
of functions $t\mapsto(u-t)^{+}(t-v)^{+}$ with real numbers $u$ and $v$, and the right-hand side is
a finite linear combination of functions $t\mapsto1_{\{p\le t<q\}}$ with real numbers $p$ and $q$.

These functions are piecewise polynomial, and they vanish outside a bounded interval. Their
integrals over $\R$, written $\int_{\R}\dots\,dt$, are therefore integrals of polynomials over
bounded intervals, and an inequality between two such functions which holds for every $t$ may be
integrated.

The next lemma contains the three computations that are needed.

\begin{lemma}\label{lem:integrals}
The following identities hold.
\begin{itemize}
\item[\textup{(i)}] For all real numbers $u$ and $v$,
\[
   \int_{\R}(u-t)^{+}\,(t-v)^{+}\,dt\;=\;\frac{\bigl((u-v)^{+}\bigr)^{3}}{6}.
\]
\item[\textup{(ii)}] For all real numbers $p$ and $q$,
\[
   \int_{\R}1_{\{p\le t<q\}}\,dt\;=\;(q-p)^{+}.
\]
\item[\textup{(iii)}] For every $j\in[n]$,
\[
   \sum_{x\in Q_n}\bigl(f(x)-f(x^{j})\bigr)^{+}\;=\;2^{n-1}\,L_j(f).
\]
\end{itemize}
\end{lemma}

\begin{proof}
(i) Let $u\le v$. Then the integrand vanishes for every $t$: its first factor is $0$ if $t\ge u$,
and its second factor is $0$ if $t<u$, because then $t<v$. The right-hand side is $0$ as well.

Let $v<u$, and put $\delta=u-v$. The integrand vanishes for $t\le v$ and for $t\ge u$, and it is
equal to $(u-t)(t-v)$ for $v\le t\le u$. With the substitution $t=v+\rho$ the integral becomes
\[
   \int_{0}^{\delta}(\delta-\rho)\,\rho\,d\rho\;=\;\frac{\delta^{3}}{2}-\frac{\delta^{3}}{3}
   \;=\;\frac{\delta^{3}}{6}.
\]

(ii) The set of all $t$ with $p\le t<q$ is the interval $[p,q)$ if $p<q$, and it is empty
otherwise.

(iii) The map $x\mapsto x^{j}$ is a bijection of $Q_n$, and $(x^{j})^{j}=x$. Hence
\[
   \sum_{x\in Q_n}\bigl(f(x^{j})-f(x)\bigr)^{+}\;=\;\sum_{x\in Q_n}\bigl(f(x)-f(x^{j})\bigr)^{+},
\]
and by \eqref{eq:pospart}
\begin{align*}
   2\sum_{x\in Q_n}\bigl(f(x)-f(x^{j})\bigr)^{+}
   &=\sum_{x\in Q_n}\Bigl(\bigl(f(x)-f(x^{j})\bigr)^{+}+\bigl(f(x^{j})-f(x)\bigr)^{+}\Bigr)\\
   &=\sum_{x\in Q_n}\abs{f(x)-f(x^{j})}\;=\;2^{n}\,L_j(f). \qedhere
\end{align*}
\end{proof}

The integration gives the following inequality, which is the analogue of the all-pairs inequality (Proposition~\ref{prop:pairs})
for real functions. We state it with all its hypotheses.

\begin{proposition}[Integrated all-pairs inequality]\label{prop:pairs-real}
Let $n\ge0$, let $f\colon Q_n\to\R$, let $w\in(0,\infty)^{n}$, and for every $x\in Q_n$ let $a^{x}$
be a row of $f$ at $x$. Then the numbers
$s_{xy}(w)=\sum_{j\in\Delta(x,y)}w_ja^{x}_ja^{y}_j$ are positive for all $x,y\in Q_n$ with
$f(y)<f(x)$, and
\begin{equation}\label{eq:pairs-real}
   \sum_{\substack{x,y\in Q_n\\ f(y)<f(x)}}\frac{(f(x)-f(y))^{3}}{6\,s_{xy}(w)}
   \;\le\;4^{\,n-1}\sum_{j=1}^{n}\frac{L_j(f)}{w_j}.
\end{equation}
\end{proposition}

\begin{proof}
The positivity is Lemma~\ref{lem:level-rows}(ii). The inequality \eqref{eq:level-pairs} holds for
every real $t$, and we integrate it over $t$.

\smallskip
\emph{The left-hand side.} Let $f(y)<f(x)$. By Lemma~\ref{lem:integrals}(i) with $u=f(x)$ and
$v=f(y)$, the integral of the term of the pair $(x,y)$ in \eqref{eq:level-pairs} is
$(f(x)-f(y))^{3}/(6\,s_{xy}(w))$. So the integral of the left-hand side of \eqref{eq:level-pairs}
is the left-hand side of \eqref{eq:pairs-real}.

\smallskip
\emph{The right-hand side.} For every $j\in[n]$,
\begin{align*}
   \int_{\R}b_j(g_t)\,dt
   &\;\overset{\eqref{eq:bj-level}}{=}\;\sum_{x\in Q_n}\int_{\R}1_{\{f(x^{j})\le t<f(x)\}}\,dt
   \;\overset{\text{Lem.~\ref{lem:integrals}(ii)}}{=}\;\sum_{x\in Q_n}\bigl(f(x)-f(x^{j})\bigr)^{+}\\
   &\;\overset{\text{Lem.~\ref{lem:integrals}(iii)}}{=}\;2^{n-1}\,L_j(f).
\end{align*}
So the integral of the right-hand side of \eqref{eq:level-pairs} is
$2^{n-1}\sum_j2^{n-1}L_j(f)/w_j$, which is the right-hand side of \eqref{eq:pairs-real}.
\end{proof}

If $f$ is constant, in particular if $n=0$, there are no pairs $x,y$ with $f(y)<f(x)$, and the
left-hand side of \eqref{eq:pairs-real} is $0$.

\begin{example}\label{ex:pairs-real}
Let $n\ge1$ and $f(x)=x_1$. Then $\abs{f(x)-f(x^{S})}$ is $2$ if $1\in S$ and $0$ otherwise, so
$a^{x}=(2,0,\dots,0)$ is a row of $f$ at every point $x$.

We have $f(y)<f(x)$ if and only if
$x_1=1$ and $y_1=-1$; there are $4^{\,n-1}$ such pairs, and for each of them
$(f(x)-f(y))^{3}=8$ and $s_{xy}(w)=4w_1$. Further $L_1(f)=2$, and $L_j(f)=0$ for $j\ge2$. So the
left-hand side of \eqref{eq:pairs-real} is $4^{\,n-1}/(3w_1)$, and the right-hand side is
$2\cdot4^{\,n-1}/w_1$.
\end{example}

\subsection{The third-moment inequality}\label{subsec:third}

In the integrated all-pairs inequality (Proposition~\ref{prop:pairs-real}) the weights are removed with Lemma~\ref{lem:weights}. We
keep only the pairs $(x,y)$ in which $f(x)$ lies above a threshold $r$ and $f(y)$ does not. The
result bounds the product of the third moment of the part of $f$ above $r$ and the probability
that $f\le r$.

\begin{theorem}[Third-moment inequality]\label{thm:third}
Let $n\ge0$, let $f\colon Q_n\to\R$, and let $K$ be a real number such that for every $x\in Q_n$
there is a row $a^{x}$ of $f$ at $x$ with $\norm{a^{x}}\le K$. Let $M\ge0$ be a real number such
that $L_j(f)\le M$ for every $j\in[n]$. Then, for every real number $r$,
\begin{align}
   \E\bigl[((f-r)^{+})^{3}\bigr]\cdot\Pr[f\le r]&\;\le\;\frac32\,K^{2}M, \label{eq:third-up}\\[2pt]
   \E\bigl[((r-f)^{+})^{3}\bigr]\cdot\Pr[f\ge r]&\;\le\;\frac32\,K^{2}M. \label{eq:third-down}
\end{align}
\end{theorem}

\begin{proof}
The right-hand sides are non-negative, because $M\ge0$. We prove \eqref{eq:third-up} in three
steps, and then deduce \eqref{eq:third-down} from it.

Fix $r$, and put
\[
   A=\{x\in Q_n:\ f(x)>r\},\qquad B=\{y\in Q_n:\ f(y)\le r\}.
\]
Since $(f(x)-r)^{+}$ is equal to $f(x)-r$ for $x\in A$ and to $0$ for $x\notin A$,
\begin{equation}\label{eq:third-AB}
   \E\bigl[((f-r)^{+})^{3}\bigr]=2^{-n}\sum_{x\in A}(f(x)-r)^{3},\qquad \Pr[f\le r]=2^{-n}\abs{B}.
\end{equation}

If $A$ or $B$ is empty, then the left-hand side of \eqref{eq:third-up} is $0$, and the inequality
holds. From now on let $A$ and $B$ be non-empty.

\smallskip
\emph{Step 1: the pairs in $A\times B$.} Let $w\in(0,\infty)^{n}$, $x\in A$ and $y\in B$. Then
$f(y)\le r<f(x)$. So $s_{xy}(w)>0$ by Lemma~\ref{lem:level-rows}(ii), and
\[
   (f(x)-f(y))^{3}\;\ge\;(f(x)-r)^{3}\;>\;0 .
\]

All the terms of the sum in \eqref{eq:pairs-real} are positive. We keep those of the pairs in
$A\times B$, multiply by $6$, and use $L_j(f)\le M$. In this way
Proposition~\ref{prop:pairs-real} gives, for every $w\in(0,\infty)^{n}$,
\begin{align}
   \sum_{x\in A}\sum_{y\in B}\frac{(f(x)-r)^{3}}{s_{xy}(w)}
   &\;\le\;\sum_{x\in A}\sum_{y\in B}\frac{(f(x)-f(y))^{3}}{s_{xy}(w)}
   \;\le\;\sum_{\substack{x,y\in Q_n\\ f(y)<f(x)}}\frac{(f(x)-f(y))^{3}}{s_{xy}(w)}\notag\\
   &\;\le\;6\cdot4^{\,n-1}\sum_{j=1}^{n}\frac{L_j(f)}{w_j}
   \;\le\;6\cdot4^{\,n-1}M\sum_{j=1}^{n}\frac1{w_j}. \label{eq:third-step1}
\end{align}

\smallskip
\emph{Step 2: removing the weights.} We apply Lemma~\ref{lem:weights} to the sets $A$ and $B$, to
the vectors
\[
   \lambda^{x}=\frac{a^{x}}{(f(x)-r)^{3}}\quad(x\in A),\qquad \mu^{y}=a^{y}\quad(y\in B),
\]
to the sets $J(x,y)=\Delta(x,y)$, and to the constant $C=6\cdot4^{\,n-1}M$. For these data,
\[
   \sum_{j\in J(x,y)}w_j\,\lambda^{x}_j\,\mu^{y}_j\;=\;\frac{s_{xy}(w)}{(f(x)-r)^{3}}\;>\;0
   \qquad(x\in A,\ y\in B),
\]
so condition (i) of the lemma holds, and condition (ii) is \eqref{eq:third-step1}.

The cost of
the vector $\lambda^{x}$ is $\norm{a^{x}}/(f(x)-r)^{3}$. The lemma therefore shows that the costs
$\norm{a^{x}}$ for $x\in A$ and $\norm{a^{y}}$ for $y\in B$ are positive, and that
\begin{equation}\label{eq:third-step2}
   \Bigl(\sum_{x\in A}\frac{(f(x)-r)^{3}}{\norm{a^{x}}}\Bigr)
   \Bigl(\sum_{y\in B}\frac1{\norm{a^{y}}}\Bigr)\;\le\;6\cdot4^{\,n-1}M .
\end{equation}

\smallskip
\emph{Step 3: the bound for the costs.} For every point $z$ of $A$ or of $B$ we have
$0<\norm{a^{z}}\le K$; so $K>0$ and $1/\norm{a^{z}}\ge1/K$. All the terms in
\eqref{eq:third-step2} are positive, and therefore
\[
   \frac1{K^{2}}\Bigl(\sum_{x\in A}(f(x)-r)^{3}\Bigr)\abs{B}\;\le\;6\cdot4^{\,n-1}M .
\]
With \eqref{eq:third-AB} this gives
\[
   \E\bigl[((f-r)^{+})^{3}\bigr]\cdot\Pr[f\le r]
   \;=\;4^{-n}\Bigl(\sum_{x\in A}(f(x)-r)^{3}\Bigr)\abs{B}
   \;\le\;4^{-n}\cdot6\cdot4^{\,n-1}K^{2}M\;=\;\frac32\,K^{2}M,
\]
which is \eqref{eq:third-up}.

\smallskip
\emph{The inequality \eqref{eq:third-down}.} For every $x$ the vector $a^{x}$ is also a row of the
function $-f$ at $x$, and $L_j(-f)=L_j(f)$ for every $j$. So \eqref{eq:third-up} holds for $-f$ and
$-r$ in place of $f$ and $r$. Since $(-f)-(-r)=r-f$, and since $-f\le-r$ if and only if $f\ge r$,
this is \eqref{eq:third-down}.
\end{proof}

A special case concerns two sets on which the values of $f$ are separated by a gap, that is,
subsets of a superlevel set and of a sublevel set of $f$. These are the sets to which Lovett and
Zhang apply their conjectured inequality (Section~\ref{subsec:earlier}).

\begin{corollary}[Separated sets]\label{cor:separated}
Let $f$, $K$ and $M$ be as in Theorem~\ref{thm:third}. Let $r$ and $\theta>0$ be real numbers, and
let $X$ and $Y$ be subsets of $Q_n$ such that $f(x)\ge r+\theta$ for all $x\in X$ and $f(y)\le r$
for all $y\in Y$. Then
\[
   2\,\theta^{3}\cdot\frac{\abs{X}}{2^{n}}\cdot\frac{\abs{Y}}{2^{n}}\;\le\;3\,K^{2}M .
\]
\end{corollary}

\begin{proof}
We have $((f-r)^{+})^{3}\ge\theta^{3}$ on $X$, and $((f-r)^{+})^{3}\ge0$ everywhere; so
$\E[((f-r)^{+})^{3}]\ge\theta^{3}\abs{X}/2^{n}$. Further $\Pr[f\le r]\ge\abs{Y}/2^{n}$. The
assertion follows from \eqref{eq:third-up}, multiplied by $2$.
\end{proof}

\subsection{The mean absolute increments}\label{subsec:incr}

\begin{sloppypar}
We apply the third-moment inequality (Theorem~\ref{thm:third}) at a median of $f$, on both sides, and compare the third
absolute moment of $f$ about the median with the variance. Two elementary facts are needed: the
existence of a median, and an inequality between the second and the third absolute moment.
\end{sloppypar}

\begin{lemma}[Median]\label{lem:median}
Let $n\ge0$ and $f\colon Q_n\to\R$. Then there is a real number $m$ such that
\[
   \Pr[f\le m]\ge\frac12\qquad\text{and}\qquad\Pr[f\ge m]\ge\frac12 .
\]
\end{lemma}

\begin{proof}
The function $f$ has finitely many values, and its largest value $v_{\max}$ satisfies
$\Pr[f\le v_{\max}]=1$. Let $m$ be the least value $v$ of $f$ with $\Pr[f\le v]\ge\frac12$. Then
$\Pr[f\le m]\ge\frac12$.

If $m$ is the least value of $f$, then $\Pr[f\ge m]=1$. Otherwise let $v'$ be the largest value of
$f$ that is less than $m$. Then $f(x)<m$ if and only if $f(x)\le v'$, and $\Pr[f\le v']<\frac12$ by
the choice of $m$. Hence $\Pr[f\ge m]=1-\Pr[f<m]>\frac12$.
\end{proof}

The second fact will be applied to the function $z=f-m$.

\begin{lemma}[Second and third moments]\label{lem:power}
Let $n\ge0$. For every function $z\colon Q_n\to\R$,
\[
   \bigl(\E[z^{2}]\bigr)^{3}\;\le\;\bigl(\E[\abs{z}^{3}]\bigr)^{2}.
\]
\end{lemma}

\begin{proof}
We use the Cauchy--Schwarz inequality of Section~\ref{sec:two} twice, for families of numbers
indexed by the points $x\in Q_n$: first for the numbers $\abs{z(x)}^{1/2}$ and
$\abs{z(x)}^{3/2}$, and then for the numbers $1$ and $\abs{z(x)}$. After division by $4^{n}$ the
two inequalities read
\[
   \bigl(\E[z^{2}]\bigr)^{2}\;\le\;\E[\abs{z}]\cdot\E[\abs{z}^{3}],\qquad
   \bigl(\E[\abs{z}]\bigr)^{2}\;\le\;\E[z^{2}].
\]

We square the first inequality and insert the second:
\[
   \bigl(\E[z^{2}]\bigr)^{4}\;\le\;\bigl(\E[\abs{z}]\bigr)^{2}\,\bigl(\E[\abs{z}^{3}]\bigr)^{2}
   \;\le\;\E[z^{2}]\cdot\bigl(\E[\abs{z}^{3}]\bigr)^{2}.
\]

If $\E[z^{2}]>0$, we divide by this number. If $\E[z^{2}]=0$, there is nothing to prove.
\end{proof}

We can now prove the main result of this section, which is Theorem~\ref{thm:incr-intro} of the
introduction.

\begin{theorem}[Mean absolute increments]\label{thm:incr}
Let $n\ge0$, let $f\colon Q_n\to\R$ be a function with $\Var(f)>0$, and let $K$ be a real number
such that for every $x\in Q_n$ there is a row $a^{x}$ of $f$ at $x$ with $\norm{a^{x}}\le K$. Then
$n\ge1$ and $K>0$, and there is a coordinate $j\in[n]$ with
\[
   \Var(f)^{3}\;\le\;36\,K^{4}\,L_j(f)^{2},\qquad\text{that is,}\qquad
   \E_x\abs{f(x)-f(x^{j})}\;\ge\;\frac{\Var(f)^{3/2}}{6\,K^{2}} .
\]
\end{theorem}

\begin{proof}
\emph{Step 0: preliminaries.} Since $\Var(f)>0$, the function $f$ is not constant. So $n\ge1$, and
for every $x\in Q_n$ there is a point $y$ with $f(y)\ne f(x)$, whence
\begin{equation}\label{eq:a-pos}
   0<\abs{f(x)-f(y)}\le a^{x}(\Delta(x,y))\le\norm{a^{x}}\le K .
\end{equation}
In particular $K>0$.

Choose $j\in[n]$ such that $L_j(f)$ is as large as possible, and put
$M=L_j(f)$. Then $M\ge0$, and $L_k(f)\le M$ for every $k\in[n]$. Let $m$ be a number with the
properties of Lemma~\ref{lem:median}.

\smallskip
\emph{Step 1: the third moment about the median.} The expectation in \eqref{eq:third-up} is
non-negative, and $\Pr[f\le m]\ge\frac12$. So \eqref{eq:third-up} with $r=m$ gives
\[
   \frac12\,\E\bigl[((f-m)^{+})^{3}\bigr]\;\le\;
   \E\bigl[((f-m)^{+})^{3}\bigr]\cdot\Pr[f\le m]\;\le\;\frac32\,K^{2}M .
\]

In the same way, \eqref{eq:third-down} with $r=m$ and $\Pr[f\ge m]\ge\frac12$ give
$\frac12\,\E[((m-f)^{+})^{3}]\le\frac32K^{2}M$.

We add the two inequalities, multiply by $2$, and
use the second relation in \eqref{eq:pospart} with $\xi=f(x)-m$:
\begin{equation}\label{eq:third-abs}
   \E\bigl[\abs{f-m}^{3}\bigr]\;=\;\E\bigl[((f-m)^{+})^{3}\bigr]+\E\bigl[((m-f)^{+})^{3}\bigr]
   \;\le\;6\,K^{2}M .
\end{equation}

\smallskip
\emph{Step 2: the variance.} By \eqref{eq:var-shift} with $c=m$ we have
$\Var(f)\le\E[(f-m)^{2}]$. By Lemma~\ref{lem:power} with $z=f-m$, and by \eqref{eq:third-abs},
\[
   \Var(f)^{3}\;\overset{\eqref{eq:var-shift}}{\le}\;\bigl(\E[(f-m)^{2}]\bigr)^{3}
   \;\overset{\text{Lem.~\ref{lem:power}}}{\le}\;\bigl(\E[\abs{f-m}^{3}]\bigr)^{2}
   \;\overset{\eqref{eq:third-abs}}{\le}\;36\,K^{4}M^{2} .
\]

Since $M=L_j(f)$, this is the first form of the assertion. The second form follows by taking
square roots, because $K>0$ and $L_j(f)\ge0$.
\end{proof}

With the row bound (Theorem~\ref{thm:row}) we obtain a statement about bounded functions of low
degree.

\begin{corollary}\label{cor:incr-degree}
Let $n\ge0$ and $d\ge0$ be integers, and let $f\colon Q_n\to[-1,1]$ be a function of degree at
most $d$ with $\Var(f)>0$. Then $n\ge1$ and $d\ge1$, and there is a coordinate $j\in[n]$ with
\[
   \E_x\abs{f(x)-f(x^{j})}\;\ge\;\frac{\Var(f)^{3/2}}{6\,\pi^{4}\,d^{4}} .
\]
\end{corollary}

\begin{proof}
A function of degree at most $0$ is constant, so $d\ge1$. By Theorem~\ref{thm:row}, the function
$f$ has a row of cost at most $\pi^{2}d^{2}$ at every point. Theorem~\ref{thm:incr} with
$K=\pi^{2}d^{2}$ gives $n\ge1$ and the assertion, because $K^{2}=\pi^{4}d^{4}$.
\end{proof}

Corollary~\ref{cor:incr-degree} is a quantitative form of the $L_1$ version of the conjecture, the
version in which Aaronson and Ambainis first stated it; see \cite[Conjecture~6.1]{AA14}, where
$\E\abs{p-\E p}$ takes the place of the variance (note that
$\Var(f)\ge(\E\abs{f-\E f})^{2}$).

They prove that the two versions are equivalent
\cite[Proposition~6.2]{AA14}. One half of their argument is the inequality \eqref{eq:L-Inf}, which
we use now.

Together with \eqref{eq:L-Inf}, Corollary~\ref{cor:incr-degree} gives a coordinate $j$ with
\[
   \Inf_j(f)\;\ge\;\frac{L_j(f)^{2}}{4}\;\ge\;\frac{\Var(f)^{3}}{144\,\pi^{8}\,d^{8}} .
\]
Here $\Var(f)\le\E[f^{2}]\le1$ by \eqref{eq:var-shift} with $c=0$, and therefore
\[
   \Inf_j(f)\;\ge\;\frac{1}{144\,\pi^{8}}\Bigl(\frac{\Var(f)}{d}\Bigr)^{8}.
\]

So a bound of the kind required in the conjecture of Section~\ref{subsec:conj} already follows
at this point. Section~\ref{sec:avg} improves the exponents $3$ and $8$ of the first of the two
bounds to the exponents $\frac52$ and $6$ of Theorem~\ref{thm:main}.

\begin{remark}\label{rem:rows-real}
(1) Every function $f\colon Q_n\to\R$ has a row at every point $x$. Indeed, let
$c=\max_{y}\abs{f(x)-f(y)}$. The vector $(c,\dots,c)$ is a row of $f$ at $x$, because
$\abs{f(x)-f(x^{S})}\le c\le c\,\abs{S}$ for every non-empty set $S$, and
$\abs{f(x)-f(x^{S})}=0$ for $S=\emptyset$.

So the hypothesis on the rows in
Proposition~\ref{prop:pairs-real} and in Theorems~\ref{thm:third} and~\ref{thm:incr} can be met for
every $f$; what matters is the cost of the rows.

(2) The cost cannot be small compared with the oscillation of $f$: as in \eqref{eq:a-pos}, every row
$a$ of $f$ at $x$ satisfies $\norm{a}\ge a(\Delta(x,y))\ge\abs{f(x)-f(y)}$ for every $y$. Hence a
number $K$ as in Theorem~\ref{thm:incr} is at least $\max f-\min f$.

(3) No bound for the values of $f$ is assumed in Theorems~\ref{thm:third} and~\ref{thm:incr}, and
their statements are compatible with scaling.

Let $c>0$. Then $c\,a^{x}$ is a row of $cf$ at $x$,
and the passage from $f$, $a^{x}$ and $K$ to $cf$, $c\,a^{x}$ and $cK$ multiplies $\Var(f)$ by
$c^{2}$ and $L_j(f)$ by $c$. So both sides of the inequality $\Var(f)^{3}\le36K^{4}L_j(f)^{2}$ are
multiplied by $c^{6}$.
\end{remark}

\subsection{Relation to the argument of Lovett and Zhang}\label{subsec:lz}

In the argument by which Lovett and Zhang
derive the conjecture from their conjectured inequality \cite[Claim~38]{LZ}, the function $f$ has
values in $[0,1]$, the number $t$ is the mean of $f$, and the two sets are
\[
   X=\{x:\ f(x)\le t-\alpha\},\qquad Y=\{x:\ f(x)\ge t+\alpha\},
\]
where the number $\alpha>0$, which is at most of the order of the variance of $f$, is chosen at
the end. The argument has four steps, and they are related to this section as follows.

The first step is that a fractional certificate at a point of $X$ bounds from below the
$L_\infty$ distance from this point to the convex hull of $Y$ \cite[Lemma~28]{LZ}.
Section~\ref{subsec:level-rows} takes its place: a row of $f$ at $x$, divided by
$\abs{f(x)-t}$, is a row of $g_t$ at $x$ if $f(x)\ne t$. This is used at every level $t$, and not only
near the mean.

The second step is the conjectured inequality \cite[Conjecture~37]{LZ}, which would give a
coordinate that is influential for $X$ or for $Y$; it is not true as stated (see
Section~\ref{subsec:cert}). Here the all-pairs inequality is applied to
$g_t$ instead (Lemma~\ref{lem:level-pairs}). The two sets are now $\{f>t\}$ and $\{f\le t\}$,
with no gap between them, and the inequality involves all the pairs of points, one from each
set.

In \cite{LZ} the sets $X$ and $Y$ are large because the variance of $f$ is not small
\cite[Claim~30]{LZ}. Here Theorem~\ref{thm:third} is applied at a median $m$ of $f$, so that
$\Pr[f\le m]$ and $\Pr[f\ge m]$ are at least $\frac12$ (Section~\ref{subsec:incr}).
Corollary~\ref{cor:separated} is the form of Theorem~\ref{thm:third} for two sets which are
separated by a gap, as $X$ and $Y$ are.

The last step of \cite{LZ} passes from the set to the function: $\alpha$ is chosen in such a
way that few values of $f$ lie just above the level $t-\alpha$, and for such an $\alpha$ the
coordinate is influential for $f$. This step has no counterpart here. The inequality of
Lemma~\ref{lem:level-pairs} is integrated over all levels (Section~\ref{subsec:integration}),
and the integral of its right-hand side is expressed by the increments of $f$ itself. No level
is chosen.

As in \cite{LZ}, the conclusion is a lower bound for the mean of $\abs{f(x)-f(x^{j})}$ for some
coordinate $j$ (Theorem~\ref{thm:incr}).

\section{Averaging and the proof of the main theorem}\label{sec:avg}

Theorem~\ref{thm:incr} gives a coordinate $j$ for which the mean absolute increment $L_j(f)$ is
large. The main theorem (Theorem~\ref{thm:main}) concerns the influences $\Inf_j(f)$, which are means of squares. In this
section we pass from the one to the other.

The idea is the following. Call a coordinate $k$ \emph{light} if $\Inf_k(f)$ is small compared
with $L_k(f)$, in the sense made precise in the proof of Theorem~\ref{thm:infl}.

The influences of
the light coordinates have a small sum, so the function $h$ obtained from $f$ by averaging over the
light coordinates retains a fixed fraction of the variance of $f$.

Averaging increases neither the cost of the rows nor the mean absolute increments, and for the
coordinates over which one has averaged the mean absolute increments vanish.

So
Theorem~\ref{thm:incr}, applied to $h$, produces a coordinate $j$ that is not light and for which
$L_j(f)$ is large. For such a coordinate $\Inf_j(f)$ is large (Theorem~\ref{thm:infl}).

Together with the row bound (Theorem~\ref{thm:row}) and Corollary~\ref{cor:sens-point}, this gives
Theorem~\ref{thm:main}.

\subsection{Averaging over coordinates}\label{subsec:avg}

For a function $f\colon Q_n\to\R$ and a coordinate $j\in[n]$, the \emph{average of $f$ over the
coordinate $j$} is the function $\Av_jf\colon Q_n\to\R$ with
\[
   (\Av_jf)(x)\;=\;\frac{f(x)+f(x^{j})}{2}\qquad(x\in Q_n).
\]
The following lemma lists the properties of this function that we need.

\begin{lemma}[Averaging over one coordinate]\label{lem:avg-one}
Let $n\ge1$, let $f\colon Q_n\to\R$ and $j\in[n]$, and put $h=\Av_jf$.
\begin{itemize}
\item[\textup{(i)}] $\E h=\E f$, and $\Var(f)=\Var(h)+\Inf_j(f)$.
\item[\textup{(ii)}] $\Inf_k(h)\le\Inf_k(f)$ and $L_k(h)\le L_k(f)$ for every $k\in[n]$.
\item[\textup{(iii)}] $L_j(h)=0$.
\item[\textup{(iv)}] Let $K$ be a real number, and for every $x\in Q_n$ let $a^{x}$ be a row of $f$
at $x$ with $\norm{a^{x}}\le K$. Then, for every $x\in Q_n$, the vector
$b^{x}=(a^{x}+a^{x^{j}})/2$ is a row of $h$ at $x$, and $\norm{b^{x}}\le K$.
\end{itemize}
\end{lemma}

\begin{proof}
The map $x\mapsto x^{j}$ is a bijection of $Q_n$ which preserves the measure; so
$\E_x[\varphi(x^{j})]=\E_x[\varphi(x)]$ for every function $\varphi\colon Q_n\to\R$. Moreover
$(x^{j})^{j}=x$, and therefore
\begin{equation}\label{eq:avg-even}
   h(x^{j})=h(x)\qquad\text{for every }x\in Q_n .
\end{equation}

(i) The first statement holds because $\E_x[f(x^{j})]=\E f$.

For the second, let
$\nu=\E f=\E h$. Fix $x\in Q_n$, and put $p=h(x)-\nu$ and $q=(f(x)-f(x^{j}))/2$. Then
$f(x)-\nu=p+q$ and $f(x^{j})-\nu=p-q$, and since $(p+q)^{2}+(p-q)^{2}=2p^{2}+2q^{2}$,
\[
   (f(x)-\nu)^{2}+(f(x^{j})-\nu)^{2}
   \;=\;2\,(h(x)-\nu)^{2}+2\,\Bigl(\frac{f(x)-f(x^{j})}{2}\Bigr)^{2}.
\]
Taking the expectation over $x$, we get $2\Var(f)=2\Var(h)+2\Inf_j(f)$.

(ii) Let $k\in[n]$. Fix $x\in Q_n$, and put
\[
   P=f(x)-f(x^{k}),\qquad Q=f(x^{j})-f\bigl((x^{j})^{k}\bigr).
\]
Since $(x^{k})^{j}=(x^{j})^{k}$, we have $h(x)-h(x^{k})=(P+Q)/2$. Hence, using
$(P+Q)^{2}\le2P^{2}+2Q^{2}$ for the second inequality,
\[
   \abs{h(x)-h(x^{k})}\;\le\;\frac{\abs{P}+\abs{Q}}{2},\qquad
   \Bigl(\frac{h(x)-h(x^{k})}{2}\Bigr)^{2}\;\le\;
   \frac12\Bigl(\frac{P}{2}\Bigr)^{2}+\frac12\Bigl(\frac{Q}{2}\Bigr)^{2}.
\]

We take the expectation over $x$. The number $Q$ is obtained from $P$ by the substitution of
$x^{j}$ for $x$; so
\[
   \E_x\abs{Q}=\E_x\abs{P}=L_k(f)\qquad\text{and}\qquad
   \E_x\bigl[(Q/2)^{2}\bigr]=\E_x\bigl[(P/2)^{2}\bigr]=\Inf_k(f).
\]
This gives $L_k(h)\le L_k(f)$ and
$\Inf_k(h)\le\Inf_k(f)$.

(iii) This follows from \eqref{eq:avg-even}.

(iv) The entries of $b^{x}$ are non-negative, and
$\norm{b^{x}}=(\norm{a^{x}}+\norm{a^{x^{j}}})/2\le K$.

Let $S\subseteq[n]$. Since
$(x^{S})^{j}=(x^{j})^{S}$,
\[
   h(x)-h(x^{S})\;=\;\frac{\bigl(f(x)-f(x^{S})\bigr)+\bigl(f(x^{j})-f((x^{j})^{S})\bigr)}{2}.
\]
The vector $a^{x}$ is a row of $f$ at $x$, and $a^{x^{j}}$ is a row of $f$ at $x^{j}$. Therefore
\[
   \abs{h(x)-h(x^{S})}\;\le\;\frac{a^{x}(S)+a^{x^{j}}(S)}{2}\;=\;b^{x}(S). \qedhere
\]
\end{proof}

For a set $R\subseteq[n]$, the \emph{average of $f$ over the coordinates in $R$} is the function
$\Av_Rf\colon Q_n\to\R$ with
\[
   (\Av_Rf)(x)\;=\;2^{-\abs{R}}\sum_{T\subseteq R}f(x^{T})\qquad(x\in Q_n).
\]
Thus $\Av_{\emptyset}f=f$.

The averages over sets are obtained by averaging over one coordinate at
a time:
\begin{equation}\label{eq:avg-step}
   \Av_{R\cup\{j\}}f\;=\;\Av_j\bigl(\Av_Rf\bigr)\qquad\text{for }R\subseteq[n]\text{ and }j\in[n]\setminus R .
\end{equation}

Indeed, the subsets of $R\cup\{j\}$ are the sets $T$ and $T\cup\{j\}$ with $T\subseteq R$, and
$x^{T\cup\{j\}}=(x^{j})^{T}$ for these $T$. So
\[
   (\Av_{R\cup\{j\}}f)(x)\;=\;2^{-\abs{R}-1}\sum_{T\subseteq R}\Bigl(f(x^{T})+f\bigl((x^{j})^{T}\bigr)\Bigr)
   \;=\;\frac{(\Av_Rf)(x)+(\Av_Rf)(x^{j})}{2}.
\]

By \eqref{eq:avg-step}, the properties of Lemma~\ref{lem:avg-one} pass to averages over sets. The
next lemma states them in the form in which they are used in the proof of
Theorem~\ref{thm:infl}.

\begin{lemma}[Averaging over a set of coordinates]\label{lem:avg-set}
Let $n\ge0$, let $f\colon Q_n\to\R$, and let $K$ be a real number such that for every $x\in Q_n$
there is a row of $f$ at $x$ of cost at most $K$. Let $R\subseteq[n]$, and put $h=\Av_Rf$. Then:
\begin{itemize}
\item[\textup{(i)}] for every $x\in Q_n$ there is a row of $h$ at $x$ of cost at most $K$;
\item[\textup{(ii)}] $\Var(f)\le\Var(h)+\sum_{i\in R}\Inf_i(f)$;
\item[\textup{(iii)}] $\Inf_k(h)\le\Inf_k(f)$ and $L_k(h)\le L_k(f)$ for every $k\in[n]$;
\item[\textup{(iv)}] $L_i(h)=0$ for every $i\in R$.
\end{itemize}
\end{lemma}

\begin{proof}
We use induction on $\abs{R}$. If $R=\emptyset$, then $h=f$, and the four statements are clear.

Let $R=R'\cup\{j\}$ with $j\notin R'$, and suppose that the lemma holds for $R'$ and the function
$h'=\Av_{R'}f$. By \eqref{eq:avg-step}, $h=\Av_jh'$. We apply Lemma~\ref{lem:avg-one} to the
function $h'$ and the coordinate $j$.

(i) By the statement (i) for $R'$, the function $h'$ has a row of cost at most $K$ at every point.
By Lemma~\ref{lem:avg-one}(iv), so has $h$.

(ii) By Lemma~\ref{lem:avg-one}(i) and by the statement (iii) for $R'$,
\[
   \Var(h')\;=\;\Var(h)+\Inf_j(h')\;\le\;\Var(h)+\Inf_j(f).
\]
Together with the statement (ii) for $R'$ this gives
\[
   \Var(f)\;\le\;\Var(h')+\sum_{i\in R'}\Inf_i(f)\;\le\;\Var(h)+\sum_{i\in R}\Inf_i(f).
\]

(iii) By Lemma~\ref{lem:avg-one}(ii) and by the statement (iii) for $R'$, we have
$\Inf_k(h)\le\Inf_k(h')\le\Inf_k(f)$ and $L_k(h)\le L_k(h')\le L_k(f)$ for every $k$.

(iv) By Lemma~\ref{lem:avg-one}(iii), $L_j(h)=0$. For $i\in R'$, the statement (iv) for $R'$ and
Lemma~\ref{lem:avg-one}(ii) give $0\le L_i(h)\le L_i(h')=0$.
\end{proof}

\subsection{From mean absolute increments to influences}\label{subsec:infl}

The following theorem is Theorem~\ref{thm:infl-intro} of the introduction. Like
Theorem~\ref{thm:incr}, it assumes no bound for the values of $f$.

\begin{theorem}[Influences]\label{thm:infl}
Let $n\ge0$, let $f\colon Q_n\to\R$ be a function with $\Var(f)>0$, and let $K$ be a real number
such that for every $x\in Q_n$ there is a row $a^{x}$ of $f$ at $x$ with $\norm{a^{x}}\le K$. Let
$L(f)=\sum_{k=1}^{n}L_k(f)$. Then $n\ge1$ and $0<L(f)\le K$, and there is a coordinate $j\in[n]$
with
\[
   3\,\Var(f)^{5}\;\le\;3125\,L(f)^{2}\,K^{4}\,\Inf_j(f)^{2},\qquad\text{that is,}\qquad
   \Inf_j(f)\;\ge\;\frac{\sqrt{15}}{125}\cdot\frac{\Var(f)^{5/2}}{L(f)\,K^{2}} .
\]
In particular, $\Inf_j(f)\ge(\sqrt{15}/125)\,\Var(f)^{5/2}/K^{3}$ for this coordinate $j$.
\end{theorem}

\begin{proof}
\emph{Step 0: the number $L(f)$.} By Theorem~\ref{thm:incr}, $n\ge1$ and $K>0$, and there is a
coordinate $j_0\in[n]$ with $\Var(f)^{3}\le36K^{4}L_{j_0}(f)^{2}$. Since $\Var(f)>0$, this shows
that $L_{j_0}(f)>0$, and so $L(f)\ge L_{j_0}(f)>0$.

For the upper bound let $x\in Q_n$. Definition~\ref{def:row} with $S=\{k\}$ gives
$\abs{f(x)-f(x^{k})}\le a^{x}_k$ for every $k\in[n]$, and therefore
\[
   \sum_{k=1}^{n}\abs{f(x)-f(x^{k})}\;\le\;\norm{a^{x}}\;\le\;K\qquad\text{for every }x\in Q_n .
\]
Taking the expectation over $x$, we get $L(f)\le K$.

\smallskip
\emph{Step 1: the light coordinates.} We write $V=\Var(f)$ and $L=L(f)$, and we put
\[
   R\;=\;\bigl\{k\in[n]:\ 5\,L\,\Inf_k(f)<2\,V\,L_k(f)\bigr\}.
\]

All the numbers $L_k(f)$ are non-negative, and their sum is $L$. Hence
\[
   5\,L\sum_{k\in R}\Inf_k(f)\;\le\;2\,V\sum_{k\in R}L_k(f)\;\le\;2\,V\,L ,
\]
and since $L>0$ it follows that
\begin{equation}\label{eq:light}
   \sum_{k\in R}\Inf_k(f)\;\le\;\frac25\,V .
\end{equation}

\smallskip
\emph{Step 2: averaging.} Let $h=\Av_Rf$. By Lemma~\ref{lem:avg-set}(ii) and \eqref{eq:light},
\[
   \Var(h)\;\ge\;V-\sum_{k\in R}\Inf_k(f)\;\ge\;\frac35\,V\;>\;0 .
\]

By Lemma~\ref{lem:avg-set}(i), the function $h$ has a row of cost at most $K$ at every point. So
Theorem~\ref{thm:incr} applies to $h$, with the same number $K$: there is a coordinate $j\in[n]$
with
\begin{equation}\label{eq:incr-h}
   \Bigl(\frac35\,V\Bigr)^{3}\;\le\;\Var(h)^{3}\;\le\;36\,K^{4}\,L_j(h)^{2}.
\end{equation}

\smallskip
\emph{Step 3: the coordinate $j$ is not light.} By \eqref{eq:incr-h} we have $L_j(h)>0$, and so
$j\notin R$ by Lemma~\ref{lem:avg-set}(iv). By the definition of $R$, this means that
$2VL_j(f)\le5L\,\Inf_j(f)$. Since $L_j(h)\le L_j(f)$ by Lemma~\ref{lem:avg-set}(iii), we get
\begin{equation}\label{eq:not-light}
   0\;<\;2\,V\,L_j(h)\;\le\;2\,V\,L_j(f)\;\le\;5\,L\,\Inf_j(f).
\end{equation}

\smallskip
\emph{Step 4: conclusion.} We multiply \eqref{eq:incr-h} by $(2V)^{2}$ and use
\eqref{eq:not-light}:
\begin{align*}
   \frac{108}{125}\,V^{5}\;=\;\Bigl(\frac35\,V\Bigr)^{3}(2V)^{2}
   &\;\le\;36\,K^{4}\,\bigl(2\,V\,L_j(h)\bigr)^{2}\\
   &\;\le\;36\,K^{4}\,\bigl(5\,L\,\Inf_j(f)\bigr)^{2}\;=\;900\,L^{2}K^{4}\,\Inf_j(f)^{2}.
\end{align*}
Multiplying by $125/36$, we obtain $3V^{5}\le3125\,L^{2}K^{4}\,\Inf_j(f)^{2}$, which is the first
form of the assertion.

The second form follows by taking square roots, because $LK^{2}>0$ and
$\Inf_j(f)\ge0$, and because $\sqrt{3/3125}=\sqrt{15}/125$. Since $L\le K$, the last statement of
the theorem follows from the second form.
\end{proof}

\subsection{Proof of the main theorem}\label{subsec:proof-main}

\begin{proof}[Proof of Theorem~\ref{thm:main}]
Let $n\ge1$ and $d\ge1$, and let $f\colon Q_n\to[-1,1]$ be a function of degree at most $d$. If
$\Var(f)=0$, then every $j\in[n]$ has the required property, because $\Inf_j(f)\ge0$; and $[n]$ is
not empty.

Let $\Var(f)>0$. By Theorem~\ref{thm:row}, for every $x\in Q_n$ there is a row $a^{x}$ of $f$ at
$x$ with $\norm{a^{x}}\le\pi^{2}d^{2}$. By Corollary~\ref{cor:sens-point}, $L(f)\le4d^{2}$.

Theorem~\ref{thm:infl} with $K=\pi^{2}d^{2}$ shows that $L(f)>0$, and it gives a coordinate
$j\in[n]$ with
\[
   \Inf_j(f)\;\ge\;\frac{\sqrt{15}}{125}\cdot\frac{\Var(f)^{5/2}}{L(f)\,(\pi^{2}d^{2})^{2}}
   \;\ge\;\frac{\sqrt{15}}{125}\cdot\frac{\Var(f)^{5/2}}{4d^{2}\cdot\pi^{4}d^{4}}
   \;=\;\frac{\sqrt{15}}{500\,\pi^{4}}\cdot\frac{\Var(f)^{5/2}}{d^{6}} . \qedhere
\]
\end{proof}

It remains to prove Corollary~\ref{cor:AA01}, the form of the theorem in the normalisation of
\cite{AA14}, which was stated in Section~\ref{sec:intro}.

\begin{proof}[Proof of Corollary~\ref{cor:AA01} from Theorem~\ref{thm:main}]
Since $\Var[p]>0$, the polynomial $p$ is not constant on $\{0,1\}^{n}$; hence $n\ge1$ and $d\ge1$.
For $x\in Q_n$ put
\[
   f(x)\;:=\;2\,p\Bigl(\frac{1-x_1}{2},\dots,\frac{1-x_n}{2}\Bigr)-1 .
\]

The map $x\mapsto X=((1-x_1)/2,\dots,(1-x_n)/2)$ is a bijection from $Q_n$ onto $\{0,1\}^{n}$ which
carries $x^{i}$ to $X^{i}$ and the uniform measure to the uniform measure. Hence $\abs{f}\le 1$ on $Q_n$,
$\Var(f)=4\Var[p]$, and
\[
   \Inf_i(f)=\E_x\Bigl[\Bigl(\frac{f(x)-f(x^{i})}{2}\Bigr)^{2}\Bigr]
           =\E_X\bigl[(p(X)-p(X^{i}))^{2}\bigr]=\Inf_i[p].
\]

The function $f$ has degree at most $d$: expanding $p$ gives $f$ as a real linear combination of
products $\prod_i x_i^{k_i}$ with $\sum_ik_i\le d$, and on $Q_n$ such a product equals $\chi_T$ with
$T=\{i: k_i \text{ odd}\}$, a set of size at most $d$.

Theorem~\ref{thm:main} therefore gives an $i$ with
\[
   \Inf_i[p]=\Inf_i(f)\;\ge\;\frac{\sqrt{15}}{500\,\pi^{4}}\cdot\frac{(4\Var[p])^{5/2}}{d^{6}}
   =\frac{8\sqrt{15}}{125\,\pi^{4}}\cdot\frac{\Var[p]^{5/2}}{d^{6}},
\]
because $4^{5/2}=32$ and $32/500=8/125$.

For the second inequality we show that
\begin{equation}\label{eq:seven}
   \frac{8\sqrt{15}}{125\,\pi^{4}}\cdot\frac{v^{5/2}}{d^{6}}\;\ge\;\Bigl(\frac{v}{d}\Bigr)^{7}
   \qquad\text{for every real number $v$ with }0<v\le\tfrac14 .
\end{equation}

Put $t=v/d$; then $0<t\le\frac14$, because $d\ge1$. We have
\[
   \frac{v^{5/2}}{d^{6}}\;=\;\frac{t^{6}}{v^{7/2}}\;\ge\;128\,t^{6},
\]
since $v^{7/2}\le4^{-7/2}=\frac1{128}$.

Moreover
$\sqrt{15}>\frac72$, and $\pi^{4}<100$ because $\pi^{2}<10$. So the left-hand side of
\eqref{eq:seven} is at least
\[
   \frac{8\cdot\frac72}{125\cdot100}\cdot128\,t^{6}\;=\;\frac{3584}{12500}\,t^{6}
   \;\ge\;\frac14\,t^{6}\;\ge\;t^{7}.
\]

Since $0\le p\le 1$ on $\{0,1\}^{n}$ we have $p^{2}\le p$ there, so
$\Var[p]\le\E[p]-\E[p]^{2}\le\frac14$. Hence \eqref{eq:seven} applies to $v=\Var[p]$, and this is
the second inequality of the corollary.

For the last sentence of the corollary let $\Var[p]\ge\eps>0$. Then
$\Var[p]^{5/2}\ge\eps^{5/2}$, which gives the first inequality, and $\eps\le\frac14$, so that
\eqref{eq:seven} applies to $v=\eps$ and gives the second.
\end{proof}

\begin{remark}[The constants]\label{rem:constant}
The proof of Theorem~\ref{thm:main} uses two properties of a function $f\colon Q_n\to[-1,1]$ of
degree at most $d$: it has a row of cost at most $\pi^{2}d^{2}$ at every point, and
$\sum_j\abs{f(x)-f(x^{j})}\le4d^{2}$ at every point $x$. For both properties better constants are
known.

(1) By \cite[Theorem~1.7]{ABD}, applied to the function $(1+f)/2$ (equivalently, by
\cite[Theorem~3.3]{ABD} and Proposition~\ref{prop:duality}), the function $f$ has, at every point, a
row of cost at most $\frac{\pi^{2}}{2}d^{2}$ (Remark~\ref{rem:ABD}).

If these rows are used in the
proof of Theorem~\ref{thm:main}, that is, if Theorem~\ref{thm:infl} is applied with
$K=\frac{\pi^{2}}{2}d^{2}$, the result is
\[
   \Inf_j(f)\;\ge\;\frac{\sqrt{15}}{125}\cdot\frac{\Var(f)^{5/2}}{4d^{2}\,(\pi^{2}d^{2}/2)^{2}}
   \;=\;\frac{\sqrt{15}}{125\,\pi^{4}}\cdot\frac{\Var(f)^{5/2}}{d^{6}},
\]
and the first bound of Corollary~\ref{cor:AA01} becomes
\[
   \Inf_i[p]\;\ge\;\frac{32\sqrt{15}}{125\,\pi^{4}}\cdot\frac{\Var[p]^{5/2}}{d^{6}}.
\]
Thus the constants of Theorem~\ref{thm:main} and of Corollary~\ref{cor:AA01} are multiplied
by $4$.

(2) The inequality of Filmus, Hatami, Keller and Lifshitz \cite[Theorem~3.3]{FHKL}, which is
recalled in Remark~\ref{rem:ABD}, gives $\sum_j\abs{f(x)-f(x^{j})}\le2d^{2}$ at every point $x$,
and hence $L(f)\le2d^{2}$. If it is used in place of Corollary~\ref{cor:sens-point}, the constants
of Theorem~\ref{thm:main}, of Corollary~\ref{cor:AA01} and of (1) are multiplied by a further
factor $2$.
\end{remark}

\begin{remark}[The exponents]\label{rem:exponents}
The exponents arise as follows.

(1) In the integrated all-pairs inequality \eqref{eq:pairs-real}, the left-hand side is of the
third order in the differences $f(x)-f(y)$, and the right-hand side is of the first order in the
increments $f(x)-f(x^{j})$.

Of the three factors on the left, two come from the rows of $g_t$ at
$x$ and at $y$, which are the rows of $f$ divided by $\abs{f(x)-t}$ and by $\abs{f(y)-t}$, and
one comes from the length of the interval of the levels $t$ between $f(y)$ and $f(x)$. The two
rows also account for the factor $K^{2}$ in \eqref{eq:third-up}.

This leads to the inequality
$\E\abs{f-m}^{3}\le6K^{2}\max_kL_k(f)$ of \eqref{eq:third-abs}, and so to the exponents $\frac32$
and $2$ of Theorem~\ref{thm:incr}.

(2) Theorem~\ref{thm:incr} and the inequality $L_j(f)^{2}\le4\Inf_j(f)$ of \eqref{eq:L-Inf} give
a coordinate $j$ with $\Var(f)^{3}\le144\,K^{4}\,\Inf_j(f)$.

The proof of
Theorem~\ref{thm:infl} does not use \eqref{eq:L-Inf}. The coordinate $j$ found there is not
light, that is,
\[
   \Inf_j(f)\;\ge\;\frac{2\,\Var(f)}{5\,L(f)}\cdot L_j(f).
\]

The lower bound for $L_j(f)$ is of the order of $\Var(f)^{3/2}/K^{2}$. In the proof of
Theorem~\ref{thm:infl} it is not squared; it is multiplied by a number of the order of
$\Var(f)/L(f)$. The result is of the order of $\Var(f)^{5/2}/(L(f)K^{2})$.

(3) The resulting bound is the stronger one. Indeed, $K\ge\max f-\min f$ by
Remark~\ref{rem:rows-real}(2).

Further $\max f-\min f\ge2\Var(f)^{1/2}$: this follows from
\eqref{eq:var-shift} with $c=(\max f+\min f)/2$, because $\abs{f(x)-c}\le(\max f-\min f)/2$ for
every $x$. Hence
\[
   \frac{\sqrt{15}}{125}\cdot\frac{\Var(f)^{5/2}}{K^{3}}
   \;\ge\;\frac{2\sqrt{15}}{125}\cdot\frac{\Var(f)^{3}}{K^{4}}
   \;>\;\frac{\Var(f)^{3}}{144\,K^{4}} .
\]

(4) For a function $f\colon Q_n\to[-1,1]$ of degree at most $d$, the number $K$ can be taken of
the order of $d^{2}$, and $L(f)$ is at most of this order. This gives the exponents $\frac52$ and
$6$ of Theorem~\ref{thm:main}.
\end{remark}

\section{Remarks and open questions}\label{sec:remarks}

This section contains a comparison of the two-sided inequality (Theorem~\ref{thm:two}) with results on certificates and
decision trees (Section~\ref{subsec:cert}), examples which concern the exponent of the variance,
the dependence on the degree and the constant of the row bound
(Sections~\ref{subsec:degone}, \ref{subsec:maj} and~\ref{subsec:cheb}), and open questions
(Section~\ref{subsec:open}).

\subsection{Certificates and decision trees}\label{subsec:cert}

Let $g\colon Q_n\to\{0,1\}$ be non-constant. A \emph{certificate} of $g$ at $x$ is a set
$C\subseteq[n]$ such that $g(y)=g(x)$ for every $y\in Q_n$ with $y_i=x_i$ for all $i\in C$.

These
points $y$ are the points $x^{S}$ with $S\cap C=\emptyset$. So $C$ is a certificate at $x$ if and only
if every sensitive block of $g$ at $x$ meets $C$, that is, by \eqref{eq:row-bool}, if and only if the
indicator vector of $C$ is a row of $g$ at $x$; the cost of this row is $\abs{C}$.

Suppose that $g$
has a certificate of size at most $c_1$ at every point of $\cO$, and a certificate of size at most
$c_0$ at every point of $\cN$. For the corresponding rows the two averages in
Theorem~\ref{thm:two} are at least $\Pr[g=1]/c_1$ and $\Pr[g=0]/c_0$, and the theorem gives
\begin{equation}\label{eq:cert}
   \max_{j\in[n]}\I_j(g)\;\ge\;\frac{4\,\Pr[g=1]\,\Pr[g=0]}{c_0\,c_1}.
\end{equation}

For the function of Example~\ref{ex:and} one can take $c_1=2$ and $c_0=1$, and \eqref{eq:cert} is
the bound $\frac38$ found there.

We compare \eqref{eq:cert} with the theorem of O'Donnell, Saks, Schramm and Servedio \cite{OSSS}.

A
\emph{decision tree} is a rooted binary tree in which every inner node is labelled by a coordinate
$i\in[n]$, its two outgoing edges are labelled by $-1$ and $1$, every leaf is labelled by $0$ or
$1$, and the labels of the inner nodes on each path from the root to a leaf are distinct.

On the
input $x\in Q_n$ one starts at the root and, at an inner node labelled $i$, follows the edge labelled
$x_i$, until a leaf is reached; the labels of the inner nodes on this path are the coordinates
\emph{queried} on $x$. The tree \emph{computes} $g$ if for every $x$ the label of the leaf reached is
$g(x)$.

For a decision tree $T$ let $\delta_j(T)$ be the probability that the coordinate $j$ is
queried on a uniformly distributed input, and let $\Delta(T)=\sum_{j}\delta_j(T)$; this is the
average number of coordinates queried.

The \emph{depth} of $T$ is the largest number of coordinates
queried on an input, and $D(g)$ is the least depth of a decision tree that computes $g$.

In this
notation, Theorem~1.1 of \cite{OSSS}, for the uniform measure, is the first of the inequalities
\begin{equation}\label{eq:osss-tree}
   4\,\Pr[g=1]\,\Pr[g=0]\;\le\;\sum_{j=1}^{n}\delta_j(T)\,\I_j(g)\;\le\;\Delta(T)\,\max_{j\in[n]}\I_j(g),
\end{equation}
valid for every decision tree $T$ that computes $g$.

(In \cite{OSSS} the functions take the values
$\pm1$; the function $2g-1$ has variance $4\Pr[g=1]\Pr[g=0]$, and its influences in the sense of
\cite{OSSS} are the numbers $\I_j(g)$.)

Since $\Delta(T)$ is at most the depth of $T$, it follows
that
\begin{equation}\label{eq:osss}
   \max_{j\in[n]}\I_j(g)\;\ge\;\frac{4\,\Pr[g=1]\,\Pr[g=0]}{D(g)} ,
\end{equation}
which is the statement quoted in Section~\ref{subsec:earlier}.

The coordinates queried on an input
$x$ form a certificate of $g$ at $x$, because every $y$ that agrees with $x$ in these coordinates
follows the same path. Hence one can take $c_0=c_1=D(g)$ in \eqref{eq:cert}; the resulting bound is
weaker than \eqref{eq:osss} by the factor $D(g)$.

In the other direction, $D(g)\le c_0c_1$ by a
classical inequality, found independently by Blum and Impagliazzo \cite{BI}, by Hartmanis and
Hemachandra \cite{HH} and by Tardos \cite{Tar} (see \cite[p.~221]{Smy02} and
\cite[Section~6]{KT}), so that
\eqref{eq:cert} follows from
\eqref{eq:osss} as well.

The bounds \eqref{eq:cert} and \eqref{eq:osss} involve the largest size of a certificate and the
largest number of queried coordinates, whereas Theorem~\ref{thm:two} and \eqref{eq:osss-tree} involve
averages. This can make a large difference.

Let $n\ge2$, and let $g(x)=1$ if $x_1=1$ or
$x_2=\dots=x_n=1$, and $g(x)=0$ otherwise; then $\Pr[g=0]=\frac12-2^{-n}$.

Let $u=(-1,1,\dots,1)$.
Every certificate of $g$ at $u$ contains $\{2,\dots,n\}$, because $g(u)=1$ and $g(u^{i})=0$ for
$i\ge2$. Hence $c_1\ge n-1$ in \eqref{eq:cert}, and $D(g)\ge n-1$, because the coordinates queried on
$u$ form a certificate.

As $\Pr[g=1]\Pr[g=0]\le\frac14$, and $c_0\ge1$ because a certificate of a
non-constant function is not empty, the right-hand sides of \eqref{eq:cert} and \eqref{eq:osss} are
at most $1/(n-1)$.

On the other hand, at a point $x$ with $x_1=1$ the set $\{1\}$ is a certificate,
and at a point $x$ with $x_1=-1$ and $x_i=-1$, where $i\ge2$, the set $\{1,i\}$ is a certificate.
With the corresponding rows, and any row at $u$, the left-hand side of the inequality of
Theorem~\ref{thm:two} is at least
\[
   4\cdot\frac12\cdot\frac12\Bigl(\frac12-2^{-n}\Bigr)\;=\;\frac12-2^{-n}.
\]

The inequality
\eqref{eq:osss-tree} gives a bound of the same order for $g$. Indeed, let $T$ be the decision tree
which queries the coordinate $1$, stops with the value $1$ if $x_1=1$, and otherwise queries the
coordinates $2,3,\dots,n$ in this order, stopping with the value $0$ at the first $i$ with $x_i=-1$,
and with the value $1$ if there is no such $i$.

It computes $g$, and $\delta_1(T)=1$ and
$\delta_j(T)=2^{1-j}$ for $j\ge2$. So $\Delta(T)=2-2^{1-n}$, and \eqref{eq:osss-tree} gives
\[
   \max_j\I_j(g)\;\ge\;\frac{1-4^{1-n}}{2-2^{1-n}}.
\]
(In fact $\max_j\I_j(g)=\I_1(g)=1-2^{1-n}$.)

Averages of reciprocal costs, in place of the largest cost, are also essential in
Section~\ref{sec:level}. There the vectors $a^{x}/(f(x){-}r)^{3}$ are inserted into
Lemma~\ref{lem:weights}, from which Theorem~\ref{thm:two} was deduced.

These vectors have arbitrary non-negative
real entries, their costs $\norm{a^{x}}/(f(x)-r)^{3}$ admit no bound in terms of $K$ (they are
large at the points $x$ where $f(x)$ is close to $r$), and the conclusion of the lemma involves
only the sum of the reciprocals of these costs.

Neither of the two averages in Theorem~\ref{thm:two} can be replaced by the corresponding probability.

Let $w\ge1$ be an integer, let $m=2^{w}$ and $n=wm$, let $T_1,\dots,T_m$ be pairwise disjoint subsets of
$[n]$ with $w$ elements each, and let $g(x)=1$ if there is a $k$ such that $x_i=1$ for all $i\in T_k$,
and $g(x)=0$ otherwise; functions of this kind are the tribes functions of Ben-Or and Linial
\cite{BL} (see also \cite{OSSS}).

Then
$\Pr[g=0]=(1-1/m)^{m}$. By Bernoulli's inequality,
$(1-1/m)^{m/2}\ge\frac12$, and $(1{-}1/m)^{m}\le e^{-1}$ because $1-s\le e^{-s}$ for real $s$; hence
\[
   \frac14\;\le\;\Pr[g=0]\;<\;\frac12 .
\]

For a coordinate $j\in T_k$ we have $g(x)\ne g(x^{j})$ if and only if
$x_i=1$ for all $i\in T_k\setminus\{j\}$ and there is no $l\ne k$ such that $x_i=1$ for all $i\in T_l$;
so
\[
   \I_j(g)\;=\;2^{1-w}(1-1/m)^{m-1}\;\le\;2^{1-w}.
\]

At a point $x$ with $g(x)=1$, a set $T_k$ such that
$x_i=1$ for all $i\in T_k$ is a certificate, and its indicator vector is a row of cost $w$. With these
rows at the points of $\cO$,
\[
   4\,\E_x\Bigl[\frac{1_{\{g(x)=1\}}}{\norm{\lambda^{x}}}\Bigr]\,\Pr[g=0]
   \;=\;\frac{4}{w}\,\Pr[g=1]\,\Pr[g=0]\;\ge\;\frac{1}{2w}\;\ge\;\frac{2^{w}}{4w}\,\max_{j\in[n]}\I_j(g),
\]
and $2^{w}/(4w)$ is unbounded.

So the inequality of Theorem~\ref{thm:two} does not remain true if its
second average is replaced by $\Pr[g=0]$. The function $1-g$ has the same rows and the same numbers
$\I_j$ as $g$, so the same holds for the first average and $\Pr[g=1]$.

This is consistent with
Theorem~\ref{thm:two}: at a point $x$ with $g(x)=0$ the $m$ sets $\{i\in T_k:\ x_i=-1\}$ are pairwise
disjoint sensitive blocks of $g$ at $x$, so every row of $g$ at $x$ has cost at least $m=2^{w}$.

The same function shows that the conjectured inequality of Lovett and Zhang
\cite[Conjecture~37]{LZ}, which is recalled in Section~\ref{subsec:earlier}, is not true as stated.
In the coordinates $(1+x_i)/2$ used in \cite{LZ}, let $X=g^{-1}(1)$ and $Y=g^{-1}(0)$. If $T_k$
consists of ones at a point $x\in X$, then every $y\in Y$ has a zero in $T_k$, so the mean of
$1-z_i$ over $i\in T_k$ is at least $1/w$ for every $z$ in the convex hull of $Y$; and the average of
the $w$ points of $Y$ obtained from $x$ by setting the $j$th coordinate of every block of ones to $0$
($j=1,\dots,w$) is at $L_\infty$ distance $1/w$ from $x$. So every point of $X$ has $L_\infty$
distance exactly $1/w$ from the convex hull of $Y$. On the other hand, the influences of
\cite[Definition~36]{LZ} are $\Pr_{x\in X}[x^{i}\notin X]=\I_i(g)/(2\Pr[g=1])\le2^{-w}$ and
$\Pr_{y\in Y}[y^{i}\notin Y]=\I_i(g)/(2\Pr[g=0])=1/(2^{w}-1)$, which tend to $0$ exponentially in $w$.
With $X$ and $Y$ exchanged the distance is $2^{-w}$, so for this pair of sets the failure is
one-sided. The example concerns the conjectured inequality for arbitrary sets, not the level sets of
functions of low degree to which \cite[Claim~38]{LZ} applies it.

\subsection{The exponent of the variance}\label{subsec:degone}

Let $n\ge1$ and let $f=\hat f(\emptyset)+\sum_{i=1}^{n}\hat f(\{i\})\,x_i$ be a function of degree at
most $1$ on $Q_n$ with $\abs{f}\le1$. Then
\[
   f(x)-f(x^{j})=2\hat f(\{j\})\,x_j,\qquad\text{so that}\qquad\Inf_j(f)=\hat f(\{j\})^{2},
\]
and $\Var(f)=\sum_{i=1}^{n}\hat f(\{i\})^{2}$ by the orthonormality of the functions $\chi_T$.

Let $x$ be the point with $x_i=1$ if $\hat f(\{i\})\ge0$ and $x_i=-1$ otherwise. Then
\[
   f(x)=\hat f(\emptyset)+\sum_i\abs{\hat f(\{i\})}\le1\qquad\text{and}\qquad
   f(\bar x)=\hat f(\emptyset)-\sum_i\abs{\hat f(\{i\})}\ge-1,
\]
whence $\sum_i\abs{\hat f(\{i\})}\le1$.

Therefore
\[
   \Var(f)=\sum_{i=1}^{n}\hat f(\{i\})^{2}
   \;\le\;\max_{j}\abs{\hat f(\{j\})}\cdot\sum_{i=1}^{n}\abs{\hat f(\{i\})}
   \;\le\;\max_{j}\abs{\hat f(\{j\})},
\]
that is,
\[
   \max_{j\in[n]}\Inf_j(f)\;\ge\;\Var(f)^{2}.
\]

For $f=(x_1+\dots+x_n)/n$ there is equality: $\Var(f)=1/n$ and $\Inf_j(f)=1/n^{2}$ for every $j$.

Consequently, if $c>0$, $\alpha$ and $\beta$ are real numbers such that
$\max_j\Inf_j(f)\ge c\,\Var(f)^{\alpha}/d^{\beta}$ for all $n,d\ge1$ and all
$f\colon Q_n\to[-1,1]$ of degree at most $d$, then $\alpha\ge2$, because the case $d=1$ gives
$n^{-2}\ge c\,n^{-\alpha}$ for all $n$.

The same function shows what Theorems~\ref{thm:incr} and~\ref{thm:infl} lose. Let
$f=(x_1+\dots+x_n)/n$. The vector $(2/n,\dots,2/n)$ is a row of $f$ of cost $2$ at every point,
since $\abs{f(x)-f(x^{S})}\le2\abs{S}/n$.

These rows have the least possible cost, because every
row $a$ of $f$ at $x$ satisfies $a_i\ge\abs{f(x)-f(x^{i})}=2/n$ for all $i$. Moreover
$L_j(f)=2/n$ for every $j$, so that $L(f)=2$.

With $K=2$, Theorem~\ref{thm:incr} yields a coordinate $j$ with $L_j(f)\ge n^{-3/2}/24$, whereas
$L_j(f)=2/n$. Theorem~\ref{thm:infl} yields a coordinate $j$ with
$\Inf_j(f)\ge(\sqrt{15}/1000)\,n^{-5/2}$, whereas $\Inf_j(f)=n^{-2}$.

In both cases the bound is
smaller than the true value by a factor of the order of $n^{1/2}$. Since the rows have the least
possible cost, none of this loss is due to the rows: it occurs in the passage from the rows to the
increments and to the influences.

\subsection{The dependence on the degree}\label{subsec:maj}

There is no lower bound for $\max_j\Inf_j(f)$ in terms of $\Var(f)$ alone.

Let $n=2k+1$, and let
$f(x)\in\{-1,1\}$ be the sign of $x_1+\dots+x_n$; this sum is an odd integer, hence not zero. Then
$f(\bar x)=-f(x)$, so that $\E f=0$ and $\Var(f)=\E[f^{2}]=1$.

Fix $j$. We have $f(x)\ne f(x^{j})$ if
and only if exactly $k$ of the $2k$ coordinates $x_i$ with $i\ne j$ are equal to $1$, and in this
case $\bigl((f(x)-f(x^{j}))/2\bigr)^{2}=1$. Hence
\[
   \Inf_j(f)\;=\;p_k:=\binom{2k}{k}\,2^{-2k}\;\le\;\frac{1}{\sqrt{2k+1}}\;=\;\frac1{\sqrt n}
   \qquad(j\in[n]).
\]
The inequality holds for $k=0$, and it passes from $k$ to $k+1$, because
$p_{k+1}=p_k\cdot\frac{2k+1}{2k+2}$ and $(2k+1)(2k+3)\le(2k+2)^{2}$.

So the influences tend to $0$ as
$n\to\infty$, while the variance is equal to $1$.

\subsection{The constant in the row bound}\label{subsec:cheb}

Let $d\ge1$ and $n\ge1$, and let $f(x)=T_d\bigl((x_1+\dots+x_n)/n\bigr)$ for $x\in Q_n$. These are
the functions with which Filmus, Hatami, Keller and Lifshitz show that their inequality
$\sum_i\abs{f(x)-f(x^{i})}/2\le d^{2}$, which is recalled in Remark~\ref{rem:ABD}, is sharp
\cite[Section~3]{FHKL}.

By (C3),
$\abs{f}\le1$. By (C1), $f$ is a linear combination of the powers $s^{k}$, $0\le k\le d$, of the
function $s=(x_1+\dots+x_n)/n$, which has degree at most $1$; so $f$ has degree at most $d$.

Let $a$
be a row of $f$ at the point $\one=(1,\dots,1)\in Q_n$. For every $i\in[n]$,
\[
   a_i\;\ge\;\abs{f(\one)-f(\one^{i})}\;=\;\abs{T_d(1)-T_d(1-2/n)}\;=\;1-T_d(1-2/n),
\]
by (C3). Hence $\norm{a}\ge n\,\bigl(1-T_d(1-2/n)\bigr)$, and the right-hand side tends to
$2\,T_d'(1)=2d^{2}$ as $n\to\infty$, by (C4).

Let us call a real number $c$ \emph{admissible} if for
all $n\ge0$ and $d\ge0$, every function $Q_n\to[-1,1]$ of degree at most $d$ has, at every point, a
row of cost at most $c\,d^{2}$.

The example shows that every admissible number is at least $2$. The
number $\pi^{2}$ is admissible by Theorem~\ref{thm:row}, and $\pi^{2}/2$ is admissible by
Remark~\ref{rem:ABD}.

\subsection{Open questions}\label{subsec:open}

\emph{The exponents.} Theorem~\ref{thm:main} is a bound of the form
$c\,\Var(f)^{\alpha}/d^{\beta}$ with $\alpha=\frac52$ and $\beta=6$. By
Section~\ref{subsec:degone} the exponent $\alpha$ in such a bound is at least $2$, and by
Section~\ref{subsec:maj} the bound has to depend on $d$.

We do not know the best exponents. In
particular, we do not know whether there is a bound of this form with $\alpha=2$.

The corresponding question for functions with rows is the following. Is there a constant $c>0$
such that every non-constant function $f\colon Q_n\to\R$ which has a row of cost at most $K$ at
every point has a coordinate $j$ with $\Inf_j(f)\ge c\,\Var(f)^{2}/K^{2}$?

By
Theorem~\ref{thm:row}, a positive answer would give a bound of the above form with $\alpha=2$ and
$\beta=4$.

For Boolean functions the answer is positive. Indeed, let $g\colon Q_n\to\{0,1\}$ be non-constant,
with a row of cost at most $K$ at every point. The two averages in Theorem~\ref{thm:two} are at
least $\Pr[g=1]/K$ and $\Pr[g=0]/K$. Since $\I_j(g)=4\Inf_j(g)$ and $\Var(g)=\Pr[g=1]\Pr[g=0]$, the
theorem gives a coordinate $j$ with $\Inf_j(g)\ge\Var(g)/K^{2}$, and
$\Var(g)\ge4\Var(g)^{2}$ because $\Var(g)\le\frac14$.

\medskip
\emph{The bijection.} The proof of Theorem~\ref{thm:bij} obtains the bijection $\tau$ from a term of
a determinant that does not vanish. It would be interesting to have a direct combinatorial
construction of a bijection with the properties (a) and (b).

\medskip
\emph{The row bound.} Let $c^{*}$ be the infimum of the admissible numbers of
Section~\ref{subsec:cheb}, so that $2\le c^{*}\le\pi^{2}/2$. What is the value of $c^{*}$?

\appendix

\section{Timeline of the discovery}\label{app:timeline}

In the run that produced the proof, before any statement or method in the table below was first
proposed, various instances of the model brought in
from the literature Reimer's inequality and his Butterfly Lemma \cite{Rei}, and the dual inequality of
Kahn, Saks and Smyth \cite{KSS}, and set as a goal a version of these inequalities for fractional
certificates.
The table gives, for the main statements of the paper, the order in which the statement or the method
was first proposed and the order in which a proof can best be identified, retrospectively, as first
written down; proofs written down together share a number.
\begin{center}
\small
\begin{tabular}{lcc}
\hline
 & order proposed & order proved\\
\hline
Two-sided inequality (Theorem~\ref{thm:two-intro}) & 1 & 3\\
Integration over all levels (Section~\ref{sec:level}) & 2 & 6\\
Antipodal inequality (Theorem~\ref{thm:anti}) & 3 & 1\\
All-pairs inequality (Proposition~\ref{prop:pairs}) & 4 & 3\\
Determinant inequality (Corollary~\ref{cor:det}) & 5 & 1\\
Bijection theorem (Theorem~\ref{thm:bij}) & 6 & 3\\
Averaging over coordinates (Section~\ref{sec:avg}) & 7 & 7\\
\hline
\end{tabular}
\end{center}

The first complete proof was written down during the third day of the run, together with the proofs of the two-sided
inequality, the all-pairs inequality and the bijection theorem, and then checked in the Lean proof
assistant. It gave the bound $\Var(f)^{8}/(2^{23}\pi^{12}d^{12})$ in place of the
bound of Theorem~\ref{thm:main}, that is, the exponent $22$ in place of $7$ in
Corollary~\ref{cor:AA01}. After the instances of the model were asked to push on open problems using
the intermediate results of the proof, they also improved the exponent of $22$ by polishing these
intermediate results: the integration over all levels gave the exponent $11$, checked in Lean, and then the averaging over coordinates,
together with the bound $L(f)\le4d^{2}$ of Corollary~\ref{cor:sens-point}, gave the exponent $7$,
checked in Lean.

\end{document}